%% file: ArXiv.tex
\documentclass{article}

\usepackage{amsmath,amssymb,amsthm,amsfonts}
\usepackage{graphicx,colortbl,geometry,authblk}
\usepackage{enumitem,dblfloatfix}
\usepackage{longtable,multirow,multicol}
\usepackage{pifont,enotez,booktabs,footmisc}
\usepackage[toc,page]{appendix}
\PassOptionsToPackage{hyphens}{url}\usepackage{hyperref}

\newcommand{\tuples}{\mathcal{T}}
\newcommand{\datasets}{\mathcal{D}}
\newcommand{\outputs}{\mathcal{O}}
\newcommand{\knowledges}{\mathcal{B}}
\newcommand{\project}{\mathcal{P}}
\newcommand{\proba}[1]{\mathbb{P}\left[#1\right]}
\newcommand{\probas}[2]{\mathbb{P}_{#1}\left[#2\right]}
\newcommand{\probac}[2]{\mathbb{P}\left[#1\;\middle|\;#2\right]}
\newcommand{\probasc}[3]{\mathbb{P}_{#1}\left[#2\;\middle|\;#3\right]}
\newcommand{\ind}{\approx}

\newcommand{\expects}[2]{\mathbb{E}_{#1}\left[#2\right]}

\newcommand{\cond}[1]{_{\vert{}#1}}

\newcommand{\mecha}{\mathcal{M}}
\newcommand{\mech}[1]{\mecha\left(#1\right)}
\newcommand{\mechone}[1]{\mecha_1\left(#1\right)}
\newcommand{\mechtwo}[1]{\mecha_2\left(#1\right)}
\newcommand{\negli}{\textsf{neg}}
\newcommand{\negl}[1]{\negli\left(#1\right)}
\newcommand{\Simu}{\textsf{Sim}}
\newcommand{\Sim}[1]{\Simu\left(#1\right)}
\newcommand{\Agg}{\textsf{Agg}}
\newcommand{\aggr}{\textsf{agg}}
\newcommand{\agg}[1]{\aggr\left(#1\right)}
\newcommand{\priv}{\textsf{priv}}
\newcommand{\pri}[1]{\priv\left(#1\right)}

\newcommand{\eps}{\varepsilon}
\newcommand{\del}{\delta}
\newcommand{\epsdel}{\left(\eps,\del\right)}
\newcommand{\PrivLoss}{{\mathcal{L}}}
\newcommand{\privloss}[2]{{\PrivLoss_{#1/#2}}}
\newcommand{\privlossMit}[2]{\PrivLoss^{\mecha,\theta}_{i\leftarrow#1/i\leftarrow#2}}
\newcommand{\Theps}{\left(\Theta,\eps\right)}
\newcommand{\Thepsdel}{\left(\Theta,\eps,\del\right)}
\newcommand{\hatB}{\hat{B}}
\newcommand{\Aggeps}{\left(\Agg,\eps\right)}
\newcommand{\indeps}{\ind_{\eps}}
\newcommand{\indepsdel}{\ind_{\eps,\del}}

\newcommand{\yes}{\ensuremath{\checkmark}}
\newcommand{\no}{\ensuremath{\text{\ding{55}}}}
\newcommand{\uk}{?}
\newcommand{\Def}{\mathrm{Def}}
\newcommand{\Defone}{\Def_1}
\newcommand{\Deftwo}{\Def_2}
\newcommand{\strongerthan}{\ensuremath{\succ}}
\newcommand{\weakerthan}{\ensuremath{\prec}}
\newcommand{\same}{\ensuremath{\sim}}
\newcommand{\extendedby}{\ensuremath{\subset}}
\newcommand{\extends}{\ensuremath{\supset}}

\newcommand{\extendsweaker}{\ensuremath{\supset^\prec}}
\newcommand{\extendssame}{\ensuremath{\supset^\sim}}
\newcommand{\specialstronger}{\ensuremath{\subset^\succ}}
\newcommand{\specialweaker}{\ensuremath{\subset^\prec}}
\newcommand{\specialsame}{\ensuremath{\subset^\sim}}

\theoremstyle{plain}

\newtheorem*{theorem*}{Theorem}

\newtheorem*{lemma*}{Lemma}

\newtheorem{definition}{Definition}
\newtheorem*{definition*}{Definition}
\newtheorem{prop}{Proposition}

\author[1]{Damien Desfontaines}
\author[2]{Bal\'azs Pej\'o}

\affil[1]{Tumult Labs \authorcr \texttt{damien@desfontain.es}}
\affil[2]{CrySyS Lab \authorcr \texttt{pejo@crysys.hu}}

\title{SoK\@: Differential Privacies \\
    \Large A taxonomy of differential privacy variants and extensions}

\date{}

\begin{document}

\maketitle

\begin{abstract}
    Shortly after it was first introduced in 2006, \emph{differential privacy}
    became the flagship data privacy definition. Since then, numerous variants and
    extensions were proposed to adapt it to different scenarios and attacker models.
    In this work, we propose a systematic taxonomy of these variants and extensions.
    We list all data privacy definitions based on differential privacy, and
    partition them into seven categories, depending on which aspect of the original
    definition is modified. \\
    These categories act like dimensions: variants from the same category cannot
    be combined, but variants from different categories can be combined to form
    new definitions. We also establish a partial ordering of relative strength
    between these notions by summarizing existing results. Furthermore, we list
    which of these definitions satisfy some desirable properties, like
    composition, post-processing, and convexity by either providing a novel
    proof or collecting existing ones.
\end{abstract}

\tableofcontents
%\listoftodos[TODOs]

\input{Introduction}
\input{Preliminaries}
\input{Quantification}
\input{Neighborhood}
\input{Variation}
\input{Background}
\input{Formalism}
\input{Relativization}
\input{Computational}
\input{Summary}
\input{Related}
\input{Conclusion}

\bibliographystyle{alpha}
\bibliography{%
	Intro,%
	Preliminary,%
	Quantify,%
	Neighbour,%
	Vary,%
	Background,%
	Formalism,%
	Relative,%
	Compute,%
	Related}

\end{document}

%% file: Introduction.tex
\section{Introduction}\label{sec:intro}

%Privacy is a fundamental problem in today's information-based era. How can this
%social concept of privacy be translated into a formal definition? In the
%scientific literature, many different approaches have been proposed, either to
%measure privacy or to establish a criterion at which some data or process is
%considered to be private~\cite{wagner2018technical}.

What does it mean for data to be anonymized? Samarati and Sweeney
discovered that removing explicit identifiers from dataset records was not
enough to prevent information from being
re-identified~\cite{samarati2001protecting,sweeney2002kanonymity}, and they
proposed the first definition of anonymization. This notion, called
\emph{$k$-anonymity}, is a property of a dataset: each combination of
re-identifying fields must be present at least $k$ times. In the following
decade, further research showed that sensitive information about individuals
could still be leaked when releasing $k$-anonymous datasets, and many variants
and definitions were proposed, such as 
\emph{$l$-diversity}~\cite{machanavajjhala2006ldiversity}, 
\emph{$t$-closeness}~\cite{li2007tcloseness}, and 
\emph{$n$-confusion}~\cite{stokes2012n}.

A common shortcoming of these approaches is that they defined anonymity as a
property of the \emph{dataset}: without knowing how the dataset is generated,
arbitrary information can be leaked. This approach was changed with the
introduction of \emph{differential
privacy}~\cite{dwork2005differential,dwork2006differential} (DP): rather than
being a property of the sanitized dataset, anonymity was instead defined as a
property of the \emph{process}. It was inspired by Dalenius' privacy goal that
``Anything about an individual that can be learned from the dataset can also be
learned without access to the dataset''~\cite{dalenius1977towards}, a goal
similar to one already used in probabilistic
encryption~\cite{shafi1984probabilistic}.

Thanks to its useful properties, DP quickly became the
flagship of data privacy definitions. Many algorithms and statistical processes
were changed to satisfy DP and were adopted by organizations
like Apple \cite{apple2017dp},
Facebook \cite{messing2020facebook} \cite{herdagalen2020protecting},
Google \cite{aktay2020google} \cite{bavadekar2020google} \cite{bavadekar2021google} \cite{wilson2020differentially},
LinkedIn \cite{rogers2020members} \cite{rogers2020linkedin},
Microsoft \cite{ding2017collecting},
OhmConnect \cite{pare2020applying},
and the US Census Bureau \cite{abowd2010formal} \cite{garfinkel2018issues} \cite{foote2019releasing}.

Since the original introduction of differential privacy, many variants and
extensions have been proposed to adapt it to different contexts or assumptions.
These new definitions enable practitioners to get privacy guarantees, even in
cases that the original DP definition does not cover well. This happens in a
variety of scenarios: the noise mandated by DP can be too large and force the
data custodian to consider a weaker alternative, the risk model might be
inappropriate for certain use cases, or the context might require the data owner
to make stronger statements on what information the privacy mechanism can
reveal.

\begin{figure}[h!]
    \centering
    \includegraphics[width=8cm]{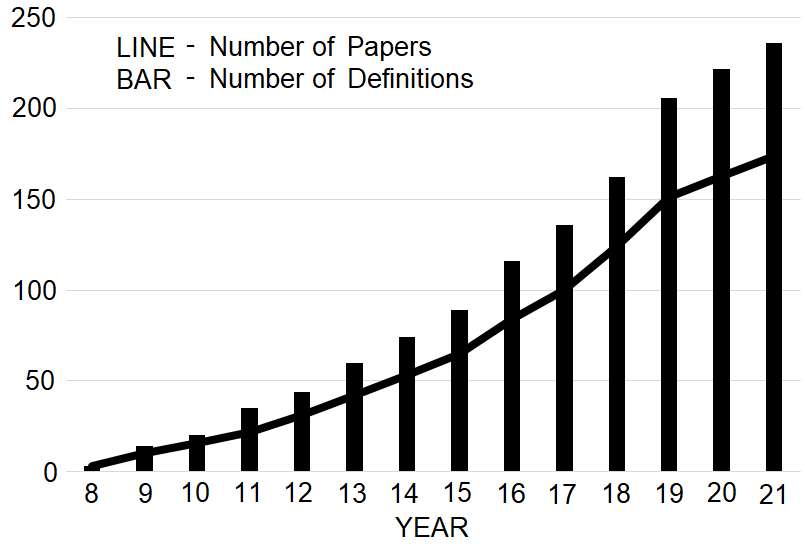}
    \caption{Accumulated number of papers which are introducing new DP notions
    	(line) and the exact number of these definitions
    	(bar) for the years from 2008 till 2021.}\label{fig:timeline}
\end{figure}

Figure~\ref{fig:timeline} shows the prevalence of this phenomenon: approximately
\emph{225} different notions\footnote{We count all the definitions which are
presented as ``new'' in the papers introducing them.}, inspired by DP, were
defined in the last years. As we show in Figure~\ref{fig:timeline}, this
phenomenon does not seem to slow down over time. These definitions can be
\emph{extensions} or \emph{variants} of DP\@. An extension encompasses the
original DP notion as a special case, while a variant changes some aspect,
typically to weaken or strengthen the original definition. 

With so many definitions, it is difficult for new practitioners to get an
overview of this research area. Many definitions have similar goals, so it is
also challenging to understand which are appropriate to use in which context.
These difficulties also affect experts: a number of definitions listed in this
work have been defined independently multiple times (often with identical
meaning but different names, or identical names but different meanings).
Finallly, variants are often introduced without a comparison to related notions.

This systematization of knowledge attempts to solve these problems. It is a
taxonomy of variants and extensions of DP, providing short explanations of the
intuition, use cases and basic properties of each. By categorizing these
definitions, we attempt to simplify the understanding of existing variants and
extensions, and of the relations between them. We hope to make it easier for new
practitioners to understand whether their use case needs an alternative
definition, and if so, which existing notions are the most appropriate, and what
their basic properties are. 

\subsection*{Contributions and organization}

We systematize the scientific literature on variants and extensions of
differential privacy, and propose a unified and comprehensive taxonomy of these
definitions. We define seven \emph{dimensions}: these are ways in which the
original definition of DP can be modified or extended. We list variants and
extensions that belong to each dimension, and we highlight representative
definitions for each. Whenever possible, we compare these definitions and
establish a partial ordering between the strengths of different notions.
Furthermore, for each definition, we specify whether it satisfies Kifer et al.'s
\emph{privacy axioms}~\cite{kifer2010towards,kifer2012axiomatic},
(post-processing and convexity), and whether they are composable.

Our survey is organized as follows:
\begin{itemize}
  \item In Section~\ref{sec:pre}, we recall the original definition of DP and
    introduce our dimensions along which DP can be modified. Moreover, we
    present the basic properties of DP, and define how definitions can related to
    each other.
  \item In the following 7 sections (Sections~\ref{sec:q} to~\ref{sec:c}), we
    introduce our dimensions, and list and compare the corresponding definitions.
  \item In Section~\ref{sec:sum}, we summarize the results from the previous
    sections into a table, showing the corresponding properties with proofs, 
    and list the known relations.
  \item In Section~\ref{sec:related} we detail the methodology and scope of this work
    and review the related literature.
  \item Finally, in~\ref{sec:conc} we conclude this work.
\end{itemize}

%% file: Preliminaries.tex
\section{Differential Privacy}\label{sec:pre}

In this chapter we recap the original DP definition with its 
basic properties, define how definitions can related to each other and
introduce our dimensions along which DP can be modified. 

Let $\tuples$ denote an arbitrary set of possible \emph{records}. We typically
use $t$ to denote the records themselves. A \emph{dataset} is a finite indexed
family of records. We denote by $\datasets$ space of possible datasets,
individuals datasets are typically called $D$, $D'$, $D_1$ or $D_2$. The indices
of a dataset are typically called $i$ and $j$, with $D(i)$ referring to the
$i$-th record of a dataset $D$. We denote by $D_{-i}$ the dataset $D$ whose
$i$-th record has been removed.

Let $\outputs$ denote an arbitrary set of possible \emph{outputs}; outputs are
typically called $O$, and sets of outputs called $S$. A \emph{mechanism} is a
randomized function which takes a dataset as input and returns an output.
Mechanisms are typically called $\mecha$ while $\mech{D}$ is usually a random variable.

Probability distributions on $\tuples$ are called $\pi$, probability
distribution on $\datasets$ are called $\theta$, and family of probability
distributions on $\datasets$ are called $\Theta$. Given some property $\phi$,
let $\mech{D}\cond{D\sim\theta,\phi}$ denote the random variable corresponding
to the output of $\mech{D}$, when $D$ is drawn from a distribution $\theta$
conditioned on $\phi$.

Table~\ref{tab:notions} summarizes the notations used throughout the paper.

{\footnotesize
\begin{table}[h!]
    \centering
    \begin{tabular}{ll}
        Notation & Description \\
        \midrule
        $\tuples$ & Set of possible records\\
        $t\in\tuples$ & A possible record\\
        $\datasets=\tuples^*$ & Set of possible datasets (sequences of records)\\
        $D\in\datasets$ & Dataset (we also use $D', D_1$, $D_2$, \ldots) \\
        $D(i)$ & $i$-th record of the dataset ($i\le|D|$) \\
        $D_{-i}$ & Dataset $D$, with its $i$-th record removed\\
        \midrule
        $\outputs$ & Set of possible outputs of privacy mechanisms \\
        $S\subseteq\outputs$ & Subset of possible outputs \\
        $O\in\outputs$ & Output of the privacy mechanism \\
        $\mecha:\datasets\rightarrow\outputs$ & Privacy mechanism (probabilistic) \\
        $\mech{D}$ & The distribution (or an instance thereof) of the outputs of $\mecha$ given input $D$ \\
        \midrule
        $d_\datasets:\datasets\times\datasets\rightarrow\mathbb{R}^+_0$ & Distance function between datasets\\
        \midrule
        $\phi\subseteq\datasets$ & Predicate on datasets \\
        $\Phi$ & Family of sensitive predicates on datasets \\
        \midrule
        $\pi$ & Probability distribution on $\mathcal{T}$ \\
        $\knowledges$ & Set of possible background knowledges \\
        $B\in\knowledges$ & Background knowledge (we also use $\hatB$) \\
        $\Theta$ & Family of probability distributions on $\datasets$, on $\datasets\times\knowledges$ \\
        $\theta\in\Theta$ & Probability distribution on $\datasets$, on $\datasets\times\knowledges$ \\
        %$\belief$ & Probability distribution on $\datasets$ according to the adversary's prior belief\\
        %$\hat{\belief}$ & Family of probability distribution on $\datasets$ according to the adversary's prior belief \\
        $\mech{D}\cond{D\sim\theta,\phi}$ & Distribution of outputs of $\mecha$ given an input drawn from $\theta$, conditioned on $\phi$ \\
        \midrule
        $\Omega$ & Probabilistic polynomial-time Turing machine, called \emph{distinguisher} \\
        \bottomrule
    \end{tabular}
    \caption{Notations used in this paper.}\label{tab:notions}
\end{table}}

\subsection{The original version}

The first DP mechanism, randomized response, was proposed in
1965~\cite{warner1965randomized}, and data privacy definitions that are a
property of a mechanism and not of the output dataset were already proposed in
as early as 2003~\cite{evfimievski2003limiting}.
However, DP and the related notion of
\emph{$\eps$-indistinguishability} were first formally defined in
2006~\cite{dwork2006differential,dwork2006calibrating,dwork2005differential}.

\begin{definition}[$\eps$-indistinguishability~\cite{dwork2006calibrating}]
    Two random variables $A$ and $B$ are $\eps$-indistinguishable, denoted
    $A\indeps B$, if for all measurable sets $X$ of possible events:
    \begin{equation*}
        \proba{A\in X} \leq e^{\eps}\cdot\proba{B\in X}
        \text{ and }
        \proba{B\in X} \leq e^{\eps}\cdot\proba{A\in X}.
    \end{equation*}
\end{definition}

Informally, $A$ and $B$ are $\eps$-indistinguishable if their distributions are
``close''. This notion originates from the cryptographic notion of
indistinguishability~\cite{goldwasser1984probabilistic}. A similar notion,
\emph{$(1,\eps)$-privacy}, was defined in~\cite{chaudhuri2006random}, where
$(1+\eps)$ used in place of $e^{\eps}$, and it was also called \emph{log-ratio
distance} in~\cite{haitner2019channels}.

The notion of $\eps$-indistinguishability is then used to define differential
privacy.
\begin{definition}[$\eps$-differential privacy~\cite{dwork2006differential}]\label{def:DP}
    A privacy mechanism $\mecha$ is $\eps$-differential private (or $\eps$-DP)
    if for all datasets $D_1$ and $D_2$ that differ only in one record,
    $\mech{D_1}\ind_\eps\mech{D_2}$.
\end{definition}

\subsubsection*{Mechanisms}\label{sec:mech}

Besides the random response mechanism (which returns the true value with 
probability $p$ and returns a random value otherwise) in general there are 
three places where noise can be injected to guarantee DP 
\cite{mireshghallah2020privacy}: it can be added to the input, to the output, 
and directly to the mechanism. For instance, in machine learning context input 
(e.g., \cite{papernot2016semi}) and output (e.g., \cite{pejo2019together})
pertubation are equivalent with sanitizing the dataset before and the 
predictions after the training respectively. Concerning mechanism pertubation, 
there are various techniques, such as loss function perturbation (e.g., 
\cite{chaudhuri2011differentially}), and gradient pertubation (e.g., 
\cite{abadi2016deep}), which insert noise to the model objective and 
update respectively. 

The most widely used distributions the noise is sampled from are Laplace, Gauss, and Exponential. Using the first and last makes any underlying mechanism to satisfy $\eps$-DP for continious and discrete cases, however, they could result in large added noise. On the contrary, the middle distribution decrease the probability of such events significantly, but the obtained differential privacy guarantee is weaker (i.e. $\epsdel$-DP, introduced in Section \ref{sec:q}. 

\subsection{Dimensions}\label{sec:dim}

Variants and extensions of differential privacy modify the original definition
in various ways. To establish a comprehensive taxonomy, a natural approach is to
partition them into \emph{categories}, depending on which aspect of the
definition they change. Unfortunately, this approach fails for privacy
definitions, many of which modify several aspects at once, so it is impossible
to have a categorization such that every definition falls neatly into only one
category.

The approach we take is to define \emph{dimensions} along which the original
definition can be modified. Each variant or extension of DP can be seen as a
point in a multidimensional space, where each coordinate corresponds to one
possible way of changing the definition along a particular dimension. To make
this representation possible, our dimensions need to satisfy two properties:

\begin{itemize}
    \item \textbf{Mutual compatibility}: definitions that vary along different
      dimensions can be combined to form a new, meaningful definition.
    \item \textbf{Inner exclusivity}: definitions in the same dimension cannot
      be combined to form a new, meaningful definition (but they can be pairwise
      comparable).
\end{itemize}

In addition, each dimension should be \emph{motivatable}: there should be an
intuitive explanation of what it means to modify DP along each dimension.
Moreover, each possible choice within a dimension should be similarly
understandable, to allow new practitioners to determine quickly which kind of
definition they should use or study, depending on their use case.

We introduce our dimensions by reformulating the guarantee offered by DP,
highlighting aspects that have been modified by its variants or extensions. Each
dimension is attributed a letter, and we note the dimension letter corresponding
to each highlight. This formulation considers the point of view of an attacker,
trying to find out some sensitive information about some input data using the
output of a mechanism.

\begin{center}
    An attacker with \textbf{perfect background knowledge} (\textbf{B}) and
    \textbf{unbounded computation power} (\textbf{C}) is \textbf{unable}
    (\textbf{R}) to \textbf{distinguish} (\textbf{F}) \textbf{anything about an
    individual} (\textbf{N}), \textbf{uniformly across users} (\textbf{V})  even
    in the \textbf{worst-case scenario} (\textbf{Q}).
\end{center}

This informal definition of DP with the seven highlighted aspects give us seven
distinct dimensions. We denote each one by a letter and summarize them in
Table~\ref{tab:dimensions_list}. Each is introduced in its corresponding
section.

\begin{table}[h!]
    \centering
    \makebox[\textwidth][c]{
        \begin{tabular}{lll}
            Dimension & Description & Typical motivations \\
            \midrule
            \midrule
            \textbf{Q}uantification of Privacy Loss
            & How is the privacy loss & Averaging risk, obtaining \\
            & quantified across outputs? & better composition properties \\
            \midrule
            \textbf{N}eighborhood Definition
            & Which properties are protected & Protecting specific values \\
            & from the attacker? & or multiple individuals\\
            \midrule
            \textbf{V}ariation of Privacy Loss
            & Can the privacy loss vary & Modeling users with different \\
            & across inputs? & privacy requirements \\
            \midrule
            \textbf{B}ackground Knowledge
            & How much prior knowledge & Using mechanisms that add \\
            & does the attacker have? & less or no noise to data \\
            \midrule
            \textbf{F}ormalism change
            & Which formalism is used to describe & Exploring other intuitive \\
            & the attacker's knowledge gain? & notions of privacy \\
            \midrule
            \textbf{R}elativization of Knowledge Gain
            & What is the knowledge gain & Guaranteeing privacy for \\
            & relative to? & correlated data \\
            \midrule
            \textbf{C}omputational Power
            & How much computational & Combining cryptography \\
            & power can the attacker use? & techniques with DP \\
            \bottomrule
        \end{tabular}
    }
    \caption{The seven dimensions and their typical motivation.}\label{tab:dimensions_list}
\end{table}

Note that the interpretation of DP is subject to some debate.
In~\cite{tschantz2017differential}, authors summarize this debate, and show that
DP can be interpreted under two possible lenses: it can be seen as an
\emph{associative} property, or as a \emph{causal} property. The difference
between the two interpretations is particularly clear when one supposes that the
input dataset is modeled as being generated by a probability distribution.

\begin{itemize}
\item In the associative view, this probability distribution is
  \emph{conditioned} upon the value of one record. If the distribution has
  correlations, this change can affect other records as well.
\item In the causal view, the dataset is first generated, and the value of 
	one record is then \emph{changed} before computing the result of the mechanism.
\end{itemize}

While the causal view does not require any additional 
assumption to capture the intuition behind DP, the associative view
requires that either all records are independent in
the original probability distribution (the \emph{independence assumption}),
or the adversary must know all data points except one (the \emph{strong
adversary assumption}, which we picked in the reformulation above).

These considerations can have a significant impact on DP
variants and extensions, either leading to distinct variants that attempt to
capture the same intuition, or to the same variant being interpreted in
different ways.

\subsection{Properties}\label{sec:prop}

In this section, we introduce three main properties of differential privacy,
that we then check against variants and extensions of DP listed in this work.

\subsubsection*{Privacy Axioms}

Two important properties of data privacy notions are called \emph{privacy
axioms}, proposed in~\cite{kifer2010towards,kifer2012axiomatic}. These are not
axioms in a sense that they assumed to be true; rather, they are consistency
checks: properties that, if not satisfied by a data privacy definition, indicate
a flaw in the definition\footnote{The necessity of these were questioned
in~\cite{huber2013defining}, where the authors showed a natural notions of
anonymity that contradict them.}.

\begin{definition}[Privacy axioms~\cite{kifer2010towards,kifer2012axiomatic}]\hspace{0cm}\\
    \begin{enumerate}
      \item \textbf{Post-processing}\footnote{This definition must be slightly
        adapted for some variants, see for example Proposition~\ref{prop:a-comp}
        in Section \ref{sec:sum}.} (or \emph{transformation invariance}): A
        privacy definition $\Def$ satisfies the post-processing axiom if, for
        any mechanism $\mecha$ satisfying $\Def$ and any probabilistic function
        $f$, the mechanism $D\rightarrow f(\mech{D})$ also satisfies $\Def$.
      \item \textbf{Convexity} (or \emph{privacy axiom of choice}): A privacy
        definition $\Def$ satisfies the convexity axiom if, for any two
        mechanisms $\mecha_1$ and $\mecha_2$ satisfying $\Def$, the mechanism
        $\mecha$ defined by $\mech{D}=\mecha_1(D)$ with probability $p$
        and $\mech{D}=\mecha_2(D)$ with probability $1-p$ also satisfies $\Def$.
    \end{enumerate}
\end{definition}

Most differential privacy variants and extensions, including the original
definition of DP, satisfy these axioms, although some do not. We highlighted
these in Table~\ref{tab:DP_variants} in Section~\ref{sec:sum}.

\subsubsection*{Composition}

A third important property is one of differential privacy's main strengths:
\emph{composability}. It guarantees that the output of two mechanisms satisfying
a privacy definition still satisfies the definition, typically with a change in
parameters. There are several types of composition: \emph{parallel composition},
\emph{sequential composition}, and \emph{adaptive composition}. We introduce
the first two below.

\begin{theorem*}[Parallel composition~\cite{dwork2006differential}]
   Let $\mecha_1$ be a $\eps_1$-differentially private mechanism, and
   $\mecha_2$ a $\eps_2$-differentially private mechanism. For any dataset $D$, let
   $D_1$ and $D_2$ be the result of an operation that separates records in two
   disjoint datasets. Then the mechanism $\mecha$ defined by
   $\mech{D}=(\mecha_1(D_1),\mecha_2(D_2))$ is
   $\max(\eps_1,\eps_2)$-differentially private.
\end{theorem*}

This property allows us to build \emph{locally differentially private}
mechanisms, in which a central server can compute global statistics without
accessing the raw data from each user. In this work, we focus on sequential
composition, which we simply call \emph{composition}.

\begin{theorem*}[Sequential composition~\cite{dwork2006differential}]
  Let $\mecha_1$ be a $\eps_1$-differentially private mechanism, and $\mecha_2$
  a $\eps_2$-differentially private mechanism. Then the mechanism $\mecha$
  defined by $\mech{D}=\left(\mecha_1(D),\mecha_2(D)\right)$ is
  $(\eps_1+\eps_2)$-differentially private.
\end{theorem*}

This theorem stays true if $\mecha_2$ depends on the value of $\mecha_1(D)$:
this variant is called \emph{adaptative composition}. This latter property
allows to quantify the gain of information over time of an attacker interacting
with a differentially private query engine.

In this work, we only consider sequential composition, in the more abstract form
formalized below.

\begin{definition}[Composability]
    A privacy definition $\Def$ with parameter $\alpha$ is \emph{composable} if
    for any two mechanisms $\mecha_1$ and $\mecha_2$ satisfying respectively
    $\alpha_1$-$\Def$ and $\alpha_2$-$\Def$, the mechanism
    $\mech{D}=\left(\mecha_1(D),\mecha_2(D)\right)$ satisfies $\alpha$-$\Def$
    for some (non-trivial) $\alpha$.
\end{definition}

\subsection{Relations between definitions}

When learning about a new data privacy notion, it is often useful to know what are
the known relations between this notion and other definitions. However,
definitions have parameters that often have different meanings, and whose value
is not directly comparable. To capture extensions, when a definition can be seen 
as a special case of another, we introduce the following definition.

\begin{definition}[Extensions]
	Let $\alpha\text{-}\Defone$ and $\beta\text{-}\Deftwo$ be data privacy
	definitions. We say that $\Defone$ is \emph{extended by} $\Deftwo$, and
	denote is as $\Defone\extendedby\Deftwo$, if for all $\alpha$, there is a
	value of $\beta$ such that $\alpha$-$\Defone$ is identical to
	$\beta$-$\Deftwo$.
\end{definition}

Concerning variants, to claim that a definition is stronger than another,
we adopt the concept of ordering established in~\cite{cuff2016differential}
using $\alpha$ and $\beta$ as tuples, encoding multiple parameters. Note that we slightly changed the original definition as that only required the 
second condition to hold, which would classify any extension as a stronger 
variant.

\begin{definition}[Relative strength of privacy definitions]
    Let $\alpha\text{-}\Defone$ and $\beta\text{-}\Deftwo$ be data privacy
    definitions. We say that $\Defone$ is \emph{stronger than} $\Deftwo$, and
    denote it $\Defone\strongerthan\Deftwo$, if:
    \begin{enumerate}
      \item for all $\alpha$, there is a $\beta$ such that
        $\alpha\text{-}\Defone\implies\beta\text{-}\Deftwo$;
	  \item for all $\beta$, there is an $\alpha$ such that
        $\alpha\text{-}\Defone\implies\beta\text{-}\Deftwo$.
    \end{enumerate}
    If $\Defone$ is both stronger than and weaker than $\Deftwo$, we say that
    the two definitions are \emph{equivalent}, and denote it
    $\Defone\sim\Deftwo$.
\end{definition}

Relative strength implies a partial ordering on the space of possible
definitions. On the other hand, if two definitions are equivalent, this does not
mean that they are equal: they could be only equal up to a change in parameters. 
Both relations are reflexive and transitive; and we define the
symmetric counterpart of these relations as well (i.e., $\weakerthan$ and
$\extends$). Moreover, for brevity, we combine these two concepts in a single
notation: if $\Defone\extendedby\Deftwo$ and $\Defone\strongerthan\Deftwo$, we
say that $\Deftwo$ is a weaker extension of $\Defone$, and denote it
$\Defone\specialstronger\Deftwo$.

A summarizing
table is presented at the end of this work, where for each definition, we also
highlight its dimensions and its relation to other notions. In
Table~\ref{tab:DP_variants}, we also
specify whether these notions satisfy the privacy axioms and the composability
property (\yes: yes, \no: no, \uk: currently unknown); in Section \ref{sec:sum} we either provide a reference or a novel proof for each of
these claims.

%% file: Quantification.tex
\section{Quantification of privacy loss (Q)}\label{sec:q}

The risk model associated to differential privacy is a \emph{worst-case}
property: it quantifies not only over all possible neighboring datasets but also
over all possible outputs. However, in many real-life risk assessments, events
with vanishingly small probability are ignored, or their risk weighted according
to their probability. It is natural to consider analogous relaxations,
especially since these relaxations often have better composition properties, and
enable natural mechanisms like the Gaussian mechanism to be considered
private~\cite{dwork2014algorithmic}.

Most of the definitions within this section can be expressed using the
\emph{privacy loss random variable}, first defined in~\cite{dinur2003revealing}
as the \emph{adversary's confidence gain}, so we first introduce this concept.
Roughly speaking, it measures how much information is revealed by the output of
a mechanism.

\begin{definition}[Privacy loss random variable~\cite{dinur2003revealing}]
    Let $\mecha$ be a mechanism, and $D_1$ and $D_2$ two datasets. The
    \emph{privacy loss random variable between $\mech{D_1}$ and $\mech{D_2}$} is
    defined as:
    \begin{equation*}
        \privloss{\mech{D_1}}{\mech{D_2}}(O)=\ln\left(\frac{\proba{\mech{D_1}=O}}{\proba{\mech{D_2}=O}}\right).
    \end{equation*}
    if neither $\proba{\mech{D_1}=O}$ nor $\proba{\mech{D_2}=O}$ is 0; in case
    only $\proba{\mech{D_2}=O}$ is zero then
    $\privloss{\mech{D_1}}{\mech{D_2}}(O)=\infty$, otherwise
    $\privloss{\mech{D_1}}{\mech{D_2}}(O)=-\infty$. When the mechanism is clear
    from context, we simply write $\privloss{D_1}{D_2}$.
\end{definition}

Differential privacy bounds the \emph{maximum value} of $\privloss{D_1}{D_2}$.
Instead of considering the maximum value, which corresponds to the worst
possible output, relaxations of this section will allow a small probability of
error, consider the average of the privacy loss random variable, or describe its
behavior in finer ways.

%\begin{definition*}[$\eps$-DP]
%    A mechanism $\mecha$ is $\eps$-DP if:
%    \begin{equation*}
%        \sup_{D,D':d(D,D')\le1}\sup_{S\subseteq\outputs}\log\frac{\proba{
%                \mech{D}\in S}}{\proba{\mech{D'}\in S}}\le\eps
%    \end{equation*}
%\end{definition*}

\subsection{Allowing a small probability of error}\label{subsec:small-error}

The first option, whose introduction is commonly attributed
to~\cite{dwork2006our}, relaxes the definition of $\eps$-indistinguishability by
allowing an additional small density of probability on which the upper $\eps$
bound does not hold. This small density, denoted $\del$, can be used to
compensate for outputs for which the privacy loss is larger than $e^\eps$. This
led to the definition of \emph{approximate differential privacy}, often simply
called $\epsdel$-DP\@. This is, by far, the most commonly used relaxation in the
scientific literature. Several of the enlisted modifications from the other 
dimensions were introduced with this additive factor $\delta$, but for 
clarity we omit these details where not crutial.

\begin{definition}[$\epsdel$-differential privacy~\cite{dwork2006our}]
    A privacy mechanism $\mecha$ is $\epsdel$-DP if for any
    datasets $D_1$ and $D_2$ that differ only on one record, and for all 
    $S\subseteq\outputs$:
    \begin{equation*}
        \proba{\mech{D_1}\in S} \le e^{\eps}\cdot\proba{\mech{D_2}\in S}+\del.
    \end{equation*}
\end{definition}

This definition is equivalent with Max-KL
stability~\cite{bassily2016algorithmic}, a special case of algorithmic
stability, which requires that one change in an algorithm's inputs does not
change its output ``too much''. 

The $\del$ in $\epsdel$-DP is sometimes explained as the probability that the
privacy loss of the output is larger than $e^\eps$ (or, equivalently, that the
$\eps$-indistinguishability formula is satisfied).
In fact, this intuition corresponds to a different definition, first introduced
in~\cite{machanavajjhala2008privacy} as \emph{probabilistic DP}
(ProDP), also called \emph{$\epsdel$-DP in distribution}
in~\cite{canard2015differential}. A detailed explanation of the distinction
between the two definitions can be found in~\cite{meiser2018approximate}.

\begin{definition}[$\epsdel$-probabilistic differential
  privacy~\cite{meiser2018approximate}]
    A privacy mechanism $\mecha$ is $\epsdel$-probabilistically DP
    (ProDP) if for any datasets $D_1$ and $D_2$ that differ only on one record
    there is a set $S_{1}\subseteq\outputs$ where
    $\proba{\mech{D_1}\in S_{1}}\le\del$, such that for all measurable sets
    $S\subseteq\outputs$:
    \begin{equation*}
        \proba{\mech{D_1}\in S\backslash S_{1}}\le 
        e^{\eps}\cdot\proba{\mech{D_2}\in S\backslash S_{1}}.
    \end{equation*}
\end{definition}

It is straightforward to show that $\epsdel$-ProDP is stronger than $\epsdel$-DP
(with no change in parameters); a proof of the reverse result (with parameter
change) is given in~\cite{zhao2019reviewing}. Both definitions can be
reformulated using the privacy loss random variable.

%\begin{definition}[$(\eps,\del_a,\del_p)$-differential privacy~\cite{zhang2015toward}
%  \footnote{Originally defined for batteries in smart grids.}]
%    A mechanism $\mecha$ satisfies $(\eps,\del_a,\del_p)$-relaxed 
%    differential privacy if for any pair data sets $D,D'$ 
%    differing in a single record and for all $S\subseteq\outputs$:
%    \begin{equation*}
%    \proba{\mech{D}\in S}\leq e^{\eps}\cdot\proba{\mech{D'}\in S}+\del_a
%    \end{equation*}
%    holds with probability $\del_p$. 
%\end{definition}

%\begin{theorem*}[Relation~\cite{zhang2015toward}]
%    $(\eps,\del, 1)$-RelDP is equivalent with $\epsdel$-DP. 
%\end{theorem*}

\begin{theorem*}%[$\eps$-DP \& $\epsdel$-DP \& $\epsdel$-ProDP using $\PrivLoss$]
    A mechanism $\mecha$ is:
    \begin{itemize}[noitemsep]
      \item $\eps$-DP $\Leftrightarrow$ $\probas{O\sim\mech{D_1}}{\PrivLoss_{D_1/D_2}(O)>\eps}=0$ for all neighboring $D_1$ and $D_2$.
      \item $\epsdel$-DP $\Leftrightarrow$ $\expects{O\sim\mech{D_1}}{\max\left(0,1-e^{\eps-\privloss{D_1}{D_2}(O)}\right)}\le\del$ for all neighboring $D_1$ and $D_2$.
      \item $\epsdel$-ProDP $\Leftrightarrow$ $\probas{O\sim\mech{D_1}}{\privloss{D_1}{D_2}(O)>\eps}\le\del$ for all neighboring $D_1$ and $D_2$. 
    \end{itemize}
\end{theorem*}

Approximate and probabilistic differential privacy can be combined to form
\emph{$(\eps,\del_a,\del_p)$-relaxed DP} (RelDP)~\cite{zhang2015toward}, which
requires $(\eps,\del_a)$-DP with probability at least $1-\del_p$.

\subsection{Averaging the privacy loss}

As $\eps$-DP corresponds to a \emph{worst-case} risk model, it is natural to
consider relaxations to allow for larger privacy loss for some outputs. It is
also natural to consider \emph{average-case} risk models: allowing larger
privacy loss values only if lower values compensate it in other cases.
One such relaxation is called \emph{Kullback-Leibler
privacy}~\cite{barber2014privacy,cuff2016differential}: it considers the
\emph{arithmetic} mean of the privacy loss random variable, which measures how
much information is revealed when the output of a private algorithm is observed.

\begin{definition}[$\eps$-Kullback-Leibler privacy~\cite{barber2014privacy,cuff2016differential}]
    A privacy mechanism $\mecha$ is $\eps$-Kullback-Leibler private (KLPr) if
    for all $D_1$, $D_2$ differing in one record: 
    \begin{equation}
        \expects{O\sim\mech{D_1}}{\privloss{D_1}{D_2}(O)}\le\eps.
    \end{equation}
    Note that this formula can be expressed as
    $D_\text{KL}\left(\mech{D_1}|\mech{D_2}\right)\le\eps$ where $D_\text{KL}$
    is the Kullback-Leibler-divergence.
\end{definition}

$\eps$-KL privacy considers the \emph{arithmetic} mean of the privacy loss
random variable or, equivalently, the \emph{geometric} mean of
$e^{\privloss{D_1}{D_2}}$. This choice of averaging function does not attribute
a lot of weight to worst-case events, where $\privloss{D_1}{D_2}$ takes high
values. \emph{R\'enyi DP} extends this idea by adding a parameter $\alpha\ge1$,
which allows controlling the choice of averaging function 
by bounding the $\alpha$th momentum of the privacy loss random variable.

\begin{definition}[$(\alpha,\eps)$-R\'enyi differential privacy~\cite{mironov2017renyi}]
    Given $\alpha>1$, a privacy mechanism $\mecha$ is $(\alpha,\eps)$-R\'enyi 
    DP (RenyiDP) if for all pairs of neighboring datasets $D_1$ and $D_2$:
    \begin{equation*}
      \expects{O\sim\mech{D_1}}{e^{(\alpha-1)\privloss{D_1}{D_2}(O)}} \le e^{(\alpha-1)\eps}.
    \end{equation*}
    Note that this formula can be expressed as
    $D_\alpha\left(\mech{D_1}|\mech{D_2}\right)\le\eps$ where $D_\alpha$ is the
    R\'enyi-divergence of order $\alpha$.
\end{definition}

This definition can be naturally extended by continuity to $\alpha=1$ (where it
is equivalent to $\eps$-KL privacy) and $\alpha=\infty$ (where it is equivalent
to $\eps$-DP). Larger values of $\alpha$ lead to more weight being assigned to
worst-case events: $(\alpha,\eps)$-R\'enyi
DP~\strongerthan~$(\alpha',\eps)$-R\'enyi DP iff $\alpha>\alpha'$. Besides
$\alpha=1$ and $\alpha=\infty$, R\'enyi DP has a simple interpretation for some
values of $\alpha$: $\alpha=2$ imposes a bound on the arithmetic mean of
$e^{\privloss{D_1}{D_2}}$, $\alpha=3$ imposes it on the quadratic mean,
$\alpha=4$ on the cubic mean, etc. 
A related technique is the moments accountant~\cite{abadi2016deep} which keeps track 
of a bound on the moments of the privacy loss random variable during composition. 
%https://arxiv.org/pdf/1902.08874.pdf

It is possible to use other divergence functions to obtain
other relaxations. For example, in~\cite{wang2018subsampled}, the authors
introduce two technical definitions, \emph{binary-$|\chi|^{\alpha}$ DP}
(b-$|\chi|^{\alpha}$ DP) and \emph{tenary-$|\chi|^{\alpha}$ DP}
(t-$|\chi|^{\alpha}$ DP), as part of a proof on amplification by sampling.
Other examples of divergences can lead to other variants, like
\emph{$\eps$-total variation privacy}~\cite{barber2014privacy} ($\eps$-TVPr,
using the total variance) and \emph{quantum DP}~\cite{colisson2016l3} (QDP,
using the quantum divergence).

\subsection{Controlling the tail distribution of the privacy loss}

Some definitions go further than simply considering a worst-case bound on the
privacy loss, or averaging it across the distribution. They try to obtain the
benefits of $\epsdel$-DP with a smaller $\eps$ which holds in most cases, but
control the behavior of the bad cases better than $\epsdel$-DP, which allows for
catastrophic privacy loss in rare cases.

The first attempt to formalize this idea was proposed
in~\cite{dwork2016concentrated}, where the authors introduce \emph{concentrated
DP} (later renamed to \emph{mean-concentrated DP} (mCoDP)
in~\cite{bun2016concentrated}). In this definition, a parameter controls the
privacy loss variable globally, and another parameter allows for some outputs to
have a greater privacy loss; while still requiring that the difference is
smaller than a Gaussian distribution.
%\begin{definition}[$(\mu,\tau)$-mean-concentrated differential 
%    privacy~\cite{dwork2016concentrated,bun2016concentrated}]
%    A mechanism $\mecha$ is $(\mu,\tau)$-mean-concentrated DP if for all
%    pairs of neighboring datasets $D_1$ and $D_2$:
%    \begin{itemize}
%      \item $\expects{O\sim\mech{D_1}}{\privloss{D_1}{D_2}(O)}\le \mu$;
%      \item for all $\alpha\ge1$,
%        $\expects{O\sim\mech{D_1}}{e^{(\alpha-1)\left(\mathcal{L}_{D_1/D_2}-
%        \expects{O\sim\mech{D_1}}{\mathcal{L}_{D_1/D_2}}\right)}}\le 
%        e^{{(\alpha-1)}^2\frac{\tau^2}{2}}$.
%    \end{itemize}
%\end{definition}
In~\cite{bun2016concentrated}, the authors show that this definition does not
satisfy the post-processing axiom, and propose another formalization of the same
idea called \emph{zero-concentrated DP} (zCoDP)~\cite{bun2016concentrated},
which requires that the privacy loss random variable is concentrated around
zero.

\begin{definition}[$(\xi,\rho)$-zero-concentrated differential privacy~\cite{bun2016concentrated}]
    A mechanism $\mecha$ is $(\xi,\rho)$-zero-concentrated DP if for
    all pairs of neighboring datasets $D_1$ and $D_2$ and all $\alpha>1$:
    \begin{equation*}
        \expects{O\sim\mech{D_1}}{e^{(\alpha-1)\privloss{D_1}{D_2}(O)}}\le e^{(\alpha-1)(\xi+\rho\alpha)}.
    \end{equation*}
\end{definition}

Four more variants of concentrated DP exist:
\begin{itemize}
\item \emph{$(\xi,\rho,\del)$-approximate zero-concentrated
  DP}~\cite{bun2016concentrated} (AzCoDP), which relaxes $(\xi,\rho)$-zCoDP by
  only taking the R\'enyi divergence on events with probability higher than
  $1-\del$ instead of on the full distribution.
\item \emph{$(\xi,\rho,\omega)$-bounded CoDP}~\cite{bun2016concentrated} (bCoDP)
  relaxes $(\xi,\rho)$-zCoDP by requiring the inequality to hold only for
  $\alpha\le\omega$.
\item \emph{$(\rho,\omega)$-truncated CoDP}~\cite{bun2018composable}
  (tCoDP\textsuperscript{\cite{bun2018composable}}) relaxes $(0,\rho)$-zCoDP in
  the same way.
\item \emph{$(\xi,\tau)$-truncated CoDP}~\cite{colisson2016l3}
  (tCoDP\textsuperscript{\cite{colisson2016l3}}) requires the R\'enyi divergence
  to be smaller than $\min(\xi,\alpha\tau)$ for all $\alpha\ge1$.
\end{itemize}
The relations between these definitions and other notions in this section is
well-understood. Besides the special cases (e.g.,
$(\rho,\infty)$-tCoDP\textsuperscript{\cite{bun2018composable}} is the same as
$(0,\rho)$-zCoDP) and the relations that are a direct consequence of the
definitions (e.g., $(\xi,\rho)$-zCoDP is the same as the condition
``$(\xi+\rho\alpha)$-R\'enDP for all $\alpha>0$''), we list known relations
below.

\begin{theorem*}
For all $\eps>0$, $\del>0$, $\mu>0$, $\tau>0$, $\xi\ge0$ and $\omega>1$:
\begin{itemize}
\item $\eps$-DP $\Longrightarrow$
  $\left(\frac{\eps\left(e^\eps-1\right)}{2},\eps\right)$-mCoDP (Theorem~3.5
  in~\cite{dwork2016concentrated})
\item $\eps$-DP $\Longrightarrow$ $\left(0,\frac{\eps^2}{2}\right)$-zCoDP (Lemma~8.3
  in~\cite{bun2016concentrated}) 
\item $\eps$-DP $\Longleftrightarrow$ $(\eps,0)$-zCoDP (Lemma 3.2
  in~\cite{bun2016concentrated})
\item $(\mu,\tau)$-mCoDP $\Longrightarrow$
  $\left(\mu-\frac{\tau^2}{2},\frac{\tau^2}{2}\right)$-zCoDP (Lemma~4.2
  in~\cite{bun2016concentrated})
\item $(\xi,\rho)$-zCoDP $\Longrightarrow$
  $\left(\xi+\rho,O(\sqrt{\xi+2\rho})\right)$-mCoDP (Lemma~4.3
  in~\cite{bun2016concentrated})
\item $(\xi,\rho)$-zCoDP $\Longrightarrow$
  $\left(\xi+\rho+\sqrt{4\rho\log\left(\frac{\min\left(1,\sqrt{\pi\rho}\right)}{\del}\right)},\del\right)$-DP
  (Lemma~3.5~and~3.6 in~\cite{bun2016concentrated})
\item $\left(\xi+\sqrt{\rho\log{\frac{1}{\del}}}\right)$-DP $\Longrightarrow$
  $\left(\xi-\frac{\rho}{4}+5\sqrt[4]{\rho},\frac{\rho}{4}\right)$-zCoDP
  (Lemma~3.7 in~\cite{bun2016concentrated})
\item $(\rho,\omega)$-tCoDP\textsuperscript{\cite{bun2018composable}} $\Rightarrow$ $\left(\hat\eps,\del\right)$-DP, where
  $\hat\eps=\rho+2\sqrt{\rho\log{\frac1\rho}}$ if
  $\log{\frac1\del}\le{(\omega-1)}^2\rho$, and
  $\hat\eps=\rho\omega+\frac{\log{\frac1\del}}{\omega-1}$ otherwise (Lemma~6
  in~\cite{bun2018composable})
%\item $(\xi,\rho)$-zCoDP $\Longrightarrow$
%  $\left(f_2,e^{\xi+2\rho}-1\right)$-DivDP (Section~2 in~\cite{duchi2018right})
\end{itemize}
\end{theorem*}

%\begin{definition}[$\del$-approximate $(\eta,\rho)$-concentrated differential privacy~\cite{bun2016concentrated}]
%    A privacy mechanism $\mecha$ is $\del$-approximate 
%    $(\eta,\rho)$-concentrated DP if for all $D,D'\in\datasets$ differing on a 
%    single record there exists $O,O'\in\outputs$ such that 
%    $\proba{O}\ge1-\del$ and $\proba{O'}\ge1-\delta$ and all 
%    $\alpha\in[1,\infty]$:
%    \begin{align*}
%        D_{\alpha}(\proba{\mech{D}|O}||\proba{\mech{D'}|O'})\le\eta+\rho\alpha\\
%        D_{\alpha}(\proba{\mech{D'}|O'}||\proba{\mech{D}|O})\le\eta+\rho\alpha
%    \end{align*}
%\end{definition}

%\begin{definition}[$(\eta,\rho,\omega)$-bounded-concentrated differential privacy~\cite{bun2016concentrated}]
%    A privacy mechanism $\mecha$ is $(\eta,\rho,\omega)$-bounded concentrated DP if 
%    for all $D,D'\in\datasets$ differing on a single record:
%    \begin{equation*}
%        \forall\alpha\in(1,\omega):D_{\alpha}(\mech{D}||\mech{D'})\le\eta+\rho\alpha
%    \end{equation*}
%\end{definition}

%\begin{definition}[$(\rho,\omega)$-truncated-concentrated differential privacy~\cite{bun2018composable}]
%    A privacy mechanism $\mecha$ is $(\rho,\omega)$-truncated concentrated DP 
%    ($\rho>0,\omega>1$) if for all $D,D'\in\datasets$ differing on a single 
%    record:
%    \begin{equation*}
%        \forall\alpha\in(1,\omega):D_{\alpha}(\mech{D}||\mech{D'})\le\rho\alpha
%    \end{equation*}
%\end{definition}

\subsection{Extension}

Most definitions of this section can be seen as bounding the divergence between
$\mech{D_1}$ and $\mech{D_2}$, for different possible divergence functions.
In~\cite{barber2014privacy}, the authors use this fact to generalize them and
define \emph{$(f,\eps)$-divergence DP} (DivDP), which takes the particular
divergence used as a parameter $f$.

\begin{definition}[$(f,\eps)$-divergence differential privacy~\cite{barber2014privacy}]
Let $f$ be a convex function such as $f(1)=0$. A privacy mechanism $\mecha$ is
$(f,\eps)$-divergence DP if for all pairs of neighboring
datasets $D_1$, $D_2$:
\begin{equation*}
    \expects{O\sim\mech{D_1}}{f\left(e^\privloss{D_1}{D_2}\right)}\le\eps.
\end{equation*}
\end{definition}

An instance of this definition was presented in~\cite{duchi2018right} as 
\emph{$(f_k,\eps)$-divergence DP}; which requires that
$\expects{O\sim\mech{D_1}}{{\left|e^\privloss{D_1}{D_2}-1\right|}^k}\le\eps^k$.
This definition is mainly used to prove technical results on privacy/utility
tradeoffs in the local model. For any $k\le1$, $\eps$-DP implies
$\left(f_k,e^\eps-1\right)$-DivDP, and when $k=2$, it is equivalent to
$\left(2,\log\left(1+\eps^2\right)\right)$-R\'enyiDP (Section~2
in~\cite{duchi2018right}).

Moreover, \emph{capacity bounded differential privacy} (CBDP) was introduced
in~\cite{chaudhuri2019dcapacity}, which uses $H$-restricted $f$-divergence:
$D_f^H(P|Q)=\sup_{h\in H}[\expects{x\sim P}{h(x)}-\expects{x\sim Q}{f^*(h(x))}]$
where $f$ is a divergence, $H$ is a family of functions, and $f^*$ is the
Fenchel conjugate\footnote{The Fenchel conjugate for a function $f$ with a
domain $R$ is $f^*(x)=\sup_{y\in R}[xy-f(y)]$.}. In other words, it requires the
supremum condition to hold only for a selected set of functions (queries)
instead of all possible ones.

Finally, most definitions in this section taking two real-valued parameters can
be extended to use a \emph{family} of parameters rather than a single pair of
parameters. As shown in~\cite{sommer2019privacy} (Theorem~2) for approximate DP,
probabilistic DP, and R\'enyi DP, finding the tightest possible family of
parameters (for either definition) for a given mechanism is equivalent to
specifying the behavior of its privacy loss random variable entirely.

\subsection{Multidimensional definitions}

Allowing a small probability of error $\del$ by using the same concept as in
$\epsdel$-DP is very common; many new DP definitions were proposed in the
literature with such a parameter. Unless it creates a particularly notable
effect, we do not mention it explicitly and present the definitions without this
parameter.

Definitions in this section can be used as standalone concepts: $\epsdel$-DP is
omnipresent in the literature, and the principle of averaging risk is natural
enough for R\'enyi privacy to be used in practical settings, like posterior
sampling~\cite{geumlek2017renyi} or resistance to adversarial inputs in machine
learning~\cite{pinot2019unified}. Most variants in this section, however, are
only used as technical tools to get better results on composition or privacy
amplification~\cite{dwork2014algorithmic,wang2018subsampled,feldman2018privacy,lee2018concentrated}.

%% file: Neighborhood.tex
\section{Neighborhood definition (N)}\label{sec:n}

The original definition of differential privacy considers datasets differing in
one record. Thus, the datasets can differ in two possible ways: either they have
the same size and differ only on one record, or one is a copy of the other with
one extra record. These two options do not protect the same thing: the former
protects the \emph{value} of the records while the latter also protects their
\emph{presence} in the data: together, they protect any property about a single
individual.

In many scenarios, it makes sense to protect a different property about their
dataset, e.g., the value of a specific sensitive field, or entire groups of
individuals. It is straightforward to adapt DP to protect different sensitive
properties: all one has to do is change the definition of neighborhood in the
original definition.

\subsection{Changing the sensitive property}\label{sec:n-changing}

The original definition states that the $\eps$-indistinguishability propery
should hold for ``any datasets $D_1$ and $D_2$ that differ only on the data of
one individual''. Modifying the set of pairs
$\left(D_1,D_2\right)$ such that $\mech{D_1}\ind_\eps\mech{D_2}$ is equivalent
to changing the protected sensitive property.

\subsubsection*{Weaker relaxations}

In DP, the difference between $D_1$ and $D_2$ is sometimes interpreted as ``one
record value is different'', or ``one record has been added or removed''.
In~\cite{kifer2011no}, the authors formalize these two options as \emph{bounded
DP} and \emph{unbounded DP}. They also introduced \emph{attribute DP} and
\emph{bit DP}, for smaller changes within the differing record.

\begin{definition}[\cite{kifer2011no}]
    If a privacy mechanism $\mecha$ satisfies $\mech{D_1}\indeps\mech{D_2}$
    for any pair $D_1,D_2$, where $D_1$ can be obtained from $D_2$ by\ldots
    \begin{itemize}[noitemsep]
        \item \ldots adding or removing one record, then $\mecha$ is
          \emph{$\eps$-unbounded DP} (uBoDP).
        \item \ldots changing exactly one record, then $\mecha$ is
          \emph{$\eps$-bounded DP} (BoDP).
        \item \ldots changing one attribute in a record, then $\mecha$ is
          \emph{$\eps$-attribute DP} (AttDP).
        \item \ldots changing one bit of an attribute in a record, then $\mecha$
          is \emph{$\eps$-bit DP} (BitDP).
    \end{itemize}
\end{definition}

In~\cite{kifer2011no}, authors show that $\eps$-unbounded DP implies
$2\eps$-bounded DP, as changing a record can be seen as deleting it and adding a
new one in its place. The original definition of $\eps$-DP is the conjunction of
$\eps$-unbounded DP and $\eps$-bounded DP\@. However, bounded DP is frequently
used in the literature, especially when using local differential privacy, and
often simply called differential privacy. It is also sometimes renamed, like
in~\cite{feldman2018privacy}, where the authors call it \emph{per-person DP}.

Another way to relax the neighborhood definition in DP is to consider that only
certain types of information are sensitive. For example, if the attacker learns
that their target has cancer, this is more problematic than if they learn that
their target does \emph{not} have cancer. This idea is captured in
\emph{one-sided DP} (OSDP)~\cite{doudalis2017one}: the neighbors of a dataset
$D$ are obtained by replacing a single sensitive record with any other record
(sensitive or not). The idea of sensitivity is formalized by a \emph{policy}
$P$, which specifies which records are sensitive. This idea cannot be captured
simply by $\eps$-indistinguishability, since one-sided DP is asymmetric.

\begin{definition}[$(P,\eps)$-one-sided differential privacy~\cite{doudalis2017one}]
    Given a policy $P\subseteq\tuples$, a privacy mechanism $\mecha$ is
    $(P,\eps)$-one-sided DP iff for all datasets $D_1$ and $D_2$, where
    $D_2$ has been obtained by replacing a record $t\in D_1\cap P$ by any other
    record and for all $S\subseteq\outputs$:
    \begin{equation*}
        \proba{\mech{D_1}\in S}\leq e^{\eps}\cdot\proba{\mech{D_2}\in S}.
    \end{equation*}
\end{definition}

When $P=\tuples$, this is equivalent to bounded DP\@. Similar ideas were
proposed in multiple papers:

\begin{itemize}
  \item In \cite{takagi2021asymmetric}, 
the authors propose \emph{asymetric DP}, which is the unbounded version of OSDP.
  \item In~\cite{asif2019accurately}, the authors propose \emph{sensitive
    privacy}, which determines which records are sensitive based on the data
    itself and a \emph{normality property} $N$ and a graph-based definition of
    $k$-neighborhood, instead of using a data-independent determination. 
  \item In~\cite{bittner2018using}, the authors introduce
    \emph{anomaly-restricted DP}, which assumes that there is only one outlier
    in the dataset, and that this outlier should not be protected.
\end{itemize}

\subsubsection*{Stronger notions}

More restrictive definitions are also possible. First, some definitions make the
definition of neighborhood more explicit when a single person can contribute
multiple times to a dataset; this is the case for~\emph{client/participant DP},
defined in~\cite{mcmahan2017learning}. Variants of differential privacy that do not protect individuals, but single
contributions (in the case where the same person can contribute multiple times
to a dataset), are also often used in practice, especially for machine learning
applications~\cite{mcmahan2017learning}. Some recent works also argue that
in-between definitions are appropriate: rather than protecting a single
contribution or entire users contributions, authors in~\cite{asi2019element}
suggest that protecting \emph{elements} that reveal information about users,
after deduplicating or clustering contributions. For example, rather than
protecting all website visits by a single user, or each visit individually, one
might choose to protect the fact that a user ever visited a website (but not
whether the user visited the same website once or many times). They call the
corresponding definition \emph{element-level DP} (ELDP).

In~\cite{dwork2008differential}, the
authors implicitly define \emph{$(c,\eps)$-group privacy} considers datasets
that do not differ in one record, but possibly several, to protect multiple
individuals. This can also be interpreted as taking correlations into account
when using DP\@: \emph{DP under correlation}~\cite{chen2014correlated} uses an
extra parameter to describe the maximum number of records that the change of one
individual can influence.

These two definitions are formally equivalent; but the implicit interpretation
of DP behind them is different. $(c,\eps)$-group privacy is compatible with the
associative view under the strong adversary assumption (the adversary knows all
records except $c$) or the causal view ($c$ records are changed after the data
is generated). Meanwhile, DP under correlation implicitly considers the
associative view with the independence assumption; and tries to relax that
assumption. This last approach was further developed via \emph{dependent
DP}~\cite{liu2016dependence}, which uses ``dependence relationships'' to
describe how much the variation in one record can influence the other records.

\begin{definition}[$(R,c,\eps)$-dependent differential privacy~\cite{liu2016dependence}]
    A privacy mechanism $\mecha$ provides $(R,c,\eps)$-dependent DP (DepDP) where $R$ is 
    the probabilistic dependence relationship and $c$ is the dependence size,
    if for any pair of datasets $D_1$ and $D_2$, where $D_2$ has been obtained
    from $D_1$ by changing one record and the corresponding at most $c-1$ other 
    records according to $R$, $\mech{D_1}\indeps\mech{D_2}$.
\end{definition}

Note that when $R$ is the empty relation, or when $c=1$, this definition is
equivalent to bounded DP: under the associative view of DP, this represents
independence between records. Similar definitions appear in \cite{wang2021correlated,wang2021differentially} as 
\emph{correlated indistinguishability} and \emph{correlated tuple DP} respectively, 
in which correlations are defined by a correlation matrix. 
Furthermore, in~\cite{wu2017game1,wu2017game2} as \emph{correlated DP} (CorDP), in which
correlations are defined by an observation on other datasets, and in
in~\cite{yang2015bayesian} as \emph{bayesian DP}\footnote{There are two other
notions with the same name: introduced in~\cite{triastcyn2019bayesian,leung2012bayesian}, 
we mention them in Section~\ref{sec:v} and \ref{sec:b} respectively.}
(BayDP\textsuperscript{\cite{yang2015bayesian}}), where the neighborhood
relation is defined by an adversary having some knowledge about correlations in
the data. An extension is proposed in~\cite{li2019impact} as \emph{prior DP}
(PriDP) which considers a family of adversaries instead of a single adversary.

The strongest possible variant is considered in~\cite{kifer2011no}, where the
authors define \emph{free lunch privacy} (FLPr), in which the attacker must be
unable to distinguish between any two datasets, even if they are completely
different. This guarantee is a reformulation of Dalenius' privacy
goal~\cite{dalenius1977towards}; as such, all mechanisms that satisfy free lunch
privacy have a near-total lack of utility.

\begin{definition}[$\eps$-free lunch privacy~\cite{kifer2011no}]
    A privacy mechanism $\mecha$ satisfies $\eps$-free lunch privacy if
    $\mech{D_1}\indeps\mech{D_2}$ for any pair of datasets $D_1,D_2$.
\end{definition}

%The partial ordering of these definitions free-lunch-Pr~\extendsstronger~group-DP~\extendsstronger~client-DP~\extendsstronger~unbounded-DP~\extendsstronger~bounded-DP~\extendsstronger~attribute-DP~\extendsstronger~bit-DP.

\subsection{Limiting the scope of the definition}\label{sec:n-scope}

Redefining the neighborhood property can also be used to reduce the scope of the
definitions. In~\cite{soria2017individual}, the authors note that DP requires
$\eps$-indistinguishability of results between any pair of neighboring data
sets, but in practice, the data custodian has only \emph{one} data set $D$ they
want to protect. Thus, they only require $\eps$-indistinguishability between
this data set $D$ and all its neighbors, calling the resulting definition
\emph{individual DP} (InDP). An equivalent definition was proposed
in~\cite{charest2016meaning} as \emph{conditioned DP}.

\begin{definition}[$(D,\eps)$-individual differential privacy~\cite{soria2017individual}]
  Given a dataset $D\in\datasets$, a privacy mechanism $\mecha$ satisfies
  $(D,\eps)$-individual DP if for any data set $D'$ that differs in at most one
  record from $D$, $\mech{D}\indeps\mech{D'}$.
\end{definition}

This definition was further restricted in~\cite{wang2017per} where besides
fixing a dataset $D$, a record $t$ is also fixed.

%\begin{definition}[$(D,t,\eps)$-per-instance differential privacy~\cite{wang2017per}]
%    Given a dataset $D\in\datasets$ and record $t\in\tuples$, a privacy mechanism 
%    $\mecha$ satisfies $(D,t,\eps)$-per-instance DP (PIDP) if 
%    $\mech{D}\indeps\mech{D\cup\{t\}}$.
%\end{definition}

\subsection{Applying the definition to other types of input}\label{sec:n-other}

Many adaptations of DP are simply changing the neighborhood
definition to protect different types of input data than datasets. A few
examples follow.

\begin{itemize}
	\item Location \\
	In \cite{elsalamouny2016differential} the authors defined \emph{location privacy}, in 
	which neighbors are datasets which differ in at most one record, and the two differing
	reecords are at a physical distance smaller than a given threshold. This definition also 
	appears in \cite{chen2018differentially} as \emph{DP on $r$-location set}%
	\footnote{Distinct from \emph{DP on $\del$-location	set} \cite{xiao2015protecting}, 
		which we mention in Chapter \ref{sec:v}.}. Several more location-related DP variants 
	were defined in \cite{naor2020can}: \emph{untrackability} (which adopts differential
	privacy for set of locations by protecting whether they originated from a single user 
	or by two users), \emph{undetectability} and \emph{multi user untrackability}, (which 
	extend this idea further by not assuming both sets originated from the same private 
	data and to multiple users respectively). 
	\item Graph \\
	In \cite{hay2009accurate} the authors adopt DP to graph-
	structured data and present multiple alternative definitions, which protect different
	parts of the graph: the strongest is \emph{node-DP}, which protects a node
	and all its edges; the weakest is \emph{edge-DP}, only protects one edge;
	and an intermediate definition is \emph{$k$-edge-DP}, which protects 
	a total of $k$ edges and nodes. Similarly to one-sided DP, in 
	\cite{task2012guide,kearns2016private}, the authors introduce \emph{out-link privacy}, and \emph{protected DP} which protects all outgoing edges 
	from a given node and guarantees that no observer can learn much about the set of 
	edges corresponding to any protected node respectively.
	In addition, in \cite{reuben2018towards}, the author introduces
	\emph{$QL$-edged-labeled DP} which only protecting a predetermined subset of outgoing 
	edges. In \cite{pinot2018minimum}, the author introduces \emph{$l_1$-weighted DP},
	in which graph edges are weighted, and graphs are neighbors when the total
	weight of their differing edges is smaller than 1; this notion was also defined 
	implicitly in \cite{sealfon2016shortest}. In \cite{zhang2020differentially} the 
	authors defined \emph{feasible-node DP} for control-flow graphs where an addition 
	or removal od a node results in a feasible run-time behaviors. 
	In \cite{sun2019analyzing},  the authors define \emph{decentralized DP} which 
	extends the graph neighborhood to two jumps. Finally, in \cite{ding2013seamless} the 
	authors introduce \emph{seamless privacy}, which rather than protecting characteristics 
	of a specific input graph, it ensures that certain pairs of queries on this graph return 
	similar answers.
	\item Stream \\ 
	Several authors adapt DP to a streaming context, where the
	attacker can access the mechanism's internal states.
	In \cite{dwork2010differential,dwork2010pan,dwork2010new}, authors define
	\emph{pan-privacy}, which comes in two variants: \emph{event-level}
	pan-privacy (called \emph{strong DP} in \cite{wang2020differential})
	protects individual events, and \emph{user-level} pan-privacy protects all
	events associated to a single user. In \cite{kellaris2014differentially},
	the authors extend the previous idea and propose \emph{$w$-event privacy},
	which protects any event sequence occurring within a window of at most $w$
	timestamps. In \cite{farokhi2019descount,naor2020can} this was further
	extended to an infinite horizon via \emph{discounted DP}
	(which keep assigning smaller-and-smaller weights to further-and-further
	events) and \emph{everlasting privacy} (which limit the leakage of
	information users suffer, no matter how many executions a mechanism had),
	respectively. Finally, \emph{series-indistinguishability} \cite{wang2017cts} 
	%\emph{cloacking DP} \cite{zhang2021protecting} 
	captured data correlations in the streaming context and the authors in \cite{le2013differentially,parker2021spectral} adopted DP for Kalman-filters 
	and to time-independent power spectral densities respectively.
	\item RAM and PIR \\ 
	In \cite{wagh2016differentially} the authors adopt DP for 
	Random Access Memory and Private Information Retrieval. For RAM the neighborhood 
	is defined over the sequence of logical memory requests over time; the same notion 
	appears in \cite{chan2018foundations} as \emph{differential obliviousness} and in 
	\cite{allen2018algorithmic} as \emph{oblivious DP}. The adaptation of neighborhood is 
	similar in case of PIR; a similar notion appears in \cite{toledo2016lower} as
	\emph{$\eps$-private PIR} and in \cite{patel2019storage} as \emph{$\eps$-DPIR}. 
	Additionally, in \cite{kellaris2017accessing}, the authors use a similar idea 
	to define DP for outsourced database systems.
	\item Text and Images \\ 
	In \cite{jones2018towards}, the authors adapt DP for
	symbolic control systems, and introduce \emph{word-DP} and
	\emph{substitution-word-DP}, protecting respectively pairs of words whose
	Levenshtein distance is lower than a given parameter, or whose Hamming
	distance is lower than a given parameter. In \cite{zhang2018private}, the 
	authors adapt DP for text vectors, and propose 
	\emph{text indistinguishability}, in which the neighborhood relationship 
	between two word vectors depends on their Euclidean distance. 
	In \cite{ying2013linear,fan2018image,liu2021dp}, the authors define 
	\emph{refinement DP}, \emph{DP Image}, and \emph{Pixel DP} respectively, which
	adopts the definition for images with neighbors given by some
	transformation or metric. 	
	\item Miscellaneous \\ 
	Beside the already mentioned fields DP was adopted to numerous other use-cases, 
	such as for set operations in \cite{yan2017privmin}, 
	for gossip protocols in \cite{huang2019quantifying}, 
	for functions in \cite{nozari2019networked}, 
	for genomic data in \cite{simmons2016enabling}, 
	for recommendation systems in \cite{guerraoui2015d}, 
	for machine learning in \cite{long2017towards}, 
	and for bandit algorithms in \cite{tossou2016algorithms,basu2019differential}.
\end{itemize}

\subsection{Extensions}

It is natural to generalize the variants of this section to arbitrary
neighboring relationships. One example is mentioned in~\cite{kifer2011no}, under
the name \emph{generic DP}\footnote{Another definition with 
the same name is introduced in~\cite{kifer2010towards,kifer2012axiomatic}, we
mention it in Section~\ref{sec:v}.} (GcDP\textsuperscript{\cite{kifer2011no}}), where the
neighboring relation is entirely captured by a \emph{relation} $\mathcal{R}$
between datasets.

\begin{definition}[$(\mathcal{R},\eps)$-generic differential privacy~\cite{kifer2011no}]   
    Given a relation $\mathcal{R}\subseteq\datasets^2$, a privacy mechanism 
    $\mecha$ satisfies $(\mathcal{R},\eps)$-generic DP if for all
    $\left(D_1,D_2\right)\in\mathcal{R}$, $\mech{D_1}\indeps\mech{D_2}$.
\end{definition}

This definition is symmetric, but it can easily be modified to accommodate
asymmetric definitions like one-sided DP.

Other definitions use different formalizations to also generalize the concept of
changing the neighborhood relationship. Some (like pufferfish privacy, 
mentioned in Section~\ref{sec:b}) use \emph{pairs of predicates}
$\left(\phi_1,\phi_2\right)$ that $D_1$ and $D_2$ must respectively satisfy to
be neighbors. Others (like coupled-worlds privacy, mentioned in
Section~\ref{sec:r}) use \emph{private functions},
denoted $\priv$, and define neighbors to be datasets $D_1$ and $D_2$ such as
$\priv\left(D_1\right)\neq\priv\left(D_2\right)$~\cite{bassily2013coupled}.
Others use a \emph{distance function} $d$ between datasets, and neighbors are
defined as datasets a distance lower than a given threshold $\Delta$; this is
the case for \emph{DP under a neighborhood} (DPUN), introduced
in~\cite{fang2014differential}, \emph{adjacent DP} (AdjDP), introduced
in~\cite{krishnan2018distributional}\footnote{Originally simply called
``differential privacy'' by its authors.}, \emph{constrained DP} (ConsDP),
introduced in~\cite{zhou2009differential} (where the distance $d$ captures a
utility-related constraint), and \emph{distributional privacy}\footnote{Another 
definition with the same name is introduced in~\cite{roth2010new,blum2013learning}, 
we mention it in Section~\ref{sec:v}.}
(DlPr\textsuperscript{\cite{zhou2009differential}}),
also introduced in~\cite{zhou2009differential} (with additional constraints on
the neighborhood definition: neighboring datasets must be part of a fixed set
$S_\datasets$ and have elements in common). This distance can also be defined as
the sensitivity of the mechanism, like in \emph{sensitivity-induced
DP}~\cite{rubinstein2017pain} (SIDP), or implicitly defined by a set of
constraints, like what is done implicitly in~\cite{kifer2011no} via
\emph{induced neighbors DP} (INDP).

One notable instantiation of generic DP is \emph{blowfish privacy}
(BFPr)~\cite{he2014blowfish}. Its major building blocks are a \emph{policy
graph} $G$, that specifies which pairs of domain values in $\tuples$ should not
be distinguished between by an adversary; and a set of \emph{constraints} $Q$
that specifies the set $\mathcal{I}_Q$ of possible datasets that the definition
protects.
%$G=(V,E)$ such that $V\subseteq\tuples\cup\{\bot\}$ and
%$E\subseteq{(\tuples\cup\{\bot\})}^2$, where $\bot$ is a ``dummy'' value. An edge
%$(t,t')\in E$ represents the bounded variant (i.e., $t$ is replaced by $t'$)
%while $(t,\bot)\in E$ corresponds to the unbounded case (i.e., $t$ is removed
%from the dataset). 
 %In other words, the policy graph 
%restricting the set of neighboring datasets: $D$ and $D'$ are $G$-neighbors 
%if either $(t,t')\in E$ such that $t\in D$ and $t'\in D'$ or $(t,\bot)\in E$ 
%such that $t\in D$ and $t\not\in D'$. 
It was inspired by the Pufferfish framework~\cite{kifer2012rigorous} (see
Section~\ref{sec:b}), but the attacker is not assumed to have uncertainty over
the data: instead, it models an attacker whose knowledge is a set of
deterministic constraints on the data.

\begin{definition}[$(G,\mathcal{I}_Q,\eps)$-blowfish privacy~\cite{he2014blowfish}]
    Given a policy graph $G\in\tuples^2$ and a set of datasets $\mathcal{I}_Q$,
    a privacy mechanism $\mecha$ satisfies $(G,\eps)$-blowfish privacy if for
    all datasets $D_1$ and $D_2$ in $\mathcal{I}_Q$ that differ in only one
    element $i$ such that $\left(D_1(i),D_2(i)\right)\in G$,
    $\mech{D_1}\ind_{\eps}\mech{D_2}$.
\end{definition}

As noted in~\cite{he2014blowfish}, a mechanism satisfies $\eps$-bounded DP if
and only if satisfies $(K,\mathcal{I}_n,\eps)$-blowfish privacy, where $K$ is
the complete graph, and $\mathcal{I}_n$ is any datasets of size $n$.  A
particular instantiation of this idea is explored in~\cite{kifer2011no} as 
\emph{induced DP} (IndDP), where the definition of neighbors is induced by 
a set of constraints.

\subsection{Multidimensional definitions}

Modifying the protected property is orthogonal to modifying the risk model
implied by the quantification of privacy loss: it is straightforward to combine
these two dimensions. Indeed, many definitions mentioned in this section were
actually introduced with a $\del$ parameter allowing for a small probability of
error. One-one particularly general and specific example is 
\emph{adjacency relation divergence DP} \cite{kawamoto2019allerton}, 
and \emph{node-R\'enyi DP} \cite{daigavane2021node}, 
which combines an arbitrary neighborhood definition (like in generic DP) 
with an arbitrary divergence function (like in divergence DP) 
and adopts R\'enyi DP to graphs respectively.

As the examples in Section~\ref{sec:n-other} show, it is very common to change
the definition of neighborhood in practical contexts to adapt what aspect of the
data is protected. Further, local DP mechanisms like
RAPPOR~\cite{erlingsson2014rappor} implicitly use bounded DP\@: the
participation of one individual is not secret, only the value of their record is
protected. Variants that limit the scope of the definition to one particular
dataset or user, however, provide few formal guarantees and do not seem to be
used in practice.

%% file: Variation.tex
\section{Variation of privacy loss (V)}\label{sec:v}

In DP, the privacy parameter $\eps$ is \emph{uniform}: the level of protection
is the same for all protected users or attributes, or equivalently, only the
level of risk for the most at-risk user is considered. In practice, some users
might require a higher level of protection than others or a data custodian might
want to consider the level of risk across all users, rather than only
considering the worst case. Some definitions take this into account by allowing
the privacy loss to vary across inputs, either explicitly (by associating each
user to an acceptable level of risk), or implicitly (by allowing some users to
be at risk, or averaging the risk across users).

\subsection{Varying the privacy level across inputs}\label{subsec:varying}

In Section~\ref{sec:n}, we saw how changing the definition of the neighborhood
can be used to adapt the definition of privacy and protect different aspects of
the input data. However, the privacy protection in those variants is binary:
either a given property is protected, or it was not. A possible option to
generalize this idea further is to allow the privacy level to vary across
possible inputs.

One natural example is to consider that some users might have higher privacy
requirements than others, and make the $\eps$ vary according to which user
differs between the two datasets. This is done in \emph{personalized DP}
(PerDP), a notion first defined informally in~\cite{niknami2014spatialpdp}, then
independently
in~\cite{jorgensen2015conservative,ebadi2015differential,ghosh2015selling,liu2015fast}.
An equivalent notion is also defined in~\cite{alaggan2015heterogeneous} as
\emph{heterogeneous DP}, while a location-based definition is presented
in~\cite{deldar2018pldp} as \emph{personalized location DP}. Note that in these
definitions, the privacy level associated with each user is not considered
sensitive, and cannot depend on the data itself.

\begin{definition}[$\Psi$-personalized differential privacy~\cite{jorgensen2015conservative}]
    A privacy mechanism $\mecha$ provides $\Psi$-personalized DP if for every
    pair of neighboring datasets $\left(D,D_{-i}\right)$ and for all sets of
    outputs $S\subseteq\outputs$:
    \begin{equation*}
      \proba{\mech{D_{-i}}\in S}\leq e^{\Psi\left(i\right)}\proba{\mech{D}\in S}
    \end{equation*}
    where $\Psi$ is a privacy specification, which maps each index $i$ to the
    privacy level of the record $D(i)$.
\end{definition}

As shown in~\cite{jorgensen2015conservative,cummings2018individual}, $\eps$-DP
implies $\Psi$-PerDP, where $\Psi(i)=\eps$ for all indices $i$; and $\Psi$-PerDP
implies $\eps$-DP where $\eps=\max_i\Psi(i)$.

This definition can be seen as a refinement of the intuition behind one-sided
DP, which separated records into sensitive and non-sensitive ones. The idea of
making the privacy level vary across inputs can be generalized further, by also
making the privacy level depend on the entire dataset and the data of the
differing record. This is done in~\cite{lui2015outlier}, where the authors
define \emph{tailored DP} (TaiDP).

\begin{definition}[$\Psi$-tailored differential privacy~\cite{lui2015outlier}]
    A mechanism $\mecha$ satisfies $\Psi$-tailored differential privacy
    for $\Psi:\tuples\times\datasets\rightarrow\mathbb{R}_0^\infty$ if
    for any dataset $D$, $\mech{D}\ind_{\eps(D(i),D)}\mech{D_{-i}}$.
\end{definition}

A similar notion is \emph{input-discriminative DP} \cite{gu2020providing} 
where $\Psi$ takes the two $\eps$ values corresponding to the two dataset. 
The authors also defined \emph{minID-DP} where $\Psi$ is the minimum function. 

This concept can be applied to strengthen or weaken the privacy requirement for
a record depending on whether they are an outlier in the dataset.
In~\cite{lui2015outlier}, the authors formalize this idea and introduce
\emph{outlier privacy}, which tailors an individual's protection level to their
``outlierness''. Other refinements are also introduced in~\cite{lui2015outlier}:
\emph{simple outlier privacy} (SOPr), \emph{simple outlier DP} (SODP), and
\emph{staircase outlier privacy} (SCODP). A similar idea was explored
in~\cite{kartal2019differential}, which introduced \emph{pareto DP} (ParDP): it
utilizes a pareto distribution of parameters $(p,r)$ to separate a large number
of low-frequency individuals from a small number of high-frequency, and the
sensitivity is calculated based on only the low-frequency individuals.

Finally, varying the privacy level across inputs also makes sense in
\emph{continuous} scenarios, where the neighborhood relationship between two
datasets is not binary, but quantified. This is, for example, the case for
\emph{$\eps$-geo-indistinguishability}~\cite{andres2013geo}, where two datasets
$D_1$ and $D_2$ are considered \emph{$r$-neighbors} if the only different record
between $D_1$ and $D_2$ are at a distance $r$ of each other, and the $\eps$
grows linearly with $r$.

\subsection{Randomizing the variation of privacy levels}\label{subsec:scope}

Varying the privacy level across inputs can also be done in a randomized way, by
guaranteeing that some random fraction of users have a certain privacy level.
One example is proposed in~\cite{hall2011random} as \emph{random DP} (RanDP):
the authors note that rather than requiring DP to hold for any possible
datasets, it is natural to only consider \emph{realistic datasets}, and allow
``edge-case'' or very unrealistic datasets to not be protected. This is captured
by generating the data randomly, and allowing a small proportion $\gamma$ of
cases to not satisfy the $\eps$-indistinguishability property.

\begin{definition}[$(\pi,\gamma,\eps)$-random differential privacy~\cite{hall2011random}]
    Let $\pi$ be a probability distribution on $\tuples$, $D_1$ a dataset
    generated by drawing $n$ i.i.d.\ elements in $\pi$, and $D_2$ the same
    dataset as $D_1$, except one element was changed to a new element drawn from
    $\pi$. A mechanism $\mecha$ is $(\pi,\gamma,\eps)$-random DP if
    $\mech{D_1}\indeps\mech{D_2}$, with probability at least $1-\gamma$ \emph{on
    the choice of $D_1$ and $D_2$}.
\end{definition}

The exact meaning of ``with probability at least $1-\gamma$ on the choice of
$D_1$ and $D_2$'' can vary slightly. In~\cite{hall2012new} and
in~\cite{mcclure2015relaxations}, the authors introduce \emph{predictive DP}
(PredDP) and \emph{model-specific DP} respectively, which quantify over all
possible choices of $D_1$, and picks $D_2$ randomly in the neighborhood of
$D_1$. In~\cite{dixit2012testing}, $D_1$ and $D_2$ are both taken out of a set
of density larger than $1-\gamma$, and the authors call this definition
\emph{generalized DP} (GdDP). The distribution generating the dataset is also
not always assumed to be generating i.i.d.\ records; we denote the corresponding
parameter by $\theta$.

%\begin{definition}[$\epsdel$-model-specific DP~\cite{mcclure2015relaxations}]
%    A privacy mechanism $\mecha$ is $\epsdel$-model-specific DP with respect to
%    $\theta$ if for all datasets $D$ with $\probas{\theta}{D}\ge\del$ and for all its
%    neighbors $D'$ $\mech{D}\ind_\eps\mech{D'}$ holds.
%\end{definition}

Random DP might look similar to probabilistic DP\@: in both cases, there is a
small probability that the privacy loss is unbounded. On the other hand, they are 
very different: in random DP, this probability is computed inputs of the mechanisms
(i.e., users or datasets), for probabilistic DP, it is computed across mechanism
outputs. Also similarly to probabilistic DP, excluding some cases altogether
creates definitional issues: random DP does not satisfy the convexity axiom (see
Proposition~\ref{prop:a-c-rand} in Section \ref{sec:sum}). We postulate that using a
different tool to allow some inputs to not satisfy the mechanism, similar to
approximate DP or R\'enyi DP, could solve this problem.

Usually, data-generating distributions are used for other purposes: they
typically model an adversary with partial knowledge. However, definitions in
this section still compare the outputs of the mechanisms given fixed neighboring
datasets: the only randomness in the indistinguishability property comes from
the mechanism. By contrast, definitions of Section~\ref{sec:b} compare the
output of the mechanism on a random dataset, so the randomness comes both from
the data-generating distribution and the mechanism.

\subsection{Multidimensional definitions}

As varying the privacy level or limiting the considered datasets are two
distinct way of relaxing differential privacy, it is possible to combine them
with the previously mentioned dimensions.

\subsubsection*{Combination with N}

The definitions described in Section~\ref{sec:n} (e.g., generic DP or blowfish
privacy) have the same privacy constraint for all neighboring datasets. Thus,
they cannot capture definitions that vary the privacy level across inputs.
On the other hand, they can be combined quite easily. For instance, \cite{deldar2018pldp} and \cite{seres2021effect} adopted personalized DP to location and communication graphs respectively. 
%As our dimensions capture ideas, the inner exclusivity condition mentioned in Section \ref{sec:pre} not always holds. An example for such inter-dimension combination is \emph{personalized existing edge DP} \cite{seres2021effect}, which is a combination of unbounded DP, edge-DP, assymetric DP, and personalized DP.
Varying the privacy level 
across inputs also makes sense in ``continuous'' scenarios, 
where the neighborhood relationship between two datasets is not 
binary, but quantified. This is, for example, the case for 
\emph{geo-indistinguishability} \cite{andres2013geo}, 
where two datasets $D_1$ and $D_2$ are considered ``$r$-neighbors'' 
if the only different record between $D_1$ and $D_2$ are at a distance 
$r$ of each other, and the $\eps$ grows linearly with $r$.
Both ideas can be naturally captured together via distance functions.
In~\cite{chatzikokolakis2013broadening}, the authors introduce
\emph{$d_\datasets$-privacy} ($d_\datasets$-Pr), in which the function
$d_\datasets$ takes both datasets as input, and returns the corresponding
maximal privacy loss (the $\eps$) depending on the difference between the two
datasets.

% essentially what differential privacy provides is \emph{Lipschitz
% continuity}\footnote{A function $f:X\rightarrow Y$ is Lipschitz continuous if
% $\exists K\in\mathbb{R}$ such that for $x_1\not=x_2:
% \frac{d_Y(f(x_1),f(x_2))}{d_X(x_1,x_2)}\le K$.}: it limits how fast a
% mechanism's output can change in relation with its input's change.

\begin{definition}[$d_{\datasets}$-privacy~\cite{chatzikokolakis2013broadening}]
    Let $d_{\datasets}:\datasets^2\rightarrow\mathbb{R}_\infty$. A
    privacy mechanism $\mecha$ satisfies $d_{\datasets}$-privacy if for all
    pairs of datasets $D_1$, $D_2$ and all sets of outputs $S\subseteq\outputs$:
    \begin{equation*}
        \proba{\mech{D_1}\in S} \leq
            e^{d_{\datasets}\left(D_1,D_2\right)} \cdot\proba{\mech{D_2}\in S}.
    \end{equation*}
\end{definition}

When $d_{\datasets}$ is proportional to the Hamiltonian difference between
datasets, this is equivalent to $\eps$-DP\@. In $d_{\datasets}$-privacy, the
$d_{\datasets}$ function specifies both the privacy parameter \emph{and} the
definition of neighborhood: it can simply return $\infty$ on non-neighboring
datasets, and vary the privacy level across inputs for neighboring datasets. In
the original definition, the authors impose that $d_{\datasets}$ is
\emph{symmetric}, but this condition can also be relaxed to allow
$d_{\datasets}$-privacy to extend definitions like one-sided DP\@.

Equivalent definitions of $d_\datasets$-privacy also appeared
in~\cite{elsalamouny2016differential} as \emph{$l$-privacy}, and
in~\cite{kawamoto2019esorics} as \emph{extended DP}. Several other definitions,
such as \emph{weighted DP}~\cite{proserpio2014calibrating} (WeiDP), \emph{smooth
DP}~\cite{barber2014privacy} (SmoDP)%
\footnote{Distinct from \emph{Smoothed DP} \cite{liu2021smoothed} 
introduced later in this section.} 
and \emph{earth mover's
privacy}~\cite{fernandes2018generalized} (EMDP), can be seen as particular
instantiations of $d_\datasets$-privacy for specific functions $d$ measuring the
distance between datasets. This is also the case for some definitions tailored
for location privacy, like
\emph{geo-graph-indistinguishability}~\cite{takagi2019geo,takagi2020geo}, which specifically
applies to network graphs.

Random DP can also be combined with changing the neighborhood definition:
in~\cite{xiao2015protecting}, the authors define \emph{DP on a $\del$-location
set}\footnote{Distinct from \emph{DP on $r$-location 
	set}~\cite{chen2018differentially}, which we mention in Section~\ref{sec:n}.}, 
for which the neighborhood is defined by a set of
``plausible'' locations around the true location of a user. A notable definition
using the same combination of dimensions is \emph{distributional
privacy}\footnote{Another definition with the same name is introduced in
~\cite{zhou2009differential}, we mention it in Section~\ref{sec:n}.},
introduced in~\cite{roth2010new,blum2013learning}: it combines random DP (for a
large family of distributions) and free lunch privacy.

\begin{definition}[$(\eps,\gamma)$-distributional privacy~\cite{roth2010new,blum2013learning}]
    An algorithm $\mecha$ satisfies $(\eps,\gamma)$-distributional privacy
    (DlPr\textsuperscript{\cite{roth2010new,blum2013learning}}) if for any
    distribution $\pi$ over possible tuples, if $D_1,D_2\in\datasets$ are picked
    by randomly drawing a fixed number $n$ of elements from $\pi$ without
    replacement, then with probability at least $1-\gamma$ over the choice of
    $D_1$ and $D_2$, $\mech{D_1}\indeps\mech{D_2}$.
\end{definition}

Interestingly, this definition captures an intuition similar to the variants
in~\ref{sec:r}: the adversary can only learn properties of the data-generating
distribution, but not about particular samples (except with probability lower
than $\gamma$). The authors also prove that if $\mecha$ is $(\eps,\gamma)$-DlPr
and $\gamma=o(\frac1{n^2})$, where $n$ is the size of the dataset being
generated, then $\mecha$ is also $\eps$-DP.

\subsubsection*{Combination with Q}

Different risk models, like the definitions in Section~\ref{sec:q}, are also
compatible with varying the privacy parameters across inputs. For example,
in~\cite{krehbiel2019choosing}, the author proposes \emph{endogeneous DP}
(EndDP), which is a combination of $\epsdel$-DP and personalized DP\@. Similarly,
\emph{pseudo-metric DP} (PsDP), defined in~\cite{dimitrakakis2013bayesian}, is a
combination of $d_{\datasets}$-privacy and $\epsdel$-DP\@; while \emph{extended
divergence DP} (EDivDP), defined in~\cite{kawamoto2019allerton}, is a
combination of $d_\datasets$-privacy and divergence DP\@.

Randomly limiting the scope of the definition can also be combined with ideas
from the previous sections. For example, in~\cite{triastcyn2019bayesian}, the
authors introduce \emph{weak Bayesian DP} (WBDP), which combines random DP and
approximate DP\@. In \cite{liu2021smoothed} the authors introduced \emph{Smoothed DP}% 
\footnote{Distinct from \emph{Smooth DP} \cite{barber2014privacy} 
introduced earlier in this chapter.} 
which require $\epsdel$-indistinguishability to hold for all the 
expected dataset from a set of distributions. 
In~\cite{wang2016average}, the authors introduce \emph{on
average KL privacy}, which uses KL-divergence as quantification metric, but only
requires the property to hold for an ``average dataset'', like random DP\@; a
similar notion appears in~\cite{feldman2017calibrating} as \emph{average
leave-one-out KL stability}.
In~\cite{triastcyn2019bayesian,dandekar2020differential}, the authors introduce
\emph{Bayesian DP}\footnote{There are two other notions with the same name: 
introduced in~\cite{yang2015bayesian,leung2012bayesian}, we mention them in 
Section~\ref{sec:n} and \ref{sec:b} respectively.} (BayDP\textsuperscript{\cite{triastcyn2019bayesian}}) and
\emph{privacy at risk} (PAR) respectively; both definitions combine random DP
with probabilistic DP, with slightly different approaches: the former quantifies
over all possible datasets and changes one fixed record randomly, while the
latter selects both datasets randomly, conditioned on these datasets being
neighbors.

%In other word, it measures the amount of average privacy loss of an average
%data point, where the probability is defined only over the random coins of
%private algorithms.

%\begin{definition}[$(\gamma,\eps,\del)$-typical privacy~\cite{bassily2016typicality}]
%    A mechanism $\mecha$ satisfies $(\gamma,\eps,\del)$-typical privacy with
%    respect to $\Theta$ if for all $\theta\in\Theta$ exists a set
%    $\hat{\datasets}\subseteq\datasets$ such that
%    $\probas{D,D'\sim\theta}{D,D'\in\hat{\datasets}}\ge1-\gamma$ where $D$
%    and $D'$ are independent and for all $D_1,D_2\in\hat{\datasets}$ and for
%    all $S\subseteq\outputs$:
%    \begin{equation*}
%        \proba{\mech{D_1}\in S}\le e^\eps\proba{\mech{D_2}\in S}+\del
%    \end{equation*}
%\end{definition}

In~\cite{kifer2010towards}, Kifer et al.\ go further and generalize the
intuition from generic DP, introduced in Section~\ref{sec:n}, and generalize the
indistinguishability condition entirely. The resulting definition is \emph{also}
called generic differential privacy.

\begin{definition}[$(\mathcal{R},M)$-generic differential
  privacy~\cite{kifer2010towards,kifer2012axiomatic}]
    A privacy mechanism $\mecha$ satisfies $(\mathcal{R},M)$-generic DP
    (GcDP\textsuperscript{\cite{kifer2010towards}}) if for all measurable sets
    $S\subseteq\outputs$ and for all $\left(D_1,D_2\right)\in\mathcal{R}$:
    \begin{equation*}
    M_{D_1,D_2}(\proba{\mech{D_1}\in S})\ge\proba{\mech{D_2}\in S}
    \hspace{0.2cm}\text{and}\hspace{0.2cm}
    M_{D_1,D_2}(\proba{\mech{D_1}\notin S})\ge\proba{\mech{D_2}\notin S}
    \end{equation*}
    where $M_{D_1,D_2}:[0,1]\rightarrow[0,1]$ is a concave function continuous
    on $(0,1)$ such as $M_{D_1,D_2}(1)=1$.
\end{definition}

The privacy relation $\mathcal{R}$ is still the generalization of neighborhood
and the privacy predicate is the generalization of the
$\eps$-indistinguishability to arbitrary functions. In particular, it can
encompass all variants of described in Section~\ref{sec:q} in addition to the
ones in this section: for example, if 
$M_{D_1,D_2}(x)=\min\left(1,xe^{\eps}+\del,1-(1-x-\del)e^{-\eps}\right)$ 
holds for for all $D_1$ and $D_2$, then this is equivalent to $\epsdel$-DP. 
This definition was an attempt at
finding the most generic definition that still satisfies privacy axioms: another
extension defined in the same work, \emph{abstract DP} (AbsDP) is even more
generic, but no longer satisfies the privacy axioms.

%\begin{definition}[$(\mathcal{R},q)$-abstract differential privacy~\cite{kifer2010towards}]
%    A privacy mechanism $\mecha$ satisfies $(\mathcal{R},q)$-abstract DP for
%    the binary predicate
%    ${(q_{D_1,D_2})}_{(D_1,D_2)\in\mathcal{R}}:[0,1]\times[0,1]
%    \rightarrow\{True,False\}$ with
%    the binary irreflexive relation
%    $\mathcal{R}\subseteq\datasets\times\datasets$
%    if for all measurable set $S\subseteq\outputs$ and for all
%    $(D_1,D_2)\in\mathcal{R}$ we have
%    \begin{equation*}
%        q_{D_1,D_2}(\proba{\mech{D_1}\in S},\proba{\mech{D_2}\in S})=True
%        %q_{D_1,D_2}(\proba_{\mecha}[S|D_1],\proba_{\mecha}[S|D_1])=True
%    \end{equation*}
%\end{definition}

Definitions in this section are particularly used in the context of local
DP\footnote{For details, see Section~\ref{sec:local}.} and in particular for
applications to location privacy: various metrics have been discussed to
quantify how indistinguishable different places should be to provide users of a
local DP mechanism with meaningful privacy
protection~\cite{chatzikokolakis2017methods}.

%% file: Background.tex
\section{Background knowledge (B)}\label{sec:b}

In differential privacy, the attacker is implicitly assumed to have full
knowledge of the dataset:
their only uncertainty is whether the target belongs in the dataset or not. This
implicit assumption is also present for the definitions of the previous
dimensions: indeed, the attacker has to distinguish between two \emph{fixed}
datasets $D_1$ and $D_2$. The only source of randomness in
$\eps$-indistinguishability comes from the mechanism itself. In many cases, this
assumption is unrealistic, and it is natural to consider weaker adversaries, who
do not have full background knowledge. One of the main motivations to do so is
to use significantly less noise in the mechanism~\cite{duan2009privacy}.

The typical way to represent this uncertainty formally is to assume that the
input data comes from a certain probability distribution (named \emph{data
evolution scenario} in~\cite{kifer2012rigorous}): the randomness of this
distribution models the attacker's uncertainty. Informally, the more random
this probability distribution is, the less knowledge the attacker has. However,
the definition that follows depends whether DP is considered with the
associative or the causal view. In the associative view, the sensitive property
changes \emph{before} the data is generated: it conditions the data-generating
probability distribution. In the causal view, however, the sensitive property is
only changed \emph{after} the data is generated. The two options lead to very
distinct definitions.

\subsection{Conditioning the output on the sensitive property}

Using a probability distribution to generate the input data means that the
$\eps$-indistinguishability property cannot be expressed between two fixed
datasets. Instead, one natural way to express it is to condition this
distribution on some sensitive property. The corresponding notion,
\emph{noiseless privacy}\footnote{Another definition with the same name 
is introduced in~\cite{farokhi2019noiseless}, we mention
it in Section~\ref{sec:out}.} (NPr) was first introduced in~\cite{duan2009privacy} and
formalized in~\cite{bhaskar2011noiseless}.

\begin{definition}[$\Theps$-noiseless privacy~\cite{duan2009privacy,bhaskar2011noiseless}]
    Given a family $\Theta$ of probability distribution on $\datasets$, a mechanism
    $\mecha$ is $\Theps$-noiseless private if for all $\theta\in\Theta$, all $i$
    and all $t,t'\in\tuples$:
    \begin{equation*}
        \mech{D}\cond{D\sim\theta,D(i)=t}\ind_\eps\mech{D}\cond{D\sim\theta,D(i)=t'}.
    \end{equation*}
\end{definition}

In the original definition, the \emph{auxiliary knowledge} of the attacker is
explicitly pointed out in an additional parameter. In the case where there is no
$\del$ of $(\eps,0)$-DP, this syntactic add-on is not
necessary~\cite{desfontaines2019differential}, so we omitted it here.

This definition follows naturally from considering the associative view of DP
with the strong adversary assumption, and attempting to relax this assumption.
The exact way to model the adversary's uncertainty can be changed; for example
\emph{DP under sampling}~\cite{li2011provably}, an instance of noiseless
privacy, models it using random sampling.

This definition was not the first attempt at formalizing an adversary with
restricted background knowledge. In~\cite{machanavajjhala2009data}, authors
define \emph{$\eps$-privacy}, which represents the background knowledge as a
Dirichlet distribution (instead of an arbitrary distribution), whose parameters
are interpreted as characteristics of the attacker. However, this definition
imposes a condition on the output, but not on the mechanism which produced the
output. As such, it does not offer strong semantic guarantees like the other
definitions presented in this survey.

\subsection{Removing the effect of correlations in the data}

In~\cite{bassily2013coupled}, however, the authors argue that in the presence of
correlations in the data, noiseless privacy can be too strong, and make it
impossible to learn global properties of the data. Indeed, if one record can
have an arbitrarily large influence on the rest of the data, conditioning on the
value of this record can lead to very distinguishable outputs even if the
mechanism only depends on global properties of the data. To fix this problem,
they propose \emph{distributional DP} (DistDP), an alternative definition that
that only conditions the data-generating distribution on one possible value of
the target record, and quantifies the information leakage from the mechanism.%
\footnote{Note that the original formalization used in 
\cite{bassily2013coupled} was more abstract, and uses a 
simulator, similarly to variants introduced in Section \ref{sec:r}. }.
In~\cite{desfontaines2019differential}, the authors show
that this creates definitional issues in the presence of background knowledge,
and introduce \emph{causal DP} (CausDP), to capture the same intuition without
encountering the same problems.

\begin{definition}[$\Theps$-causal differential privacy~\cite{desfontaines2019differential}]
    Given a family $\Theta$ of probability distributions on $\datasets$, a
    mechanism $\mecha$ satisfies $\Theps$-causal DP (CausDP) if for all
    probability distributions $\theta\in\Theta$, for all $i$ and all
    $t,t'\in\tuples$:
    \begin{equation*}
        \mech{D}\cond{D\sim\theta,D(i)=t}\indeps\mech{D_{i\rightarrow t'}}\cond{D\sim\theta,D(i)=t}.
    \end{equation*}
    where $D_{i\rightarrow t'}$ is the dataset obtained by changing the $i$-th
    record of $D$ into $t'$.
\end{definition}

In this definition, one record is changed \emph{after} the dataset has been
generated, so it does not affect other records through dependence relationships.
These dependence relationships are the only difference between noiseless privacy
and causal DP\@: when each record is independent of all others, this definition
is equivalent to noiseless privacy~\cite{desfontaines2019differential}.

\subsection{Multidimensional definitions}

Limiting the background knowledge of an attacker is orthogonal to the dimensions
introduced previously: one can modify the risk model, introduce different
neighborhood definitions, or even vary the privacy parameters across the
protected properties along with limiting the attacker background knowledge. 

\subsubsection*{Combination with Q}

Modifying the risk model while limiting the attacker's background knowledge has
interesting consequences. In~\cite{desfontaines2019differential}, the authors
show that two options are possible: either consider the background knowledge as
additional information given to the attacker or let the attacker
\emph{influence} the background knowledge. This distinction between an
\emph{active} and a \emph{passive} attacker does not matter if only the
worst-case scenario is considered, like in noiseless privacy. However, under
different risk models, such as allowing a small probability of error, they lead
to two different definitions.

Both of these definitions use an adapted version of the privacy loss random
variable (PLRV) which explicitly models the attacker background knowledge: the
data-generating distribution not only generates a dataset $D$ but also some
auxiliary knowledge $B$, with values in a set $\knowledges$.

\begin{definition}[PLRV for partial knowledge~\cite{desfontaines2019differential}]
Given a mechanism $\mecha$, a distribution $\theta$ with values in
$\datasets\times\knowledges$, an indice $i$, and values $a,b\in\tuples$, the
PLRV of an output $O\in\outputs$ given partial knowledge $\hatB$ is:
\begin{equation*}
\privlossMit{a}{b}(O,\hatB) =
  \ln\left(\frac{
    \probasc{(D,B)\sim\theta}{\mech{D}=O}{D(i)=a,B=\hatB}
  }{
    \probasc{(D,B)\sim\theta}{\mech{D}=O}{D(i)=b,B=\hatB}
  }\right).
\end{equation*}
if the three conditions are satisfied:
\begin{enumerate}
  \item $\probas{(D,B)\sim\theta}{D(i)=a,B=\hatB}\neq0$ and
    $\probas{D\sim\theta}{D(i)=b,B=\hatB}\neq0$
  \item $\probasc{(D,B)\sim\theta}{\mech{D}=O}{D(i)=a,B=\hatB}\neq0$
  \item $\probasc{(D,B)\sim\theta}{\mech{D}=O}{D(i)=b,B=\hatB}\neq0$.
\end{enumerate}
If condition 1 does not hold, then $\privlossMit{a}{b}(O,B)=0$. Else, if
condition 2 does not hold, then $\privlossMit{a}{b}(O,B)=-\infty$. Else, if
condition 3 does not hold, then $\privlossMit{a}{b}(O,B)=\infty$.
\end{definition}

This formalization can then be used to adapt noiseless privacy to a risk model
similar to $\epsdel$-DP, in the case of an active or a passive attacker. The
active variant, \emph{active partial knowledge differential
privacy} (APKDP), quantifies over all possible values of the background
knowledge. It was first introduced
in~\cite{bhaskar2011noiseless,bassily2013coupled} as \emph{noiseless
privacy} and reformulated in~\cite{desfontaines2019differential} to clarify that
it implicitly assumes an active attacker. 

\begin{definition}[$\Thepsdel$-active partial knowledge differential
  privacy~\cite{bhaskar2011noiseless,bassily2013coupled,desfontaines2019differential}]
    Given a family $\Theta$ of probability distribution on
    $\datasets\times\knowledges$, a mechanism $\mecha$ is $\Thepsdel$-active
    partial knowledge DP (APKDP) if for all $\theta\in\Theta$, all indices $i$,
    all $t,t'\in\tuples$, and all possible values $\hatB$ of the background
    knowledge:
    \begin{equation*}
      \expects{(D,B)\sim\theta\cond{D(i)=t,B=\hatB},O\sim\mech{D}}{\max\left(0,1-e^{\eps-\privlossMit{t}{t'}(O,\hatB)}\right)}\le\del.
    \end{equation*}
\end{definition}

One specialization of this definition is \emph{DP under
sampling}~\cite{li2011provably} (DPuS), which mandates DP to be satisfied after
a random sampling is applied to the dataset. The
authors use this definition to show that applying $k$-anonymity to a randomly
sampled dataset provides differential privacy; but this definition could also be
used on its own, to model the attacker's uncertainty using a randomly sampled
distribution.

The second definition, \emph{passive partial knowledge differential
privacy}~\cite{desfontaines2019differential} (PPKDP), is strictly weaker: it
models a passive attacker, who cannot choose their background knowledge, and
thus cannot influence the data. In this context, $\del$ does not only apply to
the output of the mechanism, but also to the value of the background knowledge.

\begin{definition}[$\Thepsdel$-passive partial knowledge differential privacy~\cite{desfontaines2019differential}]
    Given a family $\Theta$ of probability distribution on
    $\datasets\times\knowledges$, a mechanism $\mecha$ is $\Thepsdel$-passive
    partial knowledge DP (PPKDP) if for all $\theta\in\Theta$, all indices $i$,
    and all $t,t'\in\tuples$:
    \begin{equation*}
      \expects{(D,B)\sim\theta\cond{D(i)=t},O\sim\mech{D}}{\max\left(0,1-e^{\eps-\privlossMit{t}{t'}(O,B)}\right)}\le\del.
    \end{equation*}
\end{definition}

Causal DP can also be adapted to a risk model similar to $\epsdel$-DP\@:
in~\cite{burchard2019empirical}, authors introduce a similar notion to causal DP
as \emph{inherit DP} (InhDP), with the small difference that the second dataset
is obtained by removing one record from the first dataset, instead of replacing
it; and $\epsdel$-indistinguishability is used. The authors also define
\emph{empirical DP}~\cite{burchard2019empirical}\footnote{Another definition with 
the same name is introduced in~\cite{abowd2013differential}, we
mention it in Section~\ref{sec:out}.}, which is identical, except the
empirical distribution is used instead of the actual data distribution, in
context where the latter is unknown. In both cases, the influence of $\del$ on
the attacker model is unclear.

\subsubsection*{Combination with N}

Modifying the neighborhood definition is simpler: it is clearly orthogonal to
the dimensions introduced in this section. In all definitions of this section so
far, the two possibilities between which the adversary must distinguish are
similar to bounded DP\@. This can easily be changed to choose other properties
to protect from the attacker. This is done in \emph{pufferfish
privacy}~\cite{kifer2012rigorous} (PFPr), which extends the concept of
neighboring datasets to neighboring \emph{distributions} of datasets.

\begin{definition}[$(\Theta,\Phi,\eps)$-pufferfish privacy~\cite{kifer2012rigorous}]
    Given a family of probability distributions $\Theta$ on $\datasets$, and a
    family of pairs of predicates $\Phi$ on datasets, a mechanism $\mecha$ verifies
    $(\Theta,\Phi,\eps)$-pufferfish privacy if for all distributions
    $\theta\in\Theta$ and all pairs of predicates
    $\left(\phi_1,\phi_2\right)\in\Phi$:
    \begin{equation*}
    \mech{D}_{|D\sim\theta,\phi_1(D)}\ind_\eps\mech{D}_{|D\sim\theta,\phi_2(D)}
    \end{equation*}
\end{definition}

Pufferfish privacy extends the concept of neighboring datasets to neighboring
\emph{distributions} of datasets; starting with a set of data-generating
distributions, then conditioning them on sensitive attributes. The result
compares pairs of distributions encompasses noiseless privacy, as well as other
notions. For example, it captures \emph{bayesian DP}\footnote{There are two 
other notions with the same name: introduced in~\cite{yang2015bayesian,triastcyn2019bayesian}, 
we mention them in Section~\ref{sec:n} and \ref{sec:v} respectively.}
(BayDP\textsuperscript{\cite{leung2012bayesian}}), introduced
in~\cite{leung2012bayesian}, in which neighboring records have up to $k$ fixed
elements in common and all other elements are generated randomly from a
distribution $\pi$.

The same idea can be formalized by comparing pairs of distributions directly.
This is done in~\cite{jelasity2014distributional,kawamoto2019esorics} via
\emph{distribution privacy} (DnPr). The two formalisms are equivalent: an
arbitrary pair of distributions can be seen as a single distribution,
conditioned on the value of a secret parameter. Distribution privacy was
instantiated in~\cite{geumlek2019profile} via \emph{profile-based DP} (PBDP), in
which the attacker tries to distinguish between different probabilistic user
profiles. A similar idea was proposed in \cite{Bansal2019ExtendingTF} as 
\emph{robust privacy}, which uses lossy Wasserstein distance over 
the corresponding outputs to define the neighbourhood of the inputs.

Further relaxations encompassing the introduced dimensions are
\emph{probabilistic distribution privacy}~\cite{kawamoto2019esorics}
(PDnPr), a combination of distribution privacy and probabilistic DP,
\emph{extended distribution privacy}~\cite{kawamoto2019esorics} (EDnPr), a
combination of distribution privacy and $d_\datasets$-privacy, \emph{divergence
distribution privacy}~\cite{kawamoto2019allerton} (DDnPr), a combination of
distribution privacy, and \emph{extended divergence distribution
privacy}~\cite{kawamoto2019allerton} (EDDnPr), which combines the latter two
definitions. Finally, \emph{divergence distribution privacy with auxiliary
inputs}~\cite{kawamoto2019allerton} considers the setting where the attacker
might not know the input probability distribution perfectly.

Definitions of this section are an active area of research; a typical question
is to quantify in which conditions deterministic mechanisms can provide some
level privacy. However, they are not used a lot in practice, likely because of
their fragility: if the assumptions about the attacker's limited background
knowledge are wrong in practice, then the definitions do not provide any
guarantee of protection.

%% file: Formalism.tex
\section{Change in formalism (F)}\label{sec:f}

The definition of differential privacy using $\eps$-indistinguishability
compares the distribution of outputs given two neighboring inputs. This is not
the only way to capture the idea that a attacker should not be able to gain too
much information on the dataset. Other formalisms have been proposed, which
model the attacker more explicitly.

One such formalism reformulates DP in terms of hypothesis testing by limiting
the type I and the type II error of the hypothesis that the output $O$ of a
mechanism originates from $D_1$ (instead of $D_2$). Other formalisms model the
attacker explicitly, by formalizing their prior belief as a probability
distribution over all possible datasets. This can be done in two distinct ways.
Some variants consider a specific prior (or family of possible priors) of the
attacker, implicitly assuming a limited background knowledge, like in
Section~\ref{sec:b}. We show that these variants can be interpreted as changing
the prior-posterior bounds of the attacker. Finally, rather than comparing prior
and posterior, a third formalism compares two possible posteriors, quantifying
over all possible priors.

Definitions in this section provide a deeper understanding of the guarantees
given by differential privacy, and some of them lead to tighter and simpler
theorems on differential privacy, like composition or amplification results.

\subsection{Hypothesis testing}

First, differential privacy can be interpreted in terms of \emph{hypothesis
	testing}~\cite{wasserman2010statistical,kairouz2017composition}. In this
context, an adversary who wants to know whether the output $O$ of a mechanism
originates from $D_1$ (the \emph{null hypothesis}) or $D_2$ (the
\emph{alternative hypothesis}). Calling $S$ the rejection region, the
probability of false alarm (type I error), when the null hypothesis is true but
rejected, is $\mathbb{P}_{FA}=\proba{\mech{D_1}\in S}$. The probability of
missed detection (type II error), when the null hypothesis is false but
retained, is $\mathbb{P}_{MD}=\proba{\mech{D_2}\in\outputs/S}$.

It is possible to use these probabilities, to reformulate DP: 

\begin{tabular}{lll}
	$\eps$-DP & $\Leftrightarrow$ & 
	\begin{tabular}{l}
		$\mathbb{P}_{FA}+e^\eps\mathbb{P}_{MD}\ge1$ \\
		$e^\eps\mathbb{P}_{FA}+\mathbb{P}_{MD}\ge1$ 
	\end{tabular} for all $S\subseteq\outputs$ \\
	$\eps$-DP & $\Leftrightarrow$ & 
	$\mathbb{P}_{FA}+\mathbb{P}_{MD}\ge\frac2{1+e^\eps}$ \\
	$\epsdel$-DP & $\Leftrightarrow$ & $(\mathbb{P}_{FA},\mathbb{P}_{MD})=$	
	\begin{tabular}{l}
		$\{(\alpha,\beta)\in[0,1]\times[0,1]:$ \\
		$(1-\alpha\le e^\eps\beta+\del)\}$
	\end{tabular}
\end{tabular}

This hypotheses testing interpretation was used in~\cite{dong2019gaussian} to
define \emph{$f$-differential privacy} ($f$-DP), which avoids difficulties
associated with divergence based relaxations. Specifically, its composition
theorem is lossless as it provides a computationally tractable tool for
analytically approximating the privacy loss. Moreover, there is a general
duality between $f$-DP and infinite collections of $\epsdel$-DP guarantees. 

\begin{definition}[$f$-differential privacy~\cite{dong2019gaussian}]
	Let $f:[0,1]\rightarrow[0,1]$ be a convex, continuous, and non-increasing
	function such that for all $x\in[0,1]$, $f(x)\le1-x$. A privacy mechanism
	$\mecha$ satisfies $f$-DP if for all neighboring $D_1,D_2$ and all
	$x\in[0,1]$:
	\begin{equation*}
	\inf_S\left\{1-\proba{\mech{D_2}\in S}\vert\proba{\mech{D_1}\in S}\le x\right\}\ge f(x).
	\end{equation*}
\end{definition}

Here, $S$ is the rejection region; and the infimum is the trade-off function
between $\mech{D_1}$ and $\mech{D_2}$. The authors also introduce \emph{Gaussian
differential privacy} (GaussDP) as an instance of $f$-differential privacy,
which tightly bounds from below the hardness of determining whether an
individual's data was used in a computation than telling apart two shifted
Gaussian distributions. Besides, \emph{weak} and \emph{strong federated f-DP} 
\cite{zheng2021federated} are also defined that adopts the definition for 
federated learning \cite{mcmahan2021advances}. They describe the privacy guarantee
against an individual and a group of adversaries respectively. 
%Moreover, ~\cite{dong2019gaussian} show the general Central Limit Theorem, which explains why all compositions look like Gaussian mechanisms.

\subsection{Changing the shape of the prior-posterior bounds}\label{subsec:shape}

Differential privacy can be interpreted as giving a bound on the
posterior of a Bayesian attacker as a function of their prior. This is exactly
the case in \emph{indistinguishable privacy} (IndPr), an equivalent
reformulation of differential privacy defined in~\cite{liu2013semantic}: suppose
that the attacker is trying to distinguish between two options $D=D_1$ and
$D=D_2$, where $D_1$ corresponds to the option ``$t\in D$'' and $D_2$ to
``$t\notin D$''. Initially, they associate a certain prior probability
$\proba{t\in D}$ to the first option. When they observe the output of the
algorithm, this becomes the posterior probability $\proba{t\in D|\mech{D}=O}$.
Combining the definition of $\eps$-DP and the Bayes Theorem, we
have\footnote{Note that the original formalization used
in~\cite{liu2013semantic} was more abstract, and it used polynomially bounded
adversaries what we introduce in Section~\ref{sec:c}.}:
\begin{equation*}
    \frac{\proba{t\in D|\mech{D}=O}}{\proba{t\notin D|\mech{D}=O}} \le e^\eps\cdot\frac{\proba{t\in D}}{\proba{t\notin D}}
    \hspace{0.2cm}\Rightarrow\hspace{0.2cm}
    \proba{t\in D|\mech{D}=O} \le \frac{e^\eps\cdot\proba{t\in D}}{1+\left(e^\eps-1\right)\proba{t\in D}}.
\end{equation*}

\begin{figure}[!h]
    \centering
    \includegraphics[width=3.7cm]{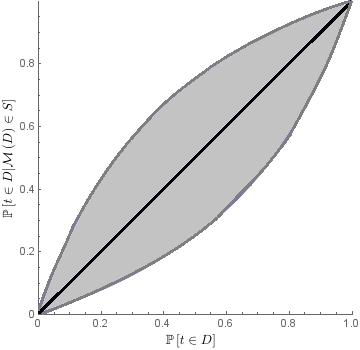}
    \includegraphics[width=3.7cm]{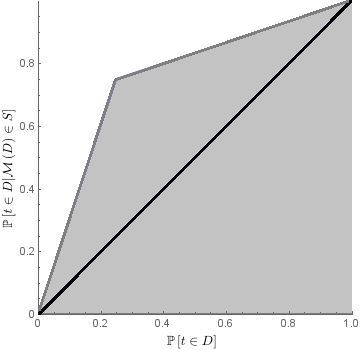}
    \includegraphics[width=3.7cm]{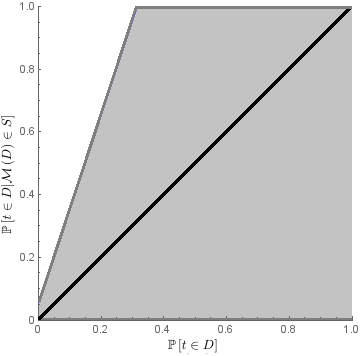}
    \includegraphics[width=3.7cm]{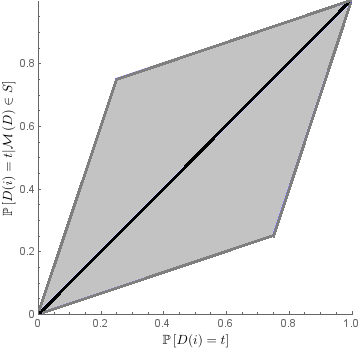}
    \caption{From left to right, using $\eps=\ln{3}$: posterior-prior
    bounds in differential privacy, positive membership privacy, adversarial
    privacy (with $\del=0.05$) and aposteriori noiseless
    privacy.}\label{fig:bounds}
\end{figure}

A similar, symmetric lower bound can be obtained. Hence, differential privacy can
be interpreted as bounding the posterior level of certainty of a Bayesian
attacker as a function of its prior. We visualize these bounds on the 
first two figures of Figure~\ref{fig:bounds}.

Some variants of differential privacy use this idea in their formalism, rather
than obtaining the posterior bound as a corollary to the classical DP
definition. For example, \emph{positive membership privacy}
(PMPr)~\cite{li2013membership} requires that the posterior does not increase too
much compared to the prior. Like noiseless privacy, it assumes an attacker with
limited background knowledge.

\begin{definition}[$\Theps$-positive membership privacy~\cite{li2013membership}]
  A privacy mechanism $\mecha$ provides $\Theps$-positive membership privacy if
  for any distribution $\theta\in\Theta$, any record $t\in\datasets$ and any
  $S\subseteq\outputs$:
  \begin{align*}
    \probas{D\sim\theta}{t\in D|\mech{D}\in S} & \leq e^\eps\probas{D\sim\theta}{t\in D}
    \hspace{0.2cm}\text{and}\hspace{0.2cm}
    \probas{D\sim\theta}{t\notin D|\mech{D}\in S} & \geq e^{-\eps}\probas{D\sim\theta}{t\notin D}.
  \end{align*}
\end{definition}

%\begin{theorem*}[Relation~\cite{li2013membership}]
	%A mechanism satisfies $\eps$-bounded DP if and only if it provides $(\mathfrak{D}_B,e^{\eps})-PMP$ where $\mathfrak{D}_B$ includes distributions that are the conditional distributions of some MI\footnote{$MI(X,Y)=H(X)+H(Y)-H(X,Y)$ where $H$ is the entropy.} distribution conditioned upon that all datasets with non-zero probability have the same size.
%\end{theorem*}

Note that this definition is \emph{asymmetric}: the posterior is bounded from
above, but not from below. In the same paper, the authors also define
\emph{negative membership privacy} (NMPr), which provides the symmetric lower
bound, and \emph{membership privacy}\footnote{Another definition with the same
name is introduced in~\cite{sablayrolles2019white}, we mention it in
Section~\ref{sec:scope}.} (MPr), which is the conjunction of positive and
negative membership privacy. They show that this definition can represent
differential privacy (in its bounded and unbounded variants), as well as other
definitions like \emph{differential identifiability}~\cite{lee2012differential}
and \emph{sampling DP}~\cite{li2011provably}, which we mention in
Section~\ref{sec:out}. Bounding the ratio between prior and posterior by
$e^\eps$ is also done in the context of location privacy:
in~\cite{dong2018limitations}, authors define \emph{$\eps$-DP location
obfuscation}, which formalizes the same intuition as membership privacy.

%\begin{definition}[$(\hat{\belief},\gamma)$-negative membership privacy~\cite{li2013membership}]
%    A privacy mechanism $\mecha$ satisfies $\gamma$-negative membership P under
%    a family of distributions $\hat{\belief}$ where $\gamma\geq1$ if and only if
%    for all $S\subset\outputs$ and any distribution $\belief\in\hat{\belief}$
%    and any record $t\in\mathcal{T}$:
%    \begin{align*}
%    \proba{\neg t|S}\leq\gamma\cdot\proba{\neg t}\\
%    \proba{t|S}\geq\frac{\proba{t}}{\gamma}\\
%    %\proba{t\not\in B|\mech{B}\in S}\leq\gamma\cdot\proba{t\not\in B}\\
%    %\proba{t\in B|\mech{B}\in S}\leq\frac{\proba{t\in B}}{\gamma}
%    \end{align*}
%    where $D\in\datasets$ is drawn according to $\belief$.
%\end{definition}

A previous attempt at formalizing the same idea was presented
in~\cite{rastogi2009relationship} as \emph{adversarial privacy} (AdvPr). This definition
is similar to positive membership privacy, except only the first relation is
used, and there is a small additive $\del$ as in approximate DP\@. We visualize
the corresponding bounds on the third figure of Figure~\ref{fig:bounds}.

\begin{definition}[$\Thepsdel$-adversarial privacy~\cite{rastogi2009relationship}]
    An algorithm $\mecha$ is $\Thepsdel$-adversarial private if for all
    $S\subseteq \outputs$, tuples $t$, and distributions $\theta\in\Theta$:
    \begin{equation*}
        \probas{D\sim\theta}{t\in D|\mech{D}\in S} \leq e^\eps\cdot\probas{D\sim\theta}{t\in D}+\del.
    \end{equation*}
\end{definition}

%\begin{theorem*}[Relation~\cite{rastogi2009relationship}]
%    \emph{$\epsdel$-bounded DP} is equivalent with $\Thepsdel$-AdvPr where $\Theta$
%    corresponds to planar total log-submodular (PTLM) adversaries\footnote{For
%    details, see Definitions 2.8, 2.9, 2.10 and 2.11
%    in~\cite{rastogi2009relationship}.}.
%\end{theorem*}

Adversarial privacy (without $\del$) was also redefined
in~\cite{wu2017information} as \emph{information privacy}\footnote{Another
definition with the same name is introduced in~\cite{du2012privacy}, we mention
it later in this section.}.

Finally, \emph{aposteriori noiseless privacy} (ANPr) is a similar variant of
noiseless privacy introduced in~\cite{bhaskar2011noiseless}; the corresponding
bounds can be seen on the last figure of Figure~\ref{fig:bounds}.

\begin{definition}[$\Theps$-aposteriori noiseless privacy~\cite{bhaskar2011noiseless}]
A mechanism $\mecha$ is said to be $\Theps$-aposteriori noiseless private (ANPr)
if for all $\theta\in\Theta$, all $S\subseteq\outputs$ and all $i$:
\begin{equation*}
    {D(i)}\cond{D\sim\theta,\mech{D}\in S}\indeps{D(i)}\cond{D\sim\theta}.
    %e^{-\eps} \leq \frac{\probas{D\sim\theta}{D(i)=t|\mech{D}\in S}}{\probas{D\sim\theta}{D(i)=t'}} \leq e^\eps.
\end{equation*}
\end{definition}

%In~\cite{bhaskar2011noiseless}, authors show that if $f$ satisfies
%$(\Theta,\eps)$-noiseless privacy, then it also satisfies
%$(\Theta,2\eps)$-aposteriori noiseless privacy; and, if $f$ satisfies
%$(\Theta,\eps)$-aposteriori noiseless privacy, then it also satisfies
%$(\Theta,\eps)$-noiseless privacy.

We visualize the prior/posterior bounds for these various definitions in
Figure~\ref{fig:bounds}.

\subsection{Comparing two posteriors}

In~\cite{ganta2008composition,kasiviswanathan2014semantics}, the authors propose an approach that
captures an intuitive idea proposed by Dwork in~\cite{dwork2006differential}:
``any conclusions drawn from the output of a private algorithm must be
similar whether or not an individual's data is present in the input or not''.
They define \emph{semantic privacy} (SemPr): instead of comparing the posterior with the
prior belief like in DP, this bounds the difference between two posterior belief
distributions, depending on which dataset was secretly chosen. The distance
chosen to represent the idea that those two posterior belief distributions are
close is the statistical distance. One important difference between the definitions
in the previous subsection is that semantic privacy quantifies over all
possible priors: like in DP, the attacker is assumed to have arbitrary
background knowledge.

\begin{definition}[$\eps$-semantic privacy~\cite{ganta2008composition,kasiviswanathan2014semantics}]
  A mechanism $\mecha$ is $\eps$-semantically private if for any distribution
  over datasets $\theta$, any index $i$, any $S\subseteq\outputs$, and any set
  of datasets $X\subseteq\datasets$:
  \begin{align*}
    \left|\probasc{D\sim\theta}{D\in X}{\mech{D}\in S}
       - \probasc{D\sim\theta}{D\in X}{\mech{D_{-i}}\in S}\right|\le\eps.
  \end{align*}
\end{definition}

%\begin{theorem*}[Relation~\cite{kasiviswanathan2014semantics}]
%    $\eps$-DP implies $(e^{\eps}-1)$-semantic privacy,
%    while $\eps$-semantic privacy implies
%    %this is from the older version of the paper from 2008
%    %$\log(4\eps+1)$-DP.
%    $6\eps$-DP if $\eps\le0.45$.
%\end{theorem*}

A couple of other definitions also compare posteriors directly:
\emph{inferential privacy}~\cite{ghosh2016inferential} is a reformulation of
noiseless privacy, and \emph{range-bounded privacy}~\cite{durfee2019practical}
(RBPr) requires that two different values of the PLRV are close to each other
(instead of being between centered aroud zero like in $\eps$-DP). It is
equivalent to $\eps$-DP up to a change in parameters, and is used as a technical
tool to prove composition results.

%\begin{definition}[$\eps$-posteriori differential privacy~\cite{wang2014tradeoff}]
%    A mechanism $\mecha$ satisfies $\eps$-posteriori differential privacy 
%    if for any pair data sets $D_1,D_2\in\datasets$ differing in a single 
%    record and for all $O\in\outputs$:
%    \begin{equation*}
%        \proba{D=D_1|\mech{D}=O}\le e^\eps\proba{D=D_2|\mech{D}=O}
%    \end{equation*}
%\end{definition}
%\begin{definition}[$\eps$-range-bounded~\cite{durfee2019practical}]
%    A privacy mechanism $\mecha$ is $\eps$-rage-bounded (RB) if for all two neighboring datasets $D_1,D_2$ and outcomes $O_1,O_2$:
%    \begin{equation*}
%        \frac{\proba{\mech{D_1}=O_1}}{\proba{\mech{D_2}=O_1}}\le e^\eps \frac{\proba{\mech{D_1}=O_2}}{\proba{\mech{D_2}=O_2}}
%    \end{equation*}
%\end{definition}

%\begin{theorem*}[Relation~\cite{durfee2019practical}]
%If a mechanism satisfies $\eps$-DP than it also satisfies $2\eps$-RB. Moreover, if it is $\eps$-RB, that it is also $\eps$-DP.
%\end{theorem*}

\subsection{Multidimensional definitions}\label{sec:d-multi}

This dimension could be combined with other dimensions fairly easily; 
several DP modifications from this section does belong to multiple dimensions, 
but to introduce the concept we overlooked these details.

Definitions that limit the background knowledge of the adversary explicitly
formulate it as a probability distribution. As such, they are natural candidates
for Bayesian reformulations. In~\cite{wu2017information}, the authors introduce
\emph{identity DP}, which is an equivalent Bayesian reformulation of noiseless
privacy.

Another example is \emph{inference-based causal
DP}~\cite{desfontaines2019differential}, similar to aposteriori noiseless DP,
except it uses causal DP instead of noiseless DP\@.

%\begin{definition}[$\eps$-identity differential privacy~\cite{wu2017information}]
%    A mechanism $\mecha$ satisfies $\eps$-identity differential privacy with respect to $X$ if for each $X_i$, any $t_i,t'_i\in\tuples_i$, and any $S\subseteq\outputs$:
%    \begin{equation*}
%        \frac{\proba{X_i=t_i|\mech{X}\in S}}{\proba{X_i=t'_i|\mech{X}\in S}}\le e^\eps\cdot\frac{\proba{X_i=t_i}}{\proba{X_i=t'_i}}
%    \end{equation*}
%    where $X_i$ is the $i$th individual's random variable with universe $\tuples_i$ and $X=\{X_1,\dots,X_n\}$ is the dataset random variable.
%\end{definition}

%\begin{theorem*}[Relation~\cite{wu2017information}]
%    If the individuals are independent then $\eps$-DP implies $\eps$-ldeDP. Moreover, if $k$ individuals are dependent, then $\frac\eps{k}$-DP would ensure $\eps$-IdeDP.
%\end{theorem*}

\begin{definition}[$\Theps$-inference-based causal differential privacy~\cite{desfontaines2019differential}]
  Given a family $\Theta$ of probability distribution on $\datasets$, a mechanism
  $\mecha$ satisfies $\Theps$-inference-based distributional DP if (IBCDP) for all
  probability distributions $\theta\in\Theta$, for all $i$, all $a,b\in\tuples$
  and all outputs $O$:
  \begin{equation*}
    {D(i)}\cond{D\sim\theta,\mech{D}=O}\indeps{D(i)}\cond{D\sim\theta,\mech{D_{i\rightarrow b}}=O}.
  \end{equation*}
  where $D_{i\rightarrow b}$ is the dataset obtained by changing the $i$-th
  record of $D$ into $b$.
\end{definition}

The authors of semantic privacy also combined it with probabilistic DP, and 
simply called it \emph{$\epsdel$-semantic privacy}. 
Further, it is possible to consider different definitions of neighborhood.
In~\cite{du2012privacy}, authors introduce \emph{information privacy}\footnote{
Another definition with the same name is introduced in~\cite{wu2017information}, 
we mention it earlier in this section.} (InfPr), which can be seen as a posteriori 
noiseless privacy combined with free lunch privacy: rather than only considering 
the knowledge gain of the adversary on one particular user, it considers 
its knowledge gain about any possible group of values of the dataset.

\begin{definition}[$\Theps$-information privacy~\cite{du2012privacy}]
  A mechanism $\mecha$ satisfies $\Theps$-information privacy if for all
  probability distributions $\theta\in\Theta$, all $D\in\datasets$ and all
  $O\in\outputs$, $D\cond{D\sim\theta}\indeps D\cond{D\sim\theta,\mech{D}=O}$.
\end{definition}

The authors further prove that if $\Theta$ contains a distribution whose support
is $\datasets$, then $\Theps$-InfPr implies $2\eps$-DP\@.

Apart from the hypothesis testing reformulations, that can be used to improve
composition and amplification results, the definitions  in this section mostly
appear in theoretical research papers, to provide a deeper understanding of
guarantees offered by DP and its alternatives. They do not seem to be used in
practical applications.

%% file: Relativization.tex
\section{Relativization of the knowledge gain (R)}\label{sec:r}

A differentially private mechanism does not reveal more than a bounded amount of
probabilistic information about a user. This view does not explicitly take into
account other ways information can leak, like side-channel functions or
knowledge about the structure of a social network. We found two approaches that
attempt to include such auxiliary functions in DP variants. One possibility is
to weaken DP by allowing or disregarding a certain amount of leakage; another option is to
explicitly forbid the mechanism to reveal more than another function, considered
to be safe for release.

\subsection{Taking into account auxiliary leakage function}

In~\cite{ligett2020bounded}, authors define \emph{bounded leakage DP}, which
quantifies the privacy that is maintained by a mechanism despite bounded,
additional leakage of information by some leakage function. Interestingly, this
leakage function $P$ shares the randomness of the privacy mechanism: it can, for
example, capture side-channel leakage from the mechanism's execution. In the
formal definition of this DP variant, the randomness is \emph{explicit}: the
privacy mechanism and the leakage takes the random bits $r\in{\{0,1\}}^*$ as an
additional parameter. 

\begin{definition}[$\left(P,\eps,\del\right)$-bounded leakage differential privacy~\cite{ligett2020bounded}]
    Let $P:\datasets\times{\{0,1\}}^*$ be a leakage function. A privacy
    mechanism $\mecha$ is $\left(P,\eps,\del\right)$-bounded leakage
    differentially private (BLDP) if for all pairs of neighboring datasets $D_1$
    and $D_2$, all outputs $O_P$ of $P$ such that
    $\proba{P\left(D_1,r\right)=O_P}\neq0$ and
    $\proba{P\left(D_2,r\right)=O_P}\neq0$, and all sets of outputs
    $S\subseteq\outputs$:
    \[
      \probac{\mech{D_1,r}\in S}{P\left(D_1,r\right)=O_P} \le e^\eps\cdot\probac{\mech{D_2,r}\in S}{P\left(D_2,r\right)=O_P}+\del
    \]
    where the randomness is taken over the random bits $r$.
\end{definition}

As expected, if there is no leakage ($P$ is a constant function), this is simply
$\epsdel$-DP\@. The authors also show that it is closed for post-processing and
composable. Furthermore, if the privacy mechanism is independent from the
leakage function, it is strictly weaker than differential privacy.

\subsection{Borrowing concepts from zero-knowledge proofs}

When using the associative interpretation with the independence assumption, it
is unclear how to adapt DP to correlated datasets like social networks: data
about someone's friends might reveal sensitive information about this person.
The causal interpretation of DP does not suffer from this problem, but how to
adapt the associative view to such correlated contexts? Changing the definition
of the neighborhood is one possibility (see Section~\ref{sec:n-changing}), but
it requires knowing in advance the exact impact of someone on other records. A
more robust option is to impose that the information released does not contain
more information than the result of some predefined algorithms on the data,
without the individual in question. The method for formalizing this intuition
borrows ideas from \emph{zero-knowledge proofs}~\cite{goldwasser1989knowledge}.

Instead of imposing that the result of the mechanism is roughly \emph{the same}
on neighboring datasets $D_1$ and $D_2$, the intuition is to impose that the
result of the mechanism on $D_1$ can be \emph{simulated} using only some
information about $D_2$. The corresponding definition, called
\emph{zero-knowledge privacy} and introduced in~\cite{gehrke2011towards},
captures the idea that the mechanism does not leak more information on a given
target than a certain class of aggregate metrics. This class, called \emph{model
of aggregate information} in~\cite{gehrke2011towards}, is formalized by a family
of (possibly randomized) family of algorithms $\Agg$.

\begin{definition}[$\Aggeps$-zero-knowledge privacy~\cite{gehrke2011towards}]
    Let $\Agg$ be a family of (possibly randomized) algorithms $\aggr$. A
    privacy mechanism $\mecha$ is $\Aggeps$-zero-knowledge private (ZKPr) if
    there exists an algorithm $\aggr\in\Agg$ and a simulator $\Simu$ such as for
    all datasets $D$ and indices $i$, $\mech{D}\indeps\Sim{\agg{D_{-i}}}$.
\end{definition}

In~\cite{gehrke2011towards}, authors show that $\Aggeps$-ZKPr implies $2\eps$-DP
for any $\Agg$, while $\eps$-DP implies $\Aggeps$-ZKPr if the identity function
is in $\Agg$. This is yet another way to formalize the intuition that
differential privacy protects against attackers who have full background
knowledge.

Another approach is to ignore some leakage instead of limiting it. This is done in 
\emph{subspace DP} \cite{gao2021subspace}, which require DP to holds for a sub-space 
within the projection of the range of $\mecha$. Furthermore, \emph{induced subspace DP} 
\cite{gao2021subspace} is also defined which ensures that the output of $\mecha$ does meet 
the invariant-linear external constraints. Essentially, they say for a mechanism to satisfy 
subspace DP, the function $D\rightarrow f(\mecha(D))$ must be DP for some function 
$f$.

\subsection{Multidimensional definitions}

Using a simulator allows making statements of the type ``this mechanism does not
leak more information on a given target than a certain class of aggregate
metrics''. Similarly to noiseless privacy, it is possible to explicitly limit
the attacker's background knowledge using a data-generating probability
distribution, as well as vary the neighborhood definitions to protect other
types of information than the presence and characteristics of individuals. This
is done in~\cite{bassily2013coupled} as \emph{coupled-worlds privacy} (CWPr), a
generalization of distributional DP, where a family of functions $\priv$
represents the protected attribute. 

\begin{definition}[$(\Theta,\Gamma,\eps)$-coupled-worlds privacy~\cite{bassily2013coupled}]
    Let $\Gamma$ be a family of pairs of functions $(\aggr,\priv)$. A mechanism
    $\mecha$ satisfies $(\Theta,\Gamma,\eps)$-coupled-worlds privacy if there is a
    simulator $\Simu$ such that for all distributions $\theta\in\Theta$, all
    $(\aggr,\priv)\in\Gamma$, and all possible values $p$:
    \begin{equation*}
      \mech{D}_{|D\sim\theta,\priv(D)=p}\ind_{\eps}\Simu{\left(\agg{D}\right)}_{|D\sim\theta,\priv(D)=p}
    \end{equation*}
\end{definition}

A special case of coupled-worlds privacy is also introduced
in~\cite{bassily2013coupled} as \emph{distributional DP}, already mentioned in
Section~\ref{sec:b}: each function $\priv$ captures the value of a single
record, and the corresponding function $\aggr$ outputs all other records.

Coupled-worlds privcay is a good example of combining variants from different
dimensions: it changes several aspects of the original definition according to
from \textbf{N}, \textbf{B} and \textbf{R}. Moreover, \textbf{Q} and \textbf{F}
can easily be integrated with this definition by using
$\epsdel$-indistinguishability with a Bayesian reformulation. This is done
explicitly in \emph{inference-based coupled-worlds
privacy}~\cite{bassily2013coupled}; the same paper also introduces
inference-based distributional differential privacy (IBDDP).

\begin{definition}[$(\Theta,\Gamma,\eps,\del)$-inference-based coupled-worlds privacy~\cite{bassily2013coupled}]
    Given a family $\Theta$ of probability distributions on
    $\datasets\times\knowledges$, and a family $\Gamma$ of pairs of functions
    $(\aggr,\priv)$, a mechanism $\mecha$ satisfies
    $(\Theta,\Gamma,\eps,\del)$-inference-based coupled-worlds privacy (IBCWPr) if
    there is a simulator $\Simu$ such that for all distributions
    $\theta\in\Theta$, and all $(\aggr,\priv)\in\Gamma$:
    \begin{equation*}
      {\pri{D}}\cond{(D,B)\sim\theta,\mech{D}=O,B=\hatB}\ind_{\epsdel}{\pri{D}}\cond{(D,B)\sim\theta,\Sim{\agg{D}}=O,B=\hatB}
    \end{equation*}
    with probability at least $1-\del$ over the choice of $O$ and $\hatB$.
\end{definition}

%Authors in~\cite{bassily2013coupled} also show that if $\mecha$ verifies
%$(\Theta,\Gamma,\eps,\del)$-CWPr, then it also verifies
%$(\Theta,\Gamma,3\eps,2\sqrt{\del|\Gamma|})$-IBCWPr.

Random DP was also combined with an idea similar to ZKPr:
in~\cite{bassily2016typical}, the authors introduce \emph{typical stability}
(TypSt), which combines random DP with approximate DP, except that rather using
$\epsdel$-indistinguishability between two outputs of the mechanism, it uses a
simulator that only knows data-generating distribution.

%TODO https://arxiv.org/pdf/1602.07726.pdf perfect generalization

\begin{definition}[$(\Theta,\gamma,\eps,\del)$-typical stability~\cite{bassily2016typical}]
  Given a family $\Theta$ of probability distributions on $\datasets$, a
  mechanism $\mecha$ satisfies $(\Theta,\gamma,\eps,\del)$-typical stability
  (TypSt) if for all distributions $\theta\in\Theta$, there is a simulator
  $\Simu$ such that with probability at least $1-\gamma$ over the choice of
  $D\sim\theta$, $\mech{D}\indepsdel\Simu(\theta)$.
\end{definition}

In the same paper, the authors introduce a variant of the same definition with
the same name, which compares two outputs of the mechanism; this is essentially
a combination between DlPr\textsuperscript{\cite{roth2010new,blum2013learning}}
and approximate DP\@.

We did not find any evidence that the variants and extensions of this section
are used outside of theoretical papers exploring the guarantees they provide.

%% file: Computational.tex
\section{Computational power (C)}\label{sec:c}

The $\eps$-indistinguishability property in DP is \emph{information-theoretic}:
the attacker is implicitly assumed to have infinite computing power. This is
unrealistic in practice, so it is natural to consider definitions where the
attacker only has polynomial computing power. Changing this assumption leads to
weaker data privacy definitions. In~\cite{mironov2009computational}, two
approaches have been proposed to formalize this idea: either modeling the
distinguisher explicitly as a polynomial Turing machine, either allow a
mechanism not to be technically differentially private, as long as one cannot
distinguish it from a truly differentially private one.

\subsubsection*{Using a polynomial distinguisher on the output}

The attacker is not explicit in the formalization of DP based on
$\eps$-indistinguishability. It is possible change the definition to make this
attacker explicit: model it as a \emph{distinguisher}, who tries to guess
whether a given output $O$ comes from a dataset $D_1$ or its neighbor $D_2$. In
doing so, it becomes straightforward to require this attacker to be
computationally bounded: simply model it as a probabilistic polynomial-time
Turing machine. In~\cite{mironov2009computational}, the authors introduce
IndCDP, short for \emph{Indistinguishability-based Computational DP}.

\begin{definition}[$\eps_\kappa$-IndCDP~\cite{mironov2009computational}]
    A family ${\left(\mecha_{\kappa}\right)}_{\kappa\in\mathbb{N}}$ of privacy
    mechanisms $\mecha_{\kappa}$ provides $\eps_{\kappa}$-IndCDP if there exists
    a negligible function $\negli$ such that for all non-uniform probabilistic
    polynomial-time Turing machines $\Omega$ (the distinguisher), all
    polynomials $p(\cdot)$, all sufficiently large $\kappa\in\mathbb{N}$, and
    all datasets $D_1,D_2\in\datasets$ of size at most $p(\kappa)$ that differ
    only one one record:
    \begin{equation*}
    \proba{\Omega(\mech{D_1})=1}\leq e^{\eps_{\kappa}}\cdot\proba{\Omega(\mech{D_2})=1}+\negl{\kappa}
    \end{equation*}
    where $\negli$ is a function that converges to zero asymptotically faster
    than the reciprocal of any polynomial.
\end{definition}

%\begin{theorem*}[Relation~\cite{he2017composing}]
%    If $\mecha$ satisfies $\frac\eps2$-IndCDP~\cite{mironov2009computational}, then it also satisfies $\epsdel$-OCDP. 
%\end{theorem*}

This definition can also be expressed using the authors define
\emph{differential indistinguishability}, a notion defined
in~\cite{backes2014differential} that adapts $\eps$-indistinguishability to a
polynomially bounded attacker.

\subsubsection*{Using a polynomial distinguisher on the mechanism}

Another natural option is to require that the mechanism ``looks like'' a truly
differentially private mechanism, at least to a computationally bounded
distinguisher. In~\cite{mironov2009computational}, the authors introduce SimCDP,
short for \emph{Simulation-based Computational DP}.

\begin{definition}[$\eps_{\kappa}$-SimCDP~\cite{mironov2009computational}]
    A family ${\left(\mecha_{\kappa}\right)}_{\kappa\in\mathbb{N}}$ of privacy
    mechanisms $\mecha_{\kappa}$ provides $\eps_{\kappa}$-SimCDP if there exists
    a family ${\left(\mecha'_{\kappa}\right)}_{\kappa\in\mathbb{N}}$ of
    $\eps_{\kappa}$-DP and a negligible function $\negli$ such that for all
    non-uniform probabilistic polynomial-time Turing machines $\Omega$, all
    polynomials $p(\cdot)$, all sufficiently large $\kappa\in\mathbb{N}$, and
    all datasets $D\in\datasets$ of size at most $p(\kappa)$:
    \begin{equation*}
        \proba{\Omega(\mech{D})=1}-\proba{\Omega(\mecha'(D))=1}\leq \negl{\kappa}
    \end{equation*}
    where $\negli$ is a function that converges to zero asymptotically faster
    than the reciprocal of any polynomial.
\end{definition}

In~\cite{mironov2009computational}, the authors show that $\eps_{\kappa}$-SimCDP
implies $\eps_{\kappa}$-IndCDP\@.

%ComDP can yield substantial accuracy improvements in various multiparty privacy
%problems. However, no such improvements are possible in the traditional
%client-server model for many natural statistical tasks without assuming the
%existence of sub-exponentially secure one-way functions and 2-message witness
%indistinguishable proofs for NP. (https://eprint.iacr.org/2016/820.pdf)

\subsection{Multidimensional definitions}

IndCDP has been adapted to different settings, and extended to arbitrary
neighborhood relationships. In \emph{output constrained DP} (OCDP), introduced 
in~\cite{he2017composing}, the setting is a two-party computation, each party
can have its own set of privacy parameters ($\eps_A$, $\eps_B$, $\del_A$, and
$\del_B$---the $\del$ parameters correspond to the $\negl{\kappa}$ term in
IndCDP), and the neighborhood relationship is determined by a function $f$. The
authors also propose \emph{DP for Record Linkage} (DPRL), an instance of 
OCDP that uses for a specific function $f$ that captures the need to protect
non-matching records during the execution of a private record linkage protocol.

Some DP variants which explicitly model an adversary with a simulator can
relatively easily be adapted to model a computationally bounded adversary,
simply by imposing that the simulator must be polynomial. This is done
explicitly in~\cite{gehrke2011towards}, where the authors define
\emph{computational zero-knowledge privacy} (CZKPr), which could also be adapted
to e.g., the two coupled-worlds privacy definitions as well.

Further, although we have not seen this done in practice in existing literature,
the idea behind SimCDP can in principle be adapted to any other definition:
rather than requiring that a given definition holds in an information-theoretic
fashion, it should be possible to require that the mechanism ``looks like'' a
mechanism which genuinely satisfies the definition.

Limiting the computational power of the attacker is a reasonable assumption, but
for a large class of queries, it cannot provide significant benefits over
classical DP in the typical client-server setting~\cite{groce2011limits}. Thus,
existing work using it focuses on multi-party
settings~\cite{bater2018shrinkwrap}.

%% file: Summary.tex
\section{Summarizing table}\label{sec:sum}

In this section we summarize the results from the previous 7 sections into a table 
where we list the known relations and show the properties with either referring to the 
original proof or creating a novel one. 

In Table~\ref{tab:DP_variants} we list the differential privacy variants and
extensions introduced in this work. For each, we specify their name, parameters
and where they were introduced (column~1), which dimensions they belong to
(column~2), which axioms they satisfy (column~3, post-processing on the left and
convexity on the right), whether they are composable (column~4) and how they
relate to other differential privacy notions (column~5). We do not list
definitions whose only difference is that they apply DP to other types of input,
like those from Section~\ref{sec:n-other}, or geolocation-specific definitions. 

\newcommand{\ftpp}{\footref{f:pp}}
\newcommand{\ftcv}{\footref{f:cv}}
\newcommand{\ftcp}{\footref{f:cp}}
\newcommand{\feqdp}{\footref{f:eq-dp}}
\newcommand{\fadiv}{\footref{f:a-div}}
\newcommand{\fcommon}{\footref{f:common}}
\newcommand{\facprob}{\footref{f:a-c-prob}}
\newcommand{\fad}{\footref{f:a-d}}
\newcommand{\facrand}{\footref{f:a-c-rand}}
\newcommand{\fadef}{\footref{f:a-def}}
\newcommand{\facomp}{\footref{f:a-comp}}
\newcommand{\fcd}{\footref{f:c-d}}
\newcommand{\fcb}{\footref{f:c-b}}
\newcommand{\fczk}{\footref{f:c-czk}}
\newcommand{\fccomp}{\footref{f:c-comp}}

% We first define and add all the endnotes, in the order that we want them to
% appear. Of course, we have to hide them, otherwise they will appear without
% context. We do that by printing them outside of the page: yes, it's gross, but
% apparently things like \phantom don't work. The % at the end of each line is
% to ignore the line breaks.
\hspace{40cm}%
\endnote{See Proposition \ref{prop:a-div}.}\label{f:a-div}%
\endnote{See Proposition \ref{prop:a-c-prob}.}\label{f:a-c-prob}%
\endnote{See Proposition \ref{prop:a-d}.}\label{f:a-d}%
\endnote{See Proposition \ref{prop:a-c-rand}.}\label{f:a-c-rand}%
\endnote{See Proposition \ref{prop:a-def}.}\label{f:a-def}%
\endnote{See Proposition \ref{prop:a-comp}.}\label{f:a-comp}%
\endnote{See Proposition \ref{prop:c-d}.}\label{f:c-d}%
\endnote{See Proposition \ref{prop:c-b}.}\label{f:c-b}%
\endnote{Post-processing}\label{f:pp}%
\endnote{Convexity}\label{f:cv}%
\endnote{Composition}\label{f:cp}%
\endnote{Follows from its \\ equivalence with $\eps$-DP.}\label{f:eq-dp}%
\endnote{A proof appears in the \\ paper introducing the \\ definition.}\label{f:common}%
\endnote{A proof for a special case appears in the paper introducing the
	definition.}\label{f:c-czk}%
\endnote{This claim appears in \cite{mironov2017renyi}, its proof is in the 
	unpublished full version.}\label{f:c-comp}%
\endnote{Abbreviations used for dimensions:
  \begin{itemize}
    \item \textbf{Q}: Quantification of privacy loss \\
    \item \textbf{N}: Neighborhood definition
    \item \textbf{V}: Variation of privacy loss
    \item \textbf{B}: Background knowledge
    \item \textbf{F}: Formalism of privacy loss
    \item \textbf{R}: Relativization of knowledge gain
    \item \textbf{C}: Computational power
  \end{itemize}
}\label{f:t-d}

\newcommand{\newres}{}

\setenotez{list-name={\vspace{-0.75cm}}}

\footnotesize
\renewcommand\arraystretch{1.33}
\setlength{\LTleft}{-20cm plus -1fill}
\setlength{\LTright}{\LTleft}
\begin{longtable}[t]{c|c|cc|c|c}
	\multirow{3}{*}{Name \& References} & \multirow{3}{*}{Dimension} &  \multicolumn{2}{c|}{Axioms} &
	\multirow{2}{*}{Cp.} & \multirow{3}{*}{Relations} \\
	& & P.P. & Cv. & & \\
	& & \ftpp & \ftcv & \ftcp & \\
	\toprule
	\endhead
	$\epsdel$-approximate DP \cite{dwork2006our}
	& {\fontfamily{pcr}\selectfont Q} & \yes\fadiv & \yes\fadiv & \yes\fcommon
	& \textbf{$\epsdel$-DP} \extendsweaker{} $\eps$-DP \\
	\midrule
	$\epsdel$-probabilistic DP \cite{meiser2018approximate}
	& {\fontfamily{pcr}\selectfont Q} & \no\fcommon & \no\facprob & \yes\fcommon
	& $\epsdel$-DP \same{} \textbf{$\epsdel$-ProDP} \\
	\midrule
	\multirow{2}{*}{$(\eps,\del_a,\del_p)$-Relaxed DP \cite{zhang2015toward}}
	& \multirow{2}{*}{{\fontfamily{pcr}\selectfont Q}} & \multirow{2}{*}{\no\fcommon} & \multirow{2}{*}{\no\facprob} & \multirow{2}{*}{\yes\fcommon} & 
	$(\eps,\del_p)$-ProDP $\specialstronger$ \textbf{$(\eps,\del_a,\del_p)$-RelDP}\\
	&&&&& $(\eps,\del_a)$-DP \specialstronger{} \textbf{$(\eps,\del_a,\del_p)$-RelDP} \\
	\midrule
	$\eps$-Kullback-Leiber Pr \cite{cuff2016differential}
	& {\fontfamily{pcr}\selectfont Q} & \yes\fadiv & \yes\fadiv & \yes\fcommon
	& $\epsdel$-DP \weakerthan{} \textbf{$\eps$-KLPr} \weakerthan{} $\eps$-DP \\
	\midrule
	$(\alpha,\eps)$-R\'enyi DP \cite{mironov2017renyi}
	& {\fontfamily{pcr}\selectfont Q} & \yes\fadiv & \yes\fadiv & \yes\fcommon
	& $\eps$-KLPr \specialweaker \textbf{$(\alpha,\eps)$-RenyiDP} \extendsweaker{} $\eps$-DP \\
	\midrule
	$(\mu,\tau)$-mean concentrated DP \cite{dwork2016concentrated}
	& {\fontfamily{pcr}\selectfont Q} & \no\fcommon & \uk & \yes\fcommon
	& $\epsdel$-DP \weakerthan{} \textbf{$(\mu,\tau)$-mCoDP} \weakerthan{} $\eps$-DP \\
	\midrule
	$(\xi,\rho)$-zero concentrated DP \cite{bun2016concentrated}
	& {\fontfamily{pcr}\selectfont Q} & \yes\fadiv & \yes\fadiv & \yes\fcommon
	& \textbf{$(\xi,\rho)$-zCoDP} $\same$ $(\mu,\tau)$-mCoDP \\
	\midrule
	$(\xi,\rho,\del)$-approximate CoDP \cite{bun2016concentrated}
	& {\fontfamily{pcr}\selectfont Q} & \no\fadiv & \uk & \yes\fcommon
	& $\epsdel$-DP \strongerthan{} \textbf{$(\xi,\rho,\del)$-ACoDP} \extendsweaker{} $(\xi,\rho)$-zCoDP \\
	\midrule
	$(\xi,\rho,\omega)$-bounded CoDP \cite{bun2016concentrated}
	& {\fontfamily{pcr}\selectfont Q} & \yes\fadiv & \yes\fadiv & \yes\fcommon
	& \textbf{$(\xi,\rho,\omega)$-BCoDP} \extendsweaker{} $(\xi,\rho)$-zCoDP \\
	\midrule
	$(\eta,\tau)$-truncated CoDP \cite{colisson2016l3}
	& {\fontfamily{pcr}\selectfont Q} & \yes\fadiv & \yes\fadiv & \yes\fcommon
	& \textbf{$(\eta,\tau)$-TCoDP\textsuperscript{\cite{colisson2016l3}}} $\same$ $\eps$-DP \\
	\midrule
	$(\rho,\omega)$-truncated CoDP \cite{bun2018composable}
	& {\fontfamily{pcr}\selectfont Q} & \yes\fadiv & \yes\fadiv & \yes\fcommon & 
	\textbf{$(\rho,\omega)$-TCoDP\textsuperscript{\cite{bun2018composable}}} $\specialweaker$ $(\xi,\rho,\omega)$-bCoDP \\
	\midrule
	$(f,\eps)$-divergence DP \cite{barber2014privacy}
	& {\fontfamily{pcr}\selectfont Q} & \yes\fadiv & \yes\fadiv & \uk
	& \textbf{$(f,\eps)$-DivDP} \extends{} most def. in {\fontfamily{pcr}\selectfont Q} \\
	\midrule
	$\left(f_k,\eps\right)$-divergence DP \cite{duchi2018right}
	& {\fontfamily{pcr}\selectfont Q} & \yes\fadiv & \yes\fadiv & \uk
	& \textbf{$\left(f_k,\eps\right)$-kDivDP} \extendedby{} $(f,\eps)$-DivDP \\
	\midrule
	$(H,f,\eps)$-capacity bounded DP \cite{chaudhuri2019dcapacity}
	& {\fontfamily{pcr}\selectfont Q} & \yes\fadiv & \yes\fadiv & \uk
	& \textbf{$(H,f,\eps)$-CBDP} \specialweaker $(f,\eps)$-DivDP \\
	\midrule
	$\eps$-unbounded DP \cite{kifer2011no}
	& {\fontfamily{pcr}\selectfont N} & \yes\fcommon & \yes\fcommon & \yes\fcd
	& $\eps$-DP \same{} \textbf{$\eps$-uBoDP} \specialsame{} $(c,\eps)$-GrDP \\
	\midrule
	$\eps$-bounded DP \cite{kifer2011no}
	& {\fontfamily{pcr}\selectfont N} & \yes\fcommon & \yes\fcommon & \yes\fcd
	& \textbf{$\eps$-BoDP} \weakerthan{} $\eps$-DP \\
	\midrule
	$(P,\eps)$-one-sided DP \cite{doudalis2017one}
	& {\fontfamily{pcr}\selectfont N} & \yes\fcommon & \yes\fcommon & \yes\fcd
	& \textbf{$(P,\eps)$-OnSDP} \extendsweaker{} $\eps$-BoDP \\
	\midrule
	$(P,\eps)$-asymetric DP \cite{takagi2021asymmetric}
	& {\fontfamily{pcr}\selectfont N} & \yes\fcommon & \yes\fcommon & \yes\fcd
	& \textbf{$(P,\eps)$-AsDP} \extendsweaker{} $\eps$-uBoDP \\
	\midrule
	$\eps$-client DP \cite{mcmahan2017learning}
	& {\fontfamily{pcr}\selectfont N} & \yes\fcommon & \yes\fcommon & \yes\fcd
	& \textbf{$\eps$-ClDP} \strongerthan{} $\eps$-DP \\
	\midrule
	$\eps$-element DP \cite{asi2019element}
	& {\fontfamily{pcr}\selectfont N} & \yes\fcommon & \yes\fcommon & \yes\fcd
	& $\eps$-ClDP \strongerthan{} \textbf{$\eps$-ElDP} \strongerthan{} $\eps$-DP \\
	\midrule
	$(c,\eps)$-group DP \cite{dwork2008differential}
	& {\fontfamily{pcr}\selectfont N} & \yes\fcommon & \yes\fcommon & \yes\fcd
	& \textbf{$(c,\eps)$-GrDP} \extendssame{} $\eps$-DP \\
	\midrule
	$(R,c,\eps)$-dependent DP \cite{liu2016dependence}
	& {\fontfamily{pcr}\selectfont N} & \yes\fcommon & \yes\fcommon & \yes\fcd
	& \textbf{$(R,c,\eps)$-DepDP} \extends{} $(c,\eps)$-GrDP\\
	\midrule
	$(\mathcal{A},\eps)$-bayesian DP \cite{yang2015bayesian}
	& {\fontfamily{pcr}\selectfont N} & \yes\fcommon & \yes\fcommon & \yes\fcd	& 
	\textbf{$(\mathcal{A},\eps)$-BayDP\textsuperscript{\cite{yang2015bayesian}}} \extends{} $(R,c,\eps)$-DepDP \\
	\midrule
	$(\mathcal{A},\eps)$-correlated DP \cite{wu2017game1,wu2017game2}
	& {\fontfamily{pcr}\selectfont N} & \yes\fcommon & \yes\fcommon & \yes\fcd & \textbf{$(\mathcal{A},\eps)$-CorDP} $\extendedby$ $(\mathcal{A},\eps)$-BayDP\textsuperscript{\cite{yang2015bayesian}}\\
	\midrule
	$(\mathcal{A},\eps)$-prior DP \cite{li2019impact}
	& {\fontfamily{pcr}\selectfont N} & \yes\fcommon & \yes\fcommon & \yes\fcd & 
	\textbf{$(\mathcal{A},\eps)$-PriDP} \extends{} $(\mathcal{A},\eps)$-BayDP\textsuperscript{\cite{yang2015bayesian}}\\
	\midrule
	$\eps$-free lunch Pr \cite{kifer2011no}
	& {\fontfamily{pcr}\selectfont N} & \yes\fcommon & \yes\fcommon & \yes\fcd
	& \textbf{$\eps$-FLPr} \strongerthan{} all def. in {\fontfamily{pcr}\selectfont N} \\
	\midrule
	$(D,\eps)$-individual DP \cite{soria2017individual}
	& {\fontfamily{pcr}\selectfont N} & \yes\fcommon & \yes\fcommon & \yes\fcd
	& \textbf{$(D,\eps)$-IndDP} \weakerthan{} $\eps$-DP \\
	\midrule
	$(D,t,\eps)$-per-instance DP \cite{wang2017per}
	& {\fontfamily{pcr}\selectfont N} & \yes\fcommon & \yes\fcommon & \yes\fcd
	& \textbf{$(D,t,\eps)$-PIDP} \weakerthan{} $(D,\eps)$-IndDP \\
	\midrule
	$(\mathcal{R},\eps)$-generic DP \cite{kifer2011no,fang2014differential}
	& {\fontfamily{pcr}\selectfont N} & \yes\fcommon & \yes\fcommon & \yes\fcd
	& \textbf{$(\mathcal{R},\eps)$-GcDP\textsuperscript{\cite{kifer2011no}}} \extends{} most def. in {\fontfamily{pcr}\selectfont N} \\
	\midrule
	$(d,\Delta,\eps)$-constrained DP \cite{zhou2009differential} & {\fontfamily{pcr}\selectfont N} & \yes\fcommon & \yes\fcommon & \yes\fcd
	& \textbf{$(d,\Delta,\eps)$-ConsDP} \same{} $(\mathcal{R},\eps)$-GcDP \\
	\midrule
	$(d,\Delta,S_\datasets,\eps)$-distributional Pr \cite{zhou2009differential}
	& {\fontfamily{pcr}\selectfont N} & \yes\fcommon & \yes\fcommon & \yes\fcd	&
	\textbf{$(d,\Delta,S_\datasets,\eps)$-DlPr\textsuperscript{\cite{zhou2009differential}}} \extendedby{} $(\mathcal{R},\eps)$-GcDP \\
	\midrule
	$(f,\eps)$-sensitivity-induced DP \cite{rubinstein2017pain}
	& {\fontfamily{pcr}\selectfont N} & \yes\fcommon & \yes\fcommon & \yes\fcd
	& \textbf{$\eps$-SIDP} \extendedby{} $(\mathcal{R},\eps)$-GcDP\\
	\midrule
	$\left(\mathcal{I}_Q,\eps\right)$-induced-neighbors DP \cite{kifer2011no}
	& {\fontfamily{pcr}\selectfont N} & \yes\fcommon & \yes\fcommon & \yes\fcd
	& \textbf{$(\mathcal{I}_Q,\eps)$-INDP} \extendedby{} $(\mathcal{R},\eps)$-GcDP\\
	\midrule
	$\left(G,\mathcal{I}_Q,\eps\right)$-blowfish Pr \cite{he2014blowfish}
	& {\fontfamily{pcr}\selectfont N} & \yes\fcommon & \yes\fcommon & \yes\fcd &
	\textbf{$(G,\mathcal{I}_Q,\eps)$-BFPr} \extendedby{} $(\mathcal{R},\eps)$-GcDP \\
	\midrule
	$\Psi$-personalized DP \cite{jorgensen2015conservative} & {\fontfamily{pcr}\selectfont V} & \yes\fad & \yes\fad & \yes\fcd
	& \textbf{$\Psi$-PerDP} \extends{} $\eps$-DP \\
	\midrule
	$\Psi$-tailored DP \cite{lui2015outlier}
	& {\fontfamily{pcr}\selectfont V} & \yes\fad & \yes\fad & \yes\fcd
	& \textbf{$\Psi$-TaiDP} \extends{} $\Psi$-PerDP \\
	\midrule
	$(\Psi,\overline{\eps})$-input-discriminative DP \cite{gu2020providing}
	& {\fontfamily{pcr}\selectfont V} & \yes\fad & \yes\fad & \yes\fcd
	& \textbf{$\Psi$-IDDP} \extends{} $\Psi$-PerDP \\
	\midrule
	$\eps(\cdot)$-outlier Pr \cite{lui2015outlier}
	& {\fontfamily{pcr}\selectfont V} & \yes\fad & \yes\fad & \yes\fcd
	& \textbf{$\eps(\cdot)$-OutPr} \extendedby{} $\Psi$-TaiDP \\
	\midrule
	$(\eps,p,r)$-Pareto DP \cite{lui2015outlier}
	& {\fontfamily{pcr}\selectfont V} & \yes\fad & \yes\fad & \yes\fcd
	& \textbf{$(\eps,p,r)$-ParDP} \extendedby{} $\Psi$-TaiDP \\
	\midrule
	$(\pi,\gamma,\eps)$-random DP \cite{hall2011random}
	& {\fontfamily{pcr}\selectfont V} & \uk & \no\facrand & \yes\fcommon
	& \textbf{$(\pi,\gamma,\eps)$-RanDP} \extendsweaker{} $\eps$-DP \\
	\midrule
	$d_{\datasets}$-Pr \cite{chatzikokolakis2013broadening}
	& {\fontfamily{pcr}\selectfont N,V} & \yes\fad & \yes\fad & \yes\fcd
	& $\eps$-DP \extendedby{} \textbf{$d_{\datasets}$-Pr} \\
	\midrule
	$(d_\datasets,\eps)$-smooth DP \cite{barber2014privacy}
	& {\fontfamily{pcr}\selectfont N,V} & \yes\fad & \yes\fad & \yes\fcd
	& \textbf{$(d_\datasets,\eps)$-SmDP} $\sim$ $d_\datasets$-Pr \\
	\midrule
	$(\eps,\gamma)$-distributional Pr \cite{zhou2009differential,roth2010new}
	& {\fontfamily{pcr}\selectfont N,V} & \uk & \uk & \uk &
	$\eps$-FLPr \extendedby{} \textbf{$(\eps,\gamma)$-DlPr\textsuperscript{\cite{roth2010new,blum2013learning}}} \\
	\midrule
	$(\pi,\eps,\del)$-weak Bayesian DP \cite{triastcyn2019bayesian}
	& {\fontfamily{pcr}\selectfont Q,V} & \yes\fcommon & \uk & \yes\fcd	&
	$\epsdel$-DP \strongerthan{} \textbf{$(\pi,\eps,\del)$-WBDP} \weakerthan{} $(\pi,\gamma,\eps)$-RanDP \\
	\midrule
	$\Theps$-on average KL Pr \cite{wang2016average}
	& {\fontfamily{pcr}\selectfont Q,V} & \yes\fcommon & \uk & \uk\fczk & 
	$\eps$-KLPr \strongerthan{} \textbf{$\Theps$-avgKLPr} \weakerthan{} $(\pi,\gamma,\eps)$-RanDP \\
	\midrule
	\multirow{2}{*}{$(\pi,\eps,\del)$-Bayesian DP \cite{triastcyn2019bayesian}}
	& \multirow{2}{*}{{\fontfamily{pcr}\selectfont Q,V}} & \multirow{2}{*}{\yes\fcommon} & \multirow{2}{*}{\no\facrand} & \multirow{2}{*}{\yes\fcd} & 
	$\epsdel$-ProDP \strongerthan{} \textbf{$(\pi,\eps,\del)$-BayDP\textsuperscript{\cite{triastcyn2019bayesian}}} \\
	&&&&& $(\pi,\gamma,\eps)$-RanDP \strongerthan{}  \textbf{$(\pi,\eps,\del)$-BayDP\textsuperscript{\cite{triastcyn2019bayesian}}} \\
	\midrule
	$(d_\datasets,\eps,\del)$-pseudo-metric DP \cite{dimitrakakis2013bayesian}
	& {\fontfamily{pcr}\selectfont Q,N,V} & \uk & \uk & \yes\fcommon & 
	$\epsdel$-DP \extendedby{} \textbf{$(d_\datasets,\eps,\del)$-PsDP} \extendsweaker{} $d_\datasets$-Pr \\
	\midrule
	$(f,d,\eps)$-extended divergence DP \cite{kawamoto2019allerton}
	& {\fontfamily{pcr}\selectfont Q,N,V} & \yes\fad & \yes\fad & \uk
	& $d_\datasets$-Pr \extendedby{} \textbf{$(f,d,\eps)$-EDivDP} \extends{} $(f,\eps)$-Div DP \\
	\midrule
	$(\mathcal{R},M)$-generic DP \cite{kifer2010towards}
	& {\fontfamily{pcr}\selectfont Q,N,V} & \yes\fcommon & \yes\fcommon & \uk
	& \textbf{$(\mathcal{R},M)$-GcDP\textsuperscript{\cite{kifer2010towards}}} \extends{} $\epsdel$-DP \\
	\midrule	 
	$(\mathcal{R},q)$-abstract DP \cite{kifer2010towards}
	& {\fontfamily{pcr}\selectfont Q,N,V} & \no\fcommon & \no\fcommon & \uk & \textbf{$(\mathcal{R},q)$-AbsDP} \extends{} $(\mathcal{R},M)$-GcDP\textsuperscript{\cite{kifer2010towards}} \\
	\midrule
	$\Theps$-noiseless Pr \cite{bhaskar2011noiseless}
	& {\fontfamily{pcr}\selectfont B} & \yes\fcommon & \yes\fcommon & \no\fcb
	& $\eps$-DP \extendsweaker{} \textbf{$\Theps$-NPr} \\
	\midrule
	$\Theps$-causal DP \cite{desfontaines2019differential}
	& {\fontfamily{pcr}\selectfont B} & \yes\fcommon & \yes\fcommon & \no\fcb
	& \textbf{$\Theps$-CausDP} \extendsweaker{} $\eps$-DP \\
	\midrule
	$(\beta,\eps)$-DP under sampling \cite{li2011provably}
	& {\fontfamily{pcr}\selectfont Q,B} & \yes\fcommon & \yes\fcommon & \no\fcb
	& $\Theps$-NPr \extends{} \textbf{$(\beta,\eps)$-SamDP} \extendsweaker{} $\eps$-DP \\
	\midrule
	$\Thepsdel$-active PK DP \cite{desfontaines2019differential} & {\fontfamily{pcr}\selectfont Q,B} & \yes\fcommon & \yes\fcommon & \no\fcb
	& \textbf{$\Thepsdel$-APKDP} \extendsweaker{} $\Theps$-NPr \\
	\midrule
	\multirow{2}{*}{$\Thepsdel$-passive PK DP \cite{desfontaines2019differential}}
	& \multirow{2}{*}{{\fontfamily{pcr}\selectfont Q,B}} & \multirow{2}{*}{\yes\fcommon} & \multirow{2}{*}{\yes\fcommon} & \multirow{2}{*}{\no\fcb}
	& $\Thepsdel$-APKDP \strongerthan{} \textbf{$\Thepsdel$-PPKDP} \\
	&&&&& $\Theps$-NPr \specialstronger{} \textbf{$\Thepsdel$-PPKDP} \\
	\midrule
	$(\Theta,\Phi,\eps)$-pufferfish Pr \cite{kifer2012rigorous}
	& {\fontfamily{pcr}\selectfont N,B} & \yes\fcommon & \yes\fcommon & \no\fcb
	& $\Theps$-NPr \extendedby{} \textbf{$(\Theta,\Phi,\eps)$-PFPr} \extends{} $(\mathcal{R},\eps)$-GcDP \\
	\midrule
	$(\pi,k,\eps)$-Bayesian DP \cite{leung2012bayesian}
	& {\fontfamily{pcr}\selectfont N,B} & \yes\fcommon & \yes\fcommon & \no\fcb
	& {\fontfamily{pcr}\selectfont BayDP\textsuperscript{\cite{leung2012bayesian}}} \extendedby{} $(\Theta,\Phi,\eps)$-PFPr \\
	\midrule
	$\Thepsdel$-distribution Pr \cite{jelasity2014distributional,kawamoto2019esorics}
	& {\fontfamily{pcr}\selectfont Q,N,B} & \yes\fad & \yes\fad & \no\fcb
	& \textbf{$\Thepsdel$-DnPr} \extends{} $\Thepsdel$-APKDP \\
	\midrule
	$\Thepsdel$-profile-based DP \cite{geumlek2019profile}
	& {\fontfamily{pcr}\selectfont Q,N,B} & \yes\fad & \yes\fad & \no\fcb
	& \textbf{$\Thepsdel$-PBDP} \extendedby{} $\Thepsdel$-DnPr \\
	\midrule
	$\Thepsdel$-probabilistic DnPr \cite{kawamoto2019esorics}
	& {\fontfamily{pcr}\selectfont Q,N,B} & \no\fcommon & \no\facprob & \no\fcb
	& $\eps$-ProDP \extendedby{} \textbf{$\Thepsdel$-PDnPr} \extends{} $\Theps$-DnPr \\
	\midrule
	$(f,\Theta,\eps)$-divergence DnPr \cite{kawamoto2019allerton}
	& {\fontfamily{pcr}\selectfont Q,N,B} & \yes\fad & \yes\fad & \no\fcb
	& $(f,\eps)$-DP \extendedby{} \textbf{$(f,\Theta,\eps)$-DDnPr} \extends{} $\Theps$-DnPr \\
	\midrule	
	$(d,\Theta,\eps)$-extended DnPr \cite{kawamoto2019esorics}
	& {\fontfamily{pcr}\selectfont N,V,B} & \yes\fad & \yes\fad & \no\fcb
	& $d_\datasets$-Pr \extendedby{} \textbf{$(d,\Theta,\eps)$-EDnPr} \extends{} $\Theps$-DnPr \\
	\midrule
	\multirow{2}{*}{$(d,f,\Theta,\eps)$-ext.\ div.\ DnPr \cite{kawamoto2019allerton}}
	& \multirow{2}{*}{{\fontfamily{pcr}\selectfont Q,N,V,B}} & \multirow{2}{*}{\yes\fad} & \multirow{2}{*}{\yes\fad} & \multirow{2}{*}{\no\fcb} & $(f,\Theta,\eps)$-DDPr \extendedby{} \textbf{$(d,f,\Theta,\eps)$-EDDnPr} \\
	&&&&& $(d,\Theta,\eps)$-EDnPr \extendedby{} \textbf{$(d,f,\Theta,\eps)$-EDDnPr} \\
	\midrule
	$\eps$-indistinguishable Pr \cite{liu2013semantic}
	& {\fontfamily{pcr}\selectfont F} & \yes\feqdp & \yes\feqdp & \yes\feqdp
	& \textbf{$\eps$-IndPr} $\same$ $\eps$-DP \\
	\midrule
	$\eps$-semantic Pr \cite{ganta2008composition,kasiviswanathan2014semantics}
	& {\fontfamily{pcr}\selectfont F} & \uk & \uk & \uk
	& \textbf{$\eps$-SemPr} $\same$ $\eps$-DP \\
	\midrule
	$\eps$-range-bounded Pr \cite{durfee2019practical}
	& {\fontfamily{pcr}\selectfont F} & \uk & \uk & \uk
	& \textbf{$\eps$-RBPr} $\same$ $\eps$-DP \\
	\midrule
	$f$-DP \cite{dong2019gaussian}
	& {\fontfamily{pcr}\selectfont Q,F} & \yes\fcommon & \uk & \yes\fcommon
	& \textbf{$f$-DP} \extends{} $\epsdel$-DP \\
	\midrule
	Gaussian DP \cite{dong2019gaussian}
	& {\fontfamily{pcr}\selectfont Q,F} & \yes\fcommon & \uk & \yes\fcommon
	& \textbf{GaussDP} \extendedby{} $f$-DP \\
	\midrule
	weak-Federated $f$-DP \cite{zheng2021federated}
	& {\fontfamily{pcr}\selectfont Q,F} & \yes\fcommon & \uk & \yes\fcommon
	& \textbf{$f$-wFDP} \extendedby{} GaussDP \\
	\midrule
	strong-Federated $f$-DP \cite{zheng2021federated}
	& {\fontfamily{pcr}\selectfont Q,F} & \yes\fcommon & \uk & \yes\fcommon
	& $f$-wFDP \weakerthan{} \textbf{$f$-sFDP} \extendedby{} GaussDP \\
	\midrule
	$\Theps$-aposteriori noiseless Pr \cite{bhaskar2011noiseless}
	& {\fontfamily{pcr}\selectfont B,F} & \yes\fadef & \yes\fadef & \uk
	& \textbf{$\Theps$-ANPr} $\same$ $\Theps$-NPr\\
	\midrule
	$\Theps$-positive membership Pr \cite{li2013membership}
	& {\fontfamily{pcr}\selectfont B,F} & \yes\fadef & \yes\fadef & \no\fcb
	& \textbf{$\Theps$-PMPr} \extends{} $\eps$-BoDP \\
	\midrule
	$\Theps$-negative membership Pr \cite{li2013membership}
	& {\fontfamily{pcr}\selectfont B,F} & \yes\fadef & \yes\fadef & \no\fcb
	& \textbf{$\Theps$-NMPr} \extends{} $\eps$-BoDP \\
	\midrule
	$\Theps$-membership Pr \cite{li2013membership}
	& {\fontfamily{pcr}\selectfont B,F} & \yes\fadef & \yes\fadef & \no\fcb
	& $\Theps$-PMPr \weakerthan{} \textbf{$\Theps$-MPr} \strongerthan{} $\Theps$-NMPr \\
	\midrule
	$\Thepsdel$-adversarial Pr \cite{rastogi2009relationship}
	& {\fontfamily{pcr}\selectfont Q,B,F} & \yes\fadef & \yes\fadef & \no\fcb
	& $\epsdel$-DP \extendedby{} \textbf{$\Thepsdel$-AdvPr} \weakerthan{} $\Theps$-PMPr \\
	\midrule
	$\Theps$-information Pr \cite{du2012privacy}
	& {\fontfamily{pcr}\selectfont N,B,F} & \uk & \uk & \uk
	& \textbf{$\Theps$-InfPr} \strongerthan{} $\eps$-DP\\
	\midrule
	$\Aggeps$-zero-knowledge Pr \cite{gehrke2011towards}
	& {\fontfamily{pcr}\selectfont R} & \yes\fcommon & \yes\fcommon & \uk\fczk
	& \textbf{$\Aggeps$-ZKPr} \strongerthan{} $\eps$-DP \\
	\midrule
	$(\project_\nu, \eps, \del)$-subspace DP \cite{gao2021subspace}
	& {\fontfamily{pcr}\selectfont Q,R} & \uk\fczk & \uk & \yes\fcommon
	& $(\project_\nu, \eps, \del)$-SubDP \weakerthan{} $\epsdel$-DP \\
	\midrule
	$(P,\eps)$-bounded leakage DP \cite{ligett2020bounded}
	& {\fontfamily{pcr}\selectfont Q,R} & \yes\fcommon & \yes\fcommon & \yes\fcommon
	& \textbf{$(P,\eps)$-BLDP} \extends{} $\epsdel$-DP \\
	\midrule
	$(\Theta,\Gamma,\eps)$-coupled-worlds Pr \cite{bassily2013coupled}
	& {\fontfamily{pcr}\selectfont N,B,R} & \yes\fcommon & \yes\fcommon & \no\fad
	& \textbf{$(\Theta,\Gamma,\eps)$-CWPr} \extends{} $\eps$-DP \\
	\midrule
	$(\Theta,\Gamma,\eps,\del)$-inference CW Pr \cite{bassily2013coupled}
	& {\fontfamily{pcr}\selectfont Q,N,B,F,R} & \uk & \uk & \no\fad
	& \textbf{$(\Theta,\Gamma,\eps,\del)$-IBCWPr} \strongerthan{} $(\Theta,\Gamma,\eps)$-CWPr \\
	\midrule
	$(\Theta,\Gamma,\eps,\del)$-inference DistDP \cite{bassily2013coupled}
	& {\fontfamily{pcr}\selectfont Q,N,B,F,R} & \uk & \uk & \no\fad & $\Theps$-DDP \strongerthan{} \textbf{$(\Theta,\Gamma,\eps,\del)$-IBDDP} \extendedby{} $(\Theta,\Gamma,\eps,\del)$-IBCWPr \\
	\midrule
	Typical Stability \cite{bassily2016typical}
	& {\fontfamily{pcr}\selectfont Q,V,R} & \uk & \uk & \no\fad &
	\textbf{ $(\Theta,\gamma,\eps,\del)$-TySt}\\
	\midrule
	$\eps_\kappa$-SIM-computational DP \cite{mironov2009computational}
	& {\fontfamily{pcr}\selectfont C} & \yes\facomp & \yes\facomp & \yes\fccomp
	& \textbf{$\eps_\kappa$-SimCDP} \weakerthan{} $\eps$-DP \\
	\midrule
	$\eps_\kappa$-IND-computational DP \cite{mironov2009computational}
	& {\fontfamily{pcr}\selectfont C} & \yes\facomp & \yes\facomp & \yes\fccomp
	& \textbf{$\eps_\kappa$-IndCDP} \weakerthan{} $\eps_\kappa$-SimCDP \\
	\midrule
	$\Aggeps$-computational ZK Pr \cite{gehrke2011towards}
	& {\fontfamily{pcr}\selectfont R,C} & \yes\facomp & \yes\facomp & \uk
	& \textbf{$\Aggeps$-CZKPr} \extendsweaker{} $\Aggeps$-ZKPr \\
	\midrule
	$(\eps,\del,f)$-output constrained DP \cite{he2017composing}
	& {\fontfamily{pcr}\selectfont N,V,C} & \yes\facomp & \yes\facomp & \yes\fccomp
	& \textbf{$(\eps,\del,f)$-OCDP} \extends{} $\eps_\kappa$-IndCDP \\
	\midrule
	$\epsdel$-DP for Record Linkage \cite{he2017composing}
	& {\fontfamily{pcr}\selectfont N,V,C} & \yes\facomp & \yes\facomp & \yes\fccomp &
	\textbf{$\epsdel$-RLDP} \extendedby{} $(\eps,\del,f)$-OCDP \\
	\bottomrule
	\caption{Summary of variants/extensions of DP representing
		the main options in each combination of dimensions.}\label{tab:DP_variants}
\end{longtable}	
\normalsize
\begin{multicols}{4}
  \printendnotes
\end{multicols}
\nopagebreak

\subsection{Proofs of properties}
  
\subsubsection{Axioms}
  
\begin{prop}\label{prop:a-div}
	All instantiations of DivDP satisfy both privacy axioms. In particular,
	approximate DP, MIDP, KLPr, RenDP, and zCoDP satisfy both axioms.
	\begin{proof}
		The post-processing axiom follows directly from the monotonicity property of
		the $f$-divergence. The convexity axiom follows directly from the joint
		convexity property of the $f$-divergence.
	\end{proof}
\end{prop}

\begin{prop}\label{prop:a-c-prob}
	ProDP and ACoDP do not satisfy the convexity axiom.
	\begin{proof}
		Consider the following mechanisms $\mecha_1$ and $\mecha_2$, with input and
		output in $\{0,1\}$.
		\begin{itemize}
			\item $\mechone{0}=0$, $\mechone{1}=1$ with probability $\del$, and
			$\mechone{1}=0$ with probability $1-\del$.
			\item $\mechtwo{0}=\mechtwo{1}=1$.
		\end{itemize}
		Both mechanisms are $\left(\frac{1}{1-\del},\del\right)$-ProDP. Now,
		consider the mechanism $\mecha$ which applies $\mecha_1$ with probability
		$1-2\del$ and $\mecha_2$ with probability $2\del$. $\mecha$ is a convex
		combination of $\mecha_1$ and $\mecha_2$, but the reader can verify that it
		is not $\left(\frac{1}{1-\del},\del\right)$-ProDP. The result for
		$(\xi,\rho,\del)$-ACoDP is a direct corollary, since is is equivalent to
		$(\xi,\del)$-ProDP when $\rho=0$.
	\end{proof}
\end{prop}

\begin{prop}\label{prop:a-d}
	$d_\datasets$-Pr satisfies both privacy axioms. Further, EDivDP also satisfies
	both privacy axioms.
	\begin{proof}
		The proof corresponding to PFPr in \cite{kifer2012rigorous} is a proof by case 
		analysis on every possible protected property. The fact that $\eps$ is 
		the same for every protected
		property has no influence on the proof, so we can directly adapt the proof
		to $d_\datasets$-Pr, and its combination with PFPr.
		Similarly, the proof can be extended to arbitrary divergence functions, like
		in Proposition \ref{prop:a-div}.
	\end{proof}
\end{prop}

\begin{prop}\label{prop:a-c-rand}
	RanDP does not satisfy the convexity axiom.
	\begin{proof}
		Let $\pi$ be the uniform distribution on $\{0,1\}$, let $D_1$ be generated
		by picking $10$ records according to $\pi$, and $D_2$ by flipping one record
		at random. Let $\mecha_0$ return $0$ if all records are $0$, and $\bot$
		otherwise. Let $\mecha_1$ return $1$ if all records are $1$, and $\bot$
		otherwise.
		
		Note that both mechanisms are $(\pi,2^{-9},0)$-RanDP. Indeed, $\mecha_0$
		will only return $0$ for $D_1$ with probability $2^{-10}$, and for $D_2$
		with probability $2^{-10}$ (if $D_1$ only has one $1$, which happens with
		probability $10\cdot2^{-10}$, and this record is flipped, which happens with
		probability $0.1$). In both cases, $\mecha_0$ will return $\bot$ for the
		other database; which will be a distinguishing event. Otherwise, $\mecha_0$
		will return $\tau$ for both databases, so $\mech{D_1}\ind_0\mech{D_2}$. The
		reasoning is the same for $\mecha_1$. 
		
		However, the mechanism $\mecha_{0.5}$ obtained by applying either $\mecha_0$
		or $\mecha_1$ uniformly randomly doesn't satisfy $(\pi,2^{-9},0)$-RanDP:
		the indinguishability property does not hold if $D_1$ or $D_2$ have all
		their records set to \emph{either} $0$ or $1$, which happens twice as often
		as either option alone.
	\end{proof}
\end{prop}

\begin{prop}\label{prop:a-def}
	All variants of MPr, AdvPr, and ANPr satisfy both axioms. As a direct
	corollary, InfPr also satisfies both axioms.
	\begin{proof}
		We prove it for AdvPr. A mechanism $\mecha$ satisfies $\Thepsdel$-AdvPr if
		for all $t\in\tuples$, $\theta\in\Theta$, and $S\subseteq\outputs$,
		$\probasc{D\sim\theta}{t\in D}{\mech{D}\in S}\le e^\eps\cdot\probas{D\sim\theta}{t\in D}+\del$.
		We first prove that it satisfies the convexity axiom. Suppose $\mecha$ is a
		convex combination of $\mecha_1$ and $\mecha_2$. Simplifying
		$\probas{D\sim\theta}{\dots}$ into $\proba{\dots}$, we have:
		\begin{align*}
			\probac{t\in D}{\mech{D}\in S} = 
			\frac{\proba{t\in D\text{ and }\mech{D}\in S\text{ and }\mecha=\mecha_1}}{\proba{\mech{D}\in S}}
			\\
			+\frac{\proba{t\in D\text{ and }\mech{D}\in S\text{ and }\mecha=\mecha_2}}{\proba{\mech{D}\in S}} \\
		\end{align*}
		Denoting $X_i=\proba{\mech{D}\in S\text{ and }\mecha=\mecha_i}$ for $i\in\{1,2\}$, this gives:
		\begin{align*}
			\probac{t\in D}{\mech{D}\in S}=\hspace{6cm}\\
			\frac{X_1\cdot\probac{t\in D}{\mechone{D}\in S}}{X_1+X_2}\cdot
			\frac{X_2\cdot\probac{t\in D}{\mechtwo{D}\in S}}{X_1+X_2}\\
			\le\frac{X_1\left(e^\eps\cdot\proba{t\in D}\right)+\del}{X_1+X_2}
			+ \frac{X_2\left(e^\eps\cdot\proba{t\in D}\right)+\del}{X_1+X_2}
			\\ 
			\qquad\qquad\qquad\qquad\le e^\eps\cdot\proba{t\in D} + \del
		\end{align*}
		The proof for the post-processing axiom is similar, summing over all
		possible outputs $\mech{D}$. It is straightforward to adapt the proof to
		all other definitions which change the shape of the prior-posterior bounds.
	\end{proof}
\end{prop}

\begin{prop}\label{prop:a-comp}
	Both versions of CDP satisfy both privacy axioms; where the post-processing
	axiom is modified to only allow post-processing with functions computable on a
  probabilistic polynomial time Turing machine. CZKPr also satisfies both
  privacy axioms.
	\begin{proof}
		For Ind-CDP and the post-processing axiom, the proof is straightforward: if
		post-processing the output could break the $\eps$-indistinguishability
		property, the attacker could do this on the original output and break the
		$\eps$-indistinguishability property of the original definition.
		
		For Ind-CDP and the convexity axiom, without loss of generality, we can
		assume that the sets of possible outputs of both mechanisms are disjoint
		(otherwise, this give strictly less information to the attacker). The proof
		is then the same as for the post-processing axiom.
		
		For SimCDP, applying the same post-processing function to the ``true''
		differentially private mechanism immediately leads to the result, since DP
		satisfies post-processing. The same reasoning holds for convexity.

    The proof that CWPr satisfies both privacy axioms can be found
    in~\cite{bassily2013coupled}; as an immediate corollary, CZKPr also
    satisfies both axioms.
	\end{proof}
\end{prop}

\subsubsection{Composition}

In this section, if $\mecha_1$ and $\mecha_2$ are two mechanisms, we denote
$\mecha_{1+2}$ the mechanism defined by
$\mecha_{1+2}(D)=\left(\mecha_1(D),\mecha_2(D)\right)$.

\begin{prop}\label{prop:c-d}
	If $\mecha_1$ is $d^1_\datasets$-private and $\mecha_2$ is
	$d^2_\datasets$-private, then $\mecha_{1+2}$ is $d^{1+2}_\datasets$-private,
	where
	$d^{1+2}_\datasets\left(D_1,D_2\right)=d^1_\datasets\left(D_1,D_2\right)+d^2_\datasets\left(D_1,D_2\right)$.
	\begin{proof}
		The proof is essentially the same as for $\eps$-DP. $\mecha_1$'s
		randomness is independent from $\mecha_2$'s, so:
		
		\begin{equation*}
			\begin{gathered}
				\proba{\begin{tabular}{c}
						$\mechone{D_1}=O_1$ \& \\
						$\mechtwo{D_1}=O_2$
				\end{tabular}}=
				\proba{\mechone{D_1}=O_1}\cdot\proba{\mechtwo{D_1}=O_2}\\
				\le e^{d^1_\datasets\left(D_1,D_2\right)}\cdot\proba{\mechtwo{D_2}=O_1}
				\cdot e^{d^2_\datasets\left(D_1,D_2\right)}\cdot\proba{\mechtwo{D_2}=O_2} \\
				\le e^{d^{1+2}_\datasets\left(D_1,D_2\right)}
				\cdot\proba{\mechone{D_2}=O_1 \text{and} \mechtwo{D_2}=O_2}
			\end{gathered}
		\end{equation*}
		
		Most definition can also be combined with $d_\datasets$-privacy, 
		and the composition proofs can be similarly adapted.
	\end{proof}
\end{prop}

\begin{prop}\label{prop:c-b}
	In general, definitions which assume limited background knowledge from the
	adversary do not compose.
	\begin{proof}
		The proof of Proposition \ref{prop:c-d} cannot be adapted to a context in
		which the attacker has limited background knowledge: as the randomness
		partially comes from the data-generating distribution, the two probabilities
		are no longer independent. A typical example considers two mechanisms which
		answer e.g., queries ``how many records satisfy property $P$'' and ``how many
		records satisfy property $P$ and have an ID different from 4217'': the
		randomness in the data might make each query private, but the combination of
		two queries trivially reveals something about a particular user. Variants of
		this proof can easily be obtained for all definitions with limited
		background knowledge.
	\end{proof}
\end{prop}

%% file: Related.tex
\section{Scope and related work}\label{sec:related}

In this section, we detail our criteria for excluding particular data privacy
definitions from our work, we list some relevant definitions that were excluded
by this criteria, and we list related works and existing surveys in the field of
data privacy.

\subsection{Methodology}\label{sec:scope}

Whether a data privacy definition fits our description is not always obvious, so we use
the following criterion: the attacker's capabilities must be clearly defined, and the
definition must prevent this attacker from learning about a protected property.
Consequently, we do not consider:
\begin{itemize}
    \item definitions which are a property of the output data and not of the
        mechanism;
    \item variants of technical notions that are not data privacy properties,
        like the different types of sensitivity;
    \item definitions whose only difference with DP is in the context and not in
        the formal property, like the distinction between local and global
    models.
\end{itemize}

In Section \ref{sec:out}, we give a list of notions that we found during our
survey, and considered to be out of scope for our book.
To find a comprehensive list of DP notions, besides the definitions we were
aware of or were suggested to us by experts, we conducted a wide literature
review using two research datasets:
BASE\footnote{\url{https://www.base-search.net/}} and Google
Scholar\footnote{\url{https://scholar.google.com/}}. The exact queries were run
several times: November 1st in 2018, August 1st in 2019, June 1st in 2020, 
and 1st of September in 2021. The final result count are summarized in
Table \ref{tab:queries}.

\begin{table}[h!]
	\centering
	\begin{tabular}{cc}	
		\toprule
		Query (BASE) & Hits \\
		\midrule
		``differential privacy'' relax & 194 \\
		``differential privacy'' variant -relax & 225 \\
		\midrule
		\midrule
		Query (Google Scholar) & Hits \\
		\midrule
		``differential privacy'' ``new notion'' & 332 \\
		``differential privacy'' ``new definition'' -``new notion'' & 248 \\
		\bottomrule
	\end{tabular}
	\caption{Queries for the literature review.}\label{tab:queries}
\end{table}

First, we manually reviewed each paper and filtered them out until we had only
papers which either contained a new definition or were applying DP in a new
setting. All papers which defined a variant or extension of DP are cited in this
work.

\subsection{Out of scope definitions}\label{sec:out}

As detailed in the previous section, we considered certain data privacy
definitions to be out of scope for our work, even when they seem to be related
to differential privacy. This section elaborates on such definitions.

\subsubsection*{Lack of semantic guarantees}

Some definitions do not provide clear semantic privacy guarantees, or are only
used as a tool in order to prove links between existing definitions. As such, we
did not include them in our survey.

\begin{itemize}
	\item \emph{$\eps$-privacy}%
	\footnote{Another definition with the same name is introduced later in this section.}, introduced in \cite{machanavajjhala2009data},
	was a first attempt at formalizing an adversary with restricted background
	knowledge by using Dirichlet distribution. This definition imposes a condition 
	on the output, but not on the mechanism, consequently it does not offer strong 
	semantic guarantees like noiseless privacy \cite{duan2009privacy,bhaskar2011noiseless} 
	(introduced in Section \ref{sec:b}).
	
	\item \emph{Relaxed indistinguishability}, introduced
	in \cite{rastogi2009relationship} is a relaxation of adversarial privacy
	that provides a plausible deniability by requiring for each tuple, that at
	least $l$ tuples must exist with $\eps$-indistinguishability. However, it 
	does not provide any guarantee against Bayesian adversaries.
	
	\item \emph{Mutual-information DP}, introduced in \cite{cuff2016differential} is 
	an alternative way to average the privacy loss, similar to Section \ref{sec:q}. 
	It formalize the intuition that any individual record should not ``give out too 
	much information'' on the output of the mechanism (or vice-versa). 
	
	\item \emph{Individual privacy}, introduce in \cite{li2021privacy} is also a 
	mutual information based extension of DP which can encapsulate the (deterministic) 
	secure multiparty computations \cite{goldwasser1984probabilistic}, hence, we exclude 
	it due to the lack of clear semantic privacy guarantees.	
	
	\item \emph{Differential identifiability}, introduced
	in \cite{lee2012differential}, bounds the probability that a given
	individual's information is included in the input datasets but does not
	measure the ``change'' in probabilities between the two alternatives.
	As such, it does not provide any guarantee against Bayesian adversaries%
	\footnote{Although it was reformulated in \cite{li2013membership} as 
		an instance of membership privacy introduced in Section \ref{sec:f}.}.
	
	\item \emph{Crowd-blending privacy}, introduced in \cite{gehrke2012crowd},
	combines differential privacy with $k$-anonymity. As it is strictly weaker
	than any mechanism which always returns a $k$-anonymous dataset, the
	guarantees it provides against a Bayesian adversary are unclear. It is
	mainly used to show that combining crowd-blending privacy with
	pre-sampling implies zero-knowledge privacy \cite{gehrke2012crowd,lui2015outlier}.
	
	\item \emph{$(k,\eps)$-anonymity}, introduced in \cite{holohan2017k}, first
	performs $k$-anonymisation on a subset of the quasi identifiers and
	then $\eps$-DP on the remaining quasi-identifiers with different settings
	for each equivalence class of the $k$-anonymous dataset. The
	semantic guarantees of this definition are not made explicit.
	
	\item \emph{Integral privacy}, introduced in \cite{torra2016integral} looks at the 
	inverse of the mechanism $\mecha$, and require all outputs $O$ to be generated by 
	a large and diverse set of databases $D$. This could capture definitions like 
	$k$-anonimity, but offer no semantic guarantee in the Bayesian sense. 
	
	\item \emph{Membership privacy}%
	\footnote{Another definition with the same name is introduced in 
		\cite{li2013membership}, we mention it in Section \ref{sec:f}}, 
	introduced in \cite{sablayrolles2019white} is tailored to membership 
	inference attacks on machine learning models; the guarantees it provides are not clear.
	
	\item \emph{Posteriori DP}, introduced in \cite{wang2014tradeoff}, compares
	two posteriors in a way similar to inferential privacy in Section \ref{sec:f}, 
	but does not make the prior (and thus, the attacker model) explicit.
	
	\item \emph{$\eps$-privacy}%
	\footnote{Another definition with the same name is introduced earlier in this section.}, 
	introduced in \cite{partovi2020ensuring}, adopts DP to location data and protect only 
	the secret locations, but it only offers absolute bounds irrespectively of the prior.
	
	\item \emph{Noiseless privacy}%
	\footnote{Another definition with the same name is introduced in 
		\cite{duan2009privacy,bhaskar2011noiseless}, we mention 
		it in Section \ref{sec:b}.}, 
	and \emph{measure of privacy}, introduced in \cite{farokhi2019noiseless} and \cite{farokhi2019development} limits the change in the number of
	possible outputs when one record in the dataset changes and determines privacy using non-stochastic information theory respectively. 
	Consequently, they do not bound the change in ``probabilities'' of the mechanism, 
	so do not seem to offer clear guarantees against a Bayesian adversary.
	
	\item \emph{Weak DP}, introduced in \cite{wang2020differential}, adapts 
	DP for streams, but it only provides a DP guarantee for the 
	``average'' of all possible mechanism outputs%
	\footnote{It also assumes that some	uncertainty comes from the data itself, 
		similarly to definitions in Section \ref{sec:f}}, rather than for the 
	mechanism itself. Thus, its semantics guarantees are also unclear.
	
	\item \emph{Data-privacy} and \emph{multi-dimensional data-privacy}, introduce in \cite{he2018preserving} and \cite{sun2020privacy} ensures that an attacker's inference accuracy and disclosure probability is below some thresholds.
	
	\item \emph{Error Preserving Privacy}, introduced in \cite{dai2018privacy},
	states that the ``variance'' of the adversary's error when trying to
	guess a given user's record does not change significantly after. 
	The exact adversary model is not specified.
\end{itemize}

\subsubsection*{Variants of sensitivity}

A important technical tool used when designing differentially 
private mechanisms is the  ``sensitivity'' of the function 
that we try to compute. There are many variants to the initial 
concept of global sensitivity \cite{dwork2006calibrating}, including
local sensitivity \cite{nissim2007smooth},
smooth sensitivity \cite{nissim2007smooth},
restricted sensitivity \cite{blocki2013differentially},
empirical sensitivity \cite{chen2013recursive},
empirical differential privacy%
\footnote{Even though it is introduced as a variant of DP, it was later 
	shown to be a measure of sensitivity \cite{charest2016meaning}.}%
\footnote{Another definition with the same name is introduced in 
	\cite{burchard2019empirical}, we mention it in Section \ref{sec:b}.} 
\cite{abowd2013differential},
recommen-dation-aware sensitivity \cite{zhu2013differential},
record and correlated sensitivity \cite{zhu2015correlated},
dependence sensitivity \cite{liu2016dependence},
per-instance sensitivity \cite{wang2017per},
individual sensitivity \cite{cummings2018individual},
elastic sensitivity \cite{johnson2018towards},
distortion sensitivity \cite{Bansal2019ExtendingTF}, and
derivative \& partial sensitivity 
\cite{laud2018achieving,mueller2021partial} respectively.
We did not consider these notions as these do not modify the actual definition
of differential privacy.

\subsection{Local model and other contexts}\label{sec:local}

In this book we focused on DP modifications typically used in the
``global model'', in which a central entity has access to the whole dataset.
It is also possible to use DP in other contexts, without formally changing the
definition. The main alternative is the ``local model'', where each
individual randomizes their own data before sending it to an aggregator. This
model is surveyed in \cite{yang2020local}, and 
formally introduced in \cite{duchi2013local}.

Many definitions we listed were initially presented in the local model, such as 
$d_\datasets$-privacy \cite{chatzikokolakis2013broadening}, 
geo-indistinguishability \cite{andres2013geo},
earth mover's Pr \cite{fernandes2018generalized}, 
location Pr \cite{elsalamouny2016differential}, 
profile-based DP \cite{geumlek2019profile}, 
input-discriminative DP \cite{gu2020providing}, 
divergence DP and smooth DP from \cite{barber2014privacy},
and extended DP, distribution Pr, and extended distribution Pr from \cite{kawamoto2019esorics}. 
Below, we list the definitions that are the same as previously listed
definitions, but used in a different attacker setting; the list also includes
alternatives to the local and global models.

\begin{itemize}
	\item In \cite{shi2011privacy}, the authors introduce \emph{distributed DP},
	which corresponds to local DP, with the additional assumption that only a
	portion of participants are honest.
	
	\item In \cite{kearns2014mechanism}, the authors define \emph{joint DP}, to
	model a game in which each player cannot learn the data from any other
	player, but are still allowed to observe the influence of their data
	on the mechanism output. 
	
	\item In \cite{wu2016inherit}, authors define a
	slightly different version of joined DP, called \emph{multiparty DP}, in which
	the view of each ``subgroup'' of players is differentially private in
	respect to other players inputs.
	
	\item In \cite{bittau2017prochlo}, the authors define \emph{DP in the
		shuffled model}, which falls in-between the global and the local model:
	the local model is augmented by an anonymous channel that randomly
	permutes a set of user-supplied messages, and differential privacy is only
	required of the output of the shuffler.
	
	\item In \cite{basu2019differential}, the authors defined \emph{local DP for bandits},
	a local version of \emph{instantaneous DP} (mentioned in Section\ref{sec:n}).
	
	\item In \cite{li2019differentially}, the authors define \emph{task-global
		DP} and \emph{task-local DP}, which are equivalents of element-level DP 
	(mentioned in Section \ref{sec:n}) in a meta-learning context.
	
	\item In \cite{murakami2018restricted}, the authors define
	\emph{utility-optimized local DP}, a local version of one-sided
	differential privacy (mentioned in Section \ref{sec:n}) which additionally
	guarantees that if the data is considered sensitive, then a certain set of
	outputs is forbidden.
	
	\item In \cite{dobbe2018customized,nie2018utility,acharya2019contex,shen2021pldp}, the
	authors define \emph{personalized local DP}, a local version of
	personalized DP (mentioned in Section \ref{sec:v}).
	
	\item In \cite{alvim2018local}, the authors define
	\emph{$d_{\datasets}$-local DP}, a local version of $d_{\datasets}$-DP
	(mentioned in Section \ref{sec:v}); this was defined as
	\emph{condensed local DP} in \cite{gursoy2019secure}.
	
	\item In \cite{jiang2018context}, the authors define \emph{localized
		information privacy}, a local version of information privacy (mentioned in
	Section \ref{sec:f}).
\end{itemize}

\subsection{Related work}

Some of the earliest surveys focusing on DP summarize algorithms achieving DP 
and applications \cite{dwork2008differential,dwork2009differential}. The more 
detailed ``privacy book'' \cite{dwork2014algorithmic} presents an in-depth 
discussion about the fundamentals of DP, techniques for achieving it, and 
applications to query-release mechanisms, distributed computations or data streams. 
Other recent surveys \cite{nelson2019chasing,rajendran2021novel} focuses on the 
release of histograms and synthetic data with DP and describing existing 
privacy metrics and patterns while providing an overall view of different 
mathematical privacy preserving framework.

In \cite{heurix2015taxonomy}, the authors classify different privacy enhancing
technologies (PETs) into 7 complementary dimensions. Indistinguishability falls
into the ``Aim'' dimension, but within this category, only $k$-anonymity and
oblivious transfer are considered; differential privacy is not mentioned.
In \cite{aghasian2018user}, the authors survey privacy concerns, measurements
and privacy-preserving techniques used in online social networks and recommender
systems. They classify privacy into 5 categories; DP falls into
``Privacy-preserving models'' along with e.g., $k$-anonymity.
In \cite{wagner2018technical} the authors classified 80+ privacy metrics into 8
categories based on the output of the privacy mechanism. One of their classes is
``Indistinguishability'', which contains DP as well as several variants. Some
variants are classified into other categories; for example R\'enyi DP is
classified into ``Uncertainty'' and mutual-information DP into
\emph{Information gain/loss}. The authors list 8 differential privacy variants;
our taxonomy can be seen as an extension of the contents of their work (and in
particular of the ``Indistinguishability'' category).

In \cite{wang2014tradeoff}, authors establish connections between 
differential privacy (seen as the additional disclosure of an 
individual's information due to the release of the data), 
``identifiability'' (seen as the posteriors of recovering the original 
data from the released data), and ``mutual-information privacy'' (which
measures the average amount of information about the original dataset contained
in the released data).

The relation between the main syntactic models of anonymity and DP was studied
in \cite{clifton2013syntactic}, in which the authors claim that the former is
designed for privacy-preserving data publishing (PPDP), while DP is more
suitable for privacy preserving data mining (PPDM).
The survey \cite{zhang2020correlated} investigate the issue of privacy loss 
due to data correlation under DP models and classify existing literature into 
three categories: using parameters to describe data correlation, using models to describe data correlation, and describing data correlation based on the framework.

Other surveys focus on location privacy.
In \cite{machanavajjhala2018analyzing}, the authors highlight privacy concerns in
this context and list mechanisms with formal provable privacy guarantees; they
describe several variants of differential privacy for streaming (e.g.,
pan-privacy) and location data (e.g., geo-indistinguishability) along with
extensions such as pufferfish and blowfish privacy.
In \cite{chatzikokolakis2017methods}, the authors analyze different kinds of
privacy breaches and compare metrics that have been proposed to protect location
data.

Finally, the appropriate selection of the privacy parameters for DP was also exhaustively
studied. This problem in not trivial, and many factors can be considered:
in \cite{hsu2014differential}, the authors used economic incentives,
in \cite{lee2011much,krehbiel2019choosing,pejo2019together}, the authors looked 
at individual preferences, and in \cite{liu2019investigating,laud2019interpreting}, the
authors took into account an adversary's capability in terms of hypothesis
testing and guessing advantage respectively. 

%% file: Conclusion.tex
\section{Conclusion}\label{sec:conc}

We proposed a classification of DP variants and extensions using the concept of
dimensions. When possible, we compared definitions from the same dimension, and
we showed that definitions from the different dimensions can be combined to form
new, meaningful definitions. In theory, it means that even if there were only
three possible ways to change a dimension (e.g., making it weaker or stronger), this 
would result in $3^7=2187$ possible definitions: the $\approx225$ existing definitions 
shown in Figure~\ref{fig:timeline} are only scratching the surface of the space of
possible notions. Using these dimensions, we unified and simplified the
different notions proposed in the literature. We highlighted their properties
such as composability and whether they satisfy the privacy axioms by either
collecting the existing results or creating new proofs, and whenever possible,
we showed their relative relations to one another. We hope that this work will
make the field of data privacy more organized and easier to navigate, especially
for new practitioners.

\subsection*{Acknowledgments}

The authors would like to thank James Bailie, David Basin, Alex Kulesza,
Esfandiar Mohammadi, Jeremy Seeman, and the anonymous reviewers for their
helpful comments. This work was partially funded by Google, and started when
Bal\'azs Pej\'o was at University of Luxembourg.